\documentstyle[twoside,fleqn,epsfig]{article}
\newcommand{\be}{\begin{equation}}
\newcommand{\ee}{\end{equation}}
\newcommand{\bea}{\begin{eqnarray}}
\newcommand{\eea}{\end{eqnarray}}

\newcommand{\AmS}{{\protect\the\textfont2
  A\kern-.1667em\lower.5ex\h\citebox{M}\kern-.125emS}}
\newcommand{\lsim}{\mathrel{\mathop{\kern 0pt \rlap
  {\raise.2ex\hbox{$<$}}}
  \lower.9ex\hbox{\kern-.190em $\sim$}}}
\newcommand{\gsim}{\mathrel{\mathop{\kern 0pt \rlap
  {\raise.2ex\hbox{$>$}}}
  \lower.9ex\hbox{\kern-.190em $\sim$}}}
\def\Journal#1#2#3#4{{#1} {#2} (#3) #4 }

\def\APP{\em Astrop. Phys.}
\def\NPA{{\em Nucl. Phys.} A}
\def\NCAA{{\em Il Nuovo Cimento} A}
\def\NCC{{\em Il Nuovo Cimento} C}

\def\NIMA{{\em Nucl. Instr. \& Methods} A}
\def\NJP{\em New Journal of Physics}

\def\NPB{{\em Nucl. Phys.} B}

\def\PLB{{\em Phys. Lett.} B}

\def\PRL{\em Phys. Rev. Lett.}
\def\PREV{\em Phys. Rev.}

\def\PRA{{\em Phys. Rev.} A}
\def\PRD{{\em Phys. Rev.} D}
\def\PRC{{\em Phys. Rev.} C}

\normalsize
\begin{document}

\begin{flushright}
{\bf ROM2F/2003/13 published on Riv. N. Cim. 26 n.1 (2003) 1-73}\\
\end{flushright}

\vspace{0.5cm}

\begin{center}
\Large \bf
Dark Matter search \\
\rm
\end{center}

\vspace{0.5cm}
\normalsize

\noindent \rm
R.\,Bernabei,~P.\,Belli,~F.\,Cappella,~R.\,Cerulli,~F.\,Montecchia\footnote{also: 
Universita' "Campus Biomedico" di Roma, 00155, Rome, Italy},~F.\,Nozzoli

\noindent {\it Dip. di Fisica, Universita' di Roma ''Tor Vergata"
and INFN, sez. Roma2, I-00133 Rome, Italy}

\vspace{3mm}

\noindent \rm A.\,Incicchitti,~D.\,Prosperi

\noindent {\it Dip. di Fisica, Universita' di Roma ''La Sapienza"
and INFN, sez. Roma, I-00185 Rome, Italy}

\vspace{3mm}

\noindent \rm C.J.\,Dai,~H.H.\,Kuang,~J.M.\,Ma,~Z.P.\,Ye\footnote{also:
University of Zhao Qing, Guang Dong, China}

\noindent {\it IHEP, Chinese Academy, P.O. Box 918/3, Beijing 100039, China}

\vspace{1cm}

\normalsize

\begin{abstract}

Main arguments on the Dark Matter particle direct detection approach 
are addressed on the basis of the work and of the results of the $\simeq$ 100 kg 
highly radiopure NaI(Tl) DAMA experiment (DAMA/NaI), which has been operative 
at the Gran Sasso National Laboratory of the I.N.F.N.
for more than one decade, including the preparation. 
The effectiveness of the WIMP model independent annual modulation signature 
is pointed out by discussing the results obtained over 7 annual cycles 
(107731 kg $\cdot$ day total exposure); the WIMP presence in the galactic halo is strongly supported 
at 6.3 $\sigma$ C.L. The complexity of the corollary model
dependent quests for a candidate particle is also addressed and several of the many possible
scenarios are examined. 
\end{abstract}

{\it Keywords:} Dark Matter; WIMPs; underground Physics

{\it PACS numbers:} 95.35.+d

\section{The physical problem}

\subsection{Evidence for Dark Matter in the Universe}

The first evidence that much more than the visible matter
should fill the Universe dates back to
1933 when F. Zwicky measured the
dispersion velocity
in the Coma galaxies \cite{Zwi}. 
This was soon after confirmed by S. Smith 
studying the Virgo cluster \cite{Smi}. Nevertheless, 
only about 50 years later the fact that Dark Matter should be 
present in large amount in our Universe finally reached a wide consensus.

Particular contribution was given in the seventies by two
groups which systematically analysed the dispersion velocity in
many spiral galaxies \cite{Spi}: in fact, the velocity curves in the galaxy
plane as a function of distance from the galactic center 
stay flat even outside the luminous disk,
crediting the presence of a dark halo.
Several other experimental evidences for the Dark Universe have been 
pointed out by the progresses -- with time passing -- in the astronomical observations,
such as: i) the Large Magellanic Cloud spins around our Galaxy faster 
than expected in case only luminous matter would be present;
ii) the observation of X-ray emitting gases surrounding elliptical galaxies;
iii) the velocity distribution of hot intergalactic plasma in clusters.
All these observations have further supported that the mass of the Universe should be much
larger than the luminous one in order to explain the observed gravitational
effects.

The existence of the Dark Universe is supported also by the standard cosmology
(based on the assumption that the Universe arose from an
initial singularity and went on expanding) in the inflationary
scenario (proposed to avoid any fine tuning in the Big Bang
initial conditions), which requires a flat Universe with density equal
to the critical one:    
$\rho_c = \frac{3H_0^2}{8 \pi G} = 1.88 h^2 \cdot 10^{-29}$ g $\cdot$ cm$^{-3}$, 
where $G$ is the Newton constant and $H_0$ is the Hubble constant equal to $100 h$ kms$^{-1}$Mpc$^{-1}$ 
and $0.55 < h < 0.75$. The uncertainty is due to the measurements 
of the actual value of the expansion rate of the Universe and 
to the considered models \cite{Cosmo}; a recent determination from the WMAP
data gives: $h = 0.72 \pm 0.05$ \cite{Wma}.

In particular, the density parameter  $\Omega= \frac{\rho}{\rho_c}$, where ${\rho}$ is
the average density of the Universe (matter + energy),  is a key parameter in the
interpretation of the data from the measurements on Cosmic Microwave Background (CMB)
since the global curvature of the Universe is related to it. The
experimental results are consistent with a flat geometry of the Universe
and, therefore, also support  $\Omega \simeq 1$  \cite{Bom}; the most recent determination 
from the WMAP gives: $\Omega = 1.02 \pm 0.02$ \cite{Wma}.
Thus, the scenario is consistent with adiabatic inflationary models and with the presence   
of acoustic oscillations in the primeval plasma and
requires the existence of Dark Matter
in the Universe since the average density of the Universe as measured
by photometric methods is: $\Omega \simeq 0.007$. 
However, the detailed composition of $\Omega$
in  term of matter, $\Omega_m$, and of energy, $\Omega_{\Lambda}$, cannot be inferred by CMB   
data alone; some information 
can be derived by introducing some other constraints \cite{Wma,Pri}.

For the sake of completeness, we also mention
that in last years studies have been performed \cite{RiPe} 
on astronomical standard candles
as supernovae type Ia, that allow to evaluate relations between redshift and distance.
These studies seem to point out an Universe whose expansion
is accelerating, crediting the possible presence of
a Dark Energy. When these
results are combined with CMB data, $\Omega_{\Lambda}$ would account for about
70\% of $\Omega$\cite{Wma,BoWm}. This form of energy,
with
repulsive gravity and possible strong implication on the future evolution of Universe,
would not be a replacement for Dark Matter and is still a mysterious task;
dedicated ground and space based experiments are planned in order to confirm
this scenario. 

Finally, as regards our Galaxy, from dynamical observations 
one can derive that it is wrapped in a dark halo, whose
density nearby the Earth has been estimated to be for example in refs. \cite{Den,Hep}:
$\rho_{halo} \simeq (0.17-1.7)$ GeV cm$^{-3}$ (see also later).

\subsection{The nature of the Dark Matter}

The investigation on the nature of the Dark Universe
has shown that large part of it should be in non-baryonic form.

In fact, as regards baryons, in the past from the theory of big-bang nucleosynthesis (BBN)
and from a
lower limit to the primordial deuterium abundance  a baryon density
$\Omega_B \lsim 0.1$ was set \cite{ScTu}. This upper limit has been  precised by recent
measurements
of primordial deuterium abundance, giving $\Omega_B h^2= 0.020\pm0.001$\cite{BuTy}, that
combined with the present determination of the Hubble constant
implies: $\Omega_B \simeq 0.04$;
the latest determination by CBM experiments: 
$\Omega_B h^2= 0.022\pm0.003$\cite{Wma,Pri},
is also in good agreement.
Recently, large efforts have been devoted to the investigation on Dark Baryonic Matter
by experiments like EROS, MACHO and OGLE, which search for massive compact
halo objects as baryonic candidates looking at microlensing effect toward
Large and Small Magellanic Clouds and toward  the Milky Way bulge.
At present, in agreement with the expectations, 
the obtained results \cite{Taup,Eros} strongly limit the possible amount of 
Galactic Dark Matter in this form.
In addition, a further argument, which also supports that
the major part of the Dark Matter in the Universe should be in non-baryonic form, is the following:
it is very difficult to build a model of galaxy formation
without the inclusion of non-baryonic Dark Matter.

Thus, a significant role should be played by non-baryonic relic particles from the 
Big Bang. They must be stable or with a lifetime comparable with the age of the Universe 
to survive up to now in a significant amount. They must be
neutral, undetectable by electromagnetic interactions and 
their cross section with ordinary matter should be weak 
(in fact, if their annihilation rate would be greater than
the Universe expansion rate, they should disappear).
The Dark Matter candidate particles are usually classified in {\it hot} Dark matter
(particles relativistic at decoupling time with 
masses $\lsim$ 30 eV)
and in {\it cold} Dark Matter (particles non relativistic at 
temperatures greater than 10$^4$ K with 
masses from few GeV to the TeV region or axions generated by
symmetry breaking during primordial Universe).
The light neutrinos are the natural candidates for {\it hot}
Dark Matter; they are strongly constrained by cosmology
and a value over the limit $\Omega_{\nu} \simeq 0.05$ gives an unacceptable
lacking of small-scale structure\cite{CrEl}.
In addition, a pure {\it hot} Dark Matter
scenario is also ruled out by the 
measurements of the CMB radiation, which does not show sufficiently
large inhomogeneity.

Thus, {\it cold} Dark Matter candidates,
which can be responsible for the initial gravitational collapse, should be present 
and in large amount, although 
a pure {\it cold} Dark Matter scenario seems to be not favoured by the
observed power spectrum of the density perturbation.
In practice, a mixed Dark Matter scenario is generally favourably considered.
However, 
other possibilities 
can be considered such as, for example, the so-called "tilted Dark matter scenario"
that introduces a significant deviation from the Zeldovich 
scale invariance of the power spectrum of the initial fluctuations. Anyhow,
in all the possible scenarios a significant fraction 
of {\it cold} Dark Matter particles is expected.

\begin{figure}[ht]
\vspace{-0.7cm} 
\centerline{\hbox{ \psfig{figure=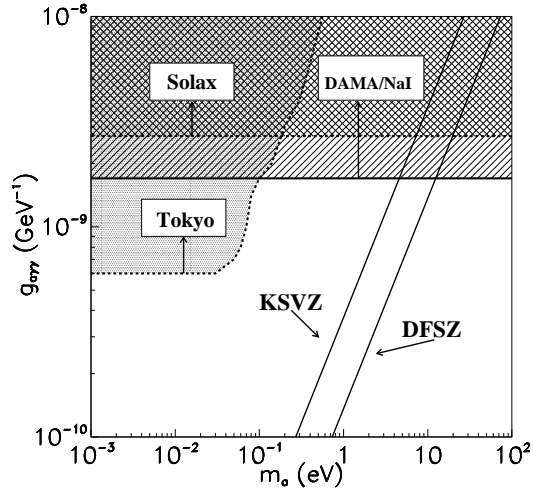,height=7.cm} }}
\vspace{-0.3cm} 
\caption{Exclusion plot in the plane axion to photon coupling constant,
$g_{a \gamma \gamma}$, versus axion mass, $m_a$, achieved by DAMA/NaI in ref. \cite{Dax}. 
The limit quoted in the paper ($g_{a \gamma \gamma}\le 1.7 \times 10^{-9}
GeV^{-1}$ at 90\% C.L.) is shown together with
the expectations of the KSVZ and DFSZ models; see ref.\cite{Dax} for details.}
\label{fig_axion}
\end{figure}

As mentioned above, {\it cold} Dark Matter can be in form of axions 
or of WIMPs (Weakly Interacting Massive Particles). 
The axions are light bosons, hypothesized 
to solve the CP problem in strong interactions.
Direct detection 
experiments are in progress since time by studying their interactions 
with strong electromagnetic fields, but
no positive evidence has been found so far \cite{Ax}.
For completeness, we mention that some experiments (including DAMA/NaI, see Fig.
\ref{fig_axion})
have also searched for possible axions produced in the Sun (see e.g. 
\cite{Dax,axion}) and that some other will be realized in near future. However, 
these latter experiments cannot be classified as experiments for Dark Matter direct
detection since they are not searching for relic axions.

For the sake of completeness, we remind that also more exotic candidates
(which generally could account for
small fraction of Dark Matter in the galactic halo)
have been considered and searched for, such as e.g. the magnetic monopoles
with mass 10$^{16}$ - 10$^{17}$ GeV \cite{Mono}, the neutral Strongly Interacting
Massive particles (SIMPs) and the neutral
nuclearities\cite{Simp,Besimp,Macnuc}, the Q-balls\cite{qballs}, etc.;
experimental
searches for such candidates have given always negative results. Some of them have also
been investigated by DAMA/NaI \cite{Besimp,qballs}.

\section{The particles searched for}
\label{sc:part}

The WIMPs are particles in thermal 
equilibrium in the early stages of the Universe, decoupled 
at freeze out temperature. 
Considering the WIMP particles as stable and 
with the same initial density for particles and antiparticles,
their annihilation cross section, $\sigma_{ann}$, should be such that 
their annihilation rate should be
lower than the expansion 
rate of the Universe:  $<\sigma_{ann} \cdot v> \simeq \frac{10^{-26}} 
{\Omega_{WIMP} \cdot h^2} cm^3s^{-1}$, where $v$ is 
the relative velocity of the particle-antiparticle pair; thus,  
the interaction cross section is of the same order 
as those known of weak interactions. In case the         
particles and antiparticles would not have the same initial  
density, this relation would represent a lower limit.

The velocity-spatial distribution of the WIMPs in our galactic halo
is not well known. So far the simplest, non-consistent and approximate 
isothermal sphere model has generally been considered in direct WIMP
searches; under this assumption the WIMPs
form a dissipationless gas trapped in the
gravitational field of our Galaxy in an equilibrium steady state and 
have a quasi-maxwellian 
velocity distribution with a cut-off at the escape velocity from 
the galactic gravitational field. 
More realistic halo models have been proposed by 
various authors such as Evans' power-law halos,
Michie models with 
an asymmetric velocity distribution,
Maxwellian halos with bulk rotation, etc. \cite{gree}. 
In particular, a devoted discussion on a wide (but still not complete)
number of consistent halo models and their implications on available experimental
data has been carried out e.g. in refs. \cite{Hep,gree}; they  will be summarized in 
\S \ref{sc:halo}.
   
At present, the most widely considered candidate for WIMP is
the lightest supersymmetric particle named neutralino, $\chi$.
In the Minimal Supersymmetric Standard Model (MSSM) where R-parity is conserved, 
the lightest SUSY particle, $\chi$, must be stable 
and can interact neither by 
electromagnetic nor by strong interactions (otherwise
it would condensate and would be detected in the galactic halo 
with the ordinary matter). The $\chi$ is defined as the
lowest-mass linear combination of photino ($\tilde{\gamma}$), 
zino ($\tilde{Z}$) and higgsinos ($\tilde{h}_{1}, 
\tilde{h}_{2}$):
$\chi = a_1 \tilde{\gamma} + a_2 \tilde{Z} + a_3 \tilde{h}_{1} 
+ a_4 \tilde{h}_{2}$
(where $\tilde{\gamma}$ and $\tilde{Z}$ 
are linear combination of U(1) and SU(2) neutral gauginos, 
$\tilde{B}$ and $\tilde{W}_3$) and
is a Majorana particle. Under some assumptions, the $\chi$ mass 
and the $a_i$ coefficients depend
on the Higgs mass mixing parameter,
$\mu$, on the $\tilde{B}$ and $\tilde{W}_3$ masses
and on $tg \beta$ (the ratio between the v.e.v's which
give masses to up and down quarks). Thus, often the theoretical estimates
and sometimes the 
experimental results are presented in terms of $\mu$, $tg \beta$ 
and wino mass, $M_2$.  
The $\chi$ cross section on ordinary matter
is described by three Feynman diagrams: i) exchange between $\chi$ and quarks
of the ordinary matter through Higgs particles (spin-independent -- SI -- interaction);
ii) exchange between $\chi$ and quarks of the
ordinary matter through Z$_0$ (spin-dependent -- SD -- interaction);
iii) exchange between $\chi$ and quarks of the
ordinary matter through squark (mixed -- SI/SD -- interaction).
The evaluation of the expected rates
for $\chi$ depends on several parameters and procedures, which are affected by significant
uncertainties, such as e.g.
the considered neutralino composition,
the present uncertainties on the measured top quark mass
and on certain sectors of the fundamental nuclear cross sections,
on some lack of information
about physical properties related to Higgs bosons and
SUSY particles, on
the possible use of
constraints from GUT schemes
and/or from
$b \rightarrow s + \gamma$ 
branching ratio, on the used rescaling procedure,
etc.; in conclusion, considering also the large number of involved parameters, 
the supersymmetric theories have unlikely no practical predictive capability.

Other candidates can also be considered as WIMPs; 
in particular, we remind an heavy neutrino of 
a 4-th family \cite{Far} and the sneutrino in the scenario described in ref. \cite{Wei01}.

The heavy neutrino of 
a 4-th family was one of the first candidate proposed 
to solve the Dark Matter problem. Still now it may be considered as
a good and realistic candidate, 
although unable to account for the whole Dark Matter missing mass.
Such a neutrino could contribute --
by its pair annihilation in the galactic halo -- to positrons, 
antiprotons and diffused gamma background
and these signatures might be better fit to the observed data \cite{Far}; moreover,
it might dominate the Higgs decay mode in near future LHC accelerator. 
\begin{figure}[!ht]
\centering
\includegraphics[width=170pt]{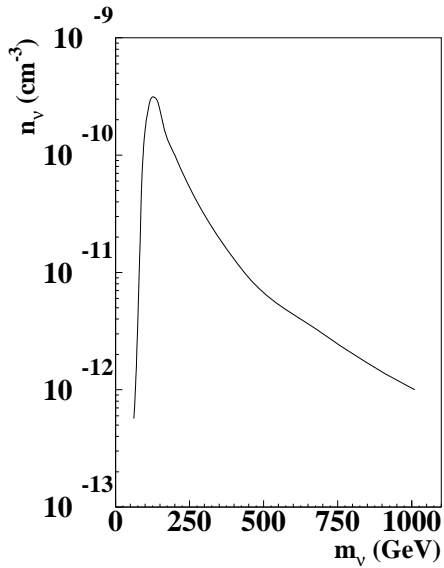}
\caption{Relic abundance of an heavy neutrino as a function of
its mass according to the calculation of ref. \cite{Fa95}; masses above 
the $Z_0$ pole are considered.}
\label{fg:relic_neut}
\end{figure}
The cosmological relic abundance of heavy neutrinos can be evaluated,
as reported in Fig. \ref{fg:relic_neut},
taking into account that the couplings are described within
the Standard Model of elementary particles. 
Applying the condition that the density of such heavy neutrinos cannot
exceed the critical density, a window in their mass can be evaluated 
\cite{neut}: 3 GeV $< m_{\nu} < $ 3 TeV. Considering the measurements of  $Z_0$ 
decay into invisible channels carried out at LEP and some implications
of the measured cosmic ray flux \cite{Fa95}, a mass range around 50 GeV
with 
a reasonable local abundance (which permits to consider it as
a Dark Matter candidate) is still open.

In some supersymmetric models the lightest supersymmetric particle (LSP)
can be the sneutrino, ${\tilde{\nu}}$, 
the spin-0 partner of the neutrino. 
In supersymmetric theories with no
violation of leptonic number, a sneutrino with mass in the range
$550 \; {\rm GeV} \lsim m_{\tilde{\nu}} \lsim 2300 \; {\rm GeV}$
could have a relevant cosmological abundance
($0.1 \lsim \Omega_{\tilde{\nu}}h^2 \lsim 1$) \cite{Fa94}; however,
because of its large interaction cross sections,
the sneutrino cannot generally be considered as major component of Cold Dark Matter. 
Anyhow, a sneutrino as a candidate remains still possible 
in supersymmetric models with violation of lepton
number \cite{Ha98}. In this framework the sneutrino 
can exist in two mass states, 
$\tilde{\nu}_{\pm}$, with a $\delta \simeq \Delta m^2/2 m_{\tilde{\nu}}$
mass splitting (for $ \Delta m^2 \ll  m_{\tilde{\nu}}^2$), being
$\Delta m^2$  a term introduced by the leptonic number violating operator.
The two mass eigenstates have off-diagonal coupling with $Z_0$ boson
and only couplings between $\tilde{\nu}_+$ e $\tilde{\nu}_-$
exist. As a consequence, the elastic scattering cross section 
on nuclei is extremely low \cite{Ha98} and sneutrinos
with mass around 40-80 GeV and $\delta$ about 5 GeV could have
cosmological relic abundance in the range 0.1-1  \cite{Ha98}. 
Moreover, whatever scalars  would be introduced in the theory, 
they can mix with sneutrinos and, consequently, the gauge 
interaction would be reduced through the mixing angle  \cite{Ar00}.
The suppression of this interaction implies a sizeable relic 
abundance of the sneutrino even for low $\delta$ values (e.g. around 
$\sim 100$ keV). 
A similar sneutrino has been proposed 
as a possible WIMP candidate providing -- through the transition
from lower to upper mass eigenstate -- inelastic 
scattering with nuclei \cite{Wei01} (see also later). 

Finally, we remind that -- in principle -- even whatever
massive and weakly interacting particle, not yet foreseen by theories, 
can be a good candidate as WIMP.

\vspace{0.4cm}

In the following we will focus our attention
on the WIMP direct detection technique in underground laboratory, where the low
environmental background allows to reach the
highest sensitivity; this is the process investigated by DAMA/NaI. We will later mention few
arguments     
on the indirect detection approach, mainly in the light of some recent analyses.

\section{Some general arguments on the WIMP direct detection approach}

The WIMP direct detection approach mainly investigates the WIMP elastic
scattering on the nuclei of a target-detector; the recoil energy is the measured quantity.
In fact, the additional possibility to 
investigate the WIMP-nucleus inelastic scattering 
producing low-lying excited nuclear states (originating successive 
de-excitation gamma rays and, thus, presence
of characteristic peaks in the measured energy spectrum) is disfavoured by the
very small expected counting rate; for this reason, 
only few preliminary efforts have been carried out so far
on this subject\cite{InJap,Bea96,Lxe20}. 

In the following subsections only few general arguments are addressed on the direct detection approach,
while we simply remind that 
most experienced detection techniques 
have already been briefly commented in ref.
\cite{erice}, mainly in the light
of a possible effective search for a WIMP signature.

\subsection{Some generalities}

A direct search for Dark Matter particles
requires: i) a suitable deep underground site to reduce at most the
background contribution from cosmic rays; ii) a suitable low background
hard shield against electromagnetic and neutron background;
iii) a deep  selection of low background materials and a suitable identification of
radio-purification techniques to build a low background set-up;
iv) severe protocols and rules for building, transporting, handling, installing 
the detectors;
v) an effective Radon removal system and control on the environment nearby the
detectors;
vi) a good model independent signature; vii) an effective monitoring of the running conditions
at the level of accuracy required by the investigated WIMP signature.

As an example of the suitable performances of a deep underground laboratory
we remind those measured 
at the Gran Sasso National Laboratory of I.N.F.N. where the DAMA/NaI experiment has been
carried out:
i) muon flux: 0.6 muons m$^{-2}$ h$^{-1}$\cite{Mac97}; ii) thermal neutron flux:
1.08 $\cdot 10^{-6}$ neutrons cm$^{-2}$ s$^{-1}$\cite{Neu89};
iii) epithermal neutron flux:
1.98 $\cdot 10^{-6}$ neutrons cm$^{-2}$ s$^{-1}$\cite{Neu89};
iv) fast ($E_n > 2.5 MeV$) neutron flux:
0.09 $\cdot 10^{-6}$ neutrons cm$^{-2}$ s$^{-1}$\cite{Cri};
v) Radon in the hall: $\simeq$ 10-30 Bq m$^{-3}$ \cite{Arpe}.

The low background technique requires very long and accurate work for 
the selection of low radioactive materials by sample
measurements with HP-Ge detectors (placed deep
underground in suitable hard shields) and/or by mass spectrometer analyses;
thus, these measurements are often 
difficult experiments themselves, depending on the required level of radiopurity.
In addition, uncertainties due to the sampling procedures  
and to the subsequent handling of the selected materials to build the
apparata also require further time and efforts. 
As an example of an investigation of materials and detector
radiopurity, one can consider 
ref. \cite{Nim98}, where the residual radioactivity measured in materials 
and detectors developed for DAMA/NaI is reported. Moreover, 
some arguments on how to further improve the radiopurity of NaI(Tl) 
detectors (largely followed e.g. in the developments of the new 
 DAMA/LIBRA set-up, now in test runs) can be found  
e.g. in ref. \cite{Valle}. An interesting paper on the low background
techniques is also e.g. ref. \cite{Heus}. 
 
Main efforts regard the reduction of 
standard contaminants:  $^{238}$U and $^{232}$Th (because of their rich chains) and 
$^{40}$K (because of its large presence in nature). When suitable radiopurity 
is reached for these components, the possible presence of
non-standard contaminants should be also seriously investigated by devoted
measurements.
As shown e.g. in ref. \cite{avi} for the case of a ionizing Ge experiment, several orders of magnitude
of rate reduction can be obtained with time and efforts in improving the
experimental conditions.

\subsection{The "traditional" model dependent approach} 
\label{trad}

Since often the used statistics in direct experiments 
is very poor, the simple comparison of the
measured energy distribution with an expectation from a given model framework is carried out.
This "traditional" approach -- the only one which can be pursued by either 
small scale or very poor duty cycle experiments -- allows only 
to calculate model dependent limits 
on WIMP-nucleus cross section at given C.L..    
In fact, although for long time the limits achieved by this approach have been 
presented as robust reference points,
similar results are quite uncertain 
not only because of possible underestimated systematics when relevant data handling and reduction is performed,
but also because the result refers only to a specific model framework.
In fact the model is identified not only by the general
astrophysical, nuclear and particle physics assumptions,
 but also by the needed theoretical and experimental parameters and by
the set of values chosen in the calculations for them.
Some of these parameters, such as the WIMP local velocity, $v_0$, and other halo
parameters, form factors' parameters, quenching factor, etc. are also affected by
significant uncertainties.
Therefore the calculation of the expected differential rate, 
which has to be compared with
the experimental one in order to evaluate an exclusion plot in the plane WIMP cross section
versus WIMP mass, is strongly model dependent. 
\begin{figure}[ht]
\centerline{\hbox{ \psfig{figure=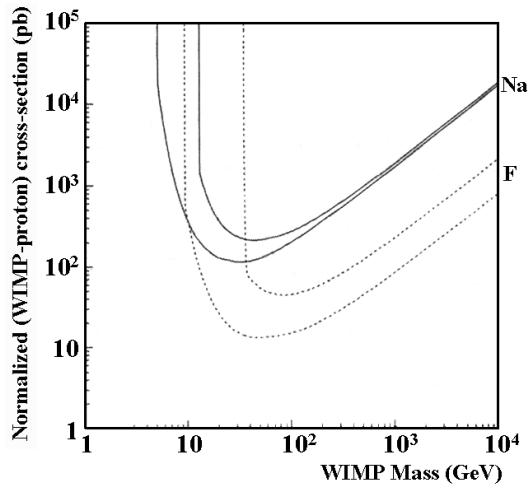,height=6.5cm} }}
\caption{Example of the effects due to the uncertainties in a given model
framework 
when calculating exclusion plots. Here the simple case for
the halo local velocity, $v_0$,  and the escape velocity, $v_{esc}$, is shown in case of
spin-dependent coupled WIMPs as from ref. \cite{Caf94}. 
The top curve
for each nucleus has been calculated -- in a given model framework -- assuming
$v_0 = 180$ km/s and $v_{esc} = 500$ km/s, while the lower one has been
calculated assuming $v_0 = 250$ km/s
and $v_{esc} = 1000$ km/s; all the considered values are possible 
at present stage of knowledge. 
Analogous effects will be found for every kind of 
experimental result when varying experimental/theoretical parameters/assumption for whatever target-nucleus.}
\label{fg:caf94}
\end{figure} 
As an example, Fig. \ref{fg:caf94} shows how an exclusion plot 
is modified by changing (within the intervals allowed by the
present determinations) the values 
of the astrophysical velocities \cite{Caf94}. Analogous effects will be
obtained when
varying -- within allowed values -- every other of the several needed 
parameters as well as when varying every one of the general assumptions considered 
in the calculations. Thus, each exclusion plot should be considered
only strictly correlated with the "cooking list" of the used experimental/theoretical 
assumptions and parameters as well as with detailed information on possible data 
reduction/selection, on efficiencies, calibration procedures, etc. 
Moreover, 
since WIMP-nucleus cross sections on different nuclei cannot directly be compared, generally 
cross sections normalized to the WIMP-nucleon one are presented; this adds further
uncertainties in the results and in the comparisons, requiring the assumptions of scaling laws
\footnote{
We take this occasion also to stress that exclusion plots given in terms of cross
sections on nucleus are not model independent as quoted sometimes "traditionally"
in literature, since they depend e.g. on the considered halo model, on the considered
nuclear form factors, etc.}.

Thus, comparisons should be very cautious since they have not an 
universal character. In addition, different experiments can have e.g. different sensitivity to
the different possible WIMP couplings. 
  
In conclusion, this model dependent approach has no general meaning, no potentiality of
discovery
and - by its nature - can give only "negative" results. 
Therefore, experiments offering model independent signature for WIMP
presence in the galactic halo are mandatory.

\subsubsection{... with electromagnetic background rejection technique}
\label{sc:rej}

In order to overcome the 
long and difficult work of developing very low background set-ups,
strategies to reject electromagnetic background from the data are sometimes
pursued. This can be realized
in several scintillators by pulse shape discrimination (since 
electrons show a different decay time respect to nuclear recoils, as carried out 
in NaI(Tl) and LXe e.g. by DAMA/NaI in ref. \cite{Psd96} and by DAMA/LXe in ref. \cite{Xe98}) 
or by comparing, for the same event, two different signals (when the
recoil/electron response ratio is expected to be different, such as heat/ionization in Ge or
Si \cite{CDMS,Edel} and
heat/light in CaWO$_4$ \cite{Cres,Rose}). The first case offers a relatively safer 
approach than the second one since basic quantities (such as e.g. the sensitive volume)
are well defined, while the second one is more uncertain. Just as an example, in case of
heat/ionization read-out  the precise  knowledge of the effective sensitive volume 
for each one of the two signals and the related efficiencies as a function of the energy
are required.
A further discrimination strategy,
which uses a two-phases gas/liquid Xenon detector with an applied electric field,
has been also suggested for future experiments; there
the light amplitudes of the primary and of the secondary scintillation pulses
are compared \cite{Clin}. However, in this case the discrimination critically depends e.g. 
on the definition of the real sensitive volume, on the dependence 
of the discrimination power with ionization position, on gas purity, etc.

In every case, whatever strategy is followed, always only a
statistical discrimination is possible (on the contrary of what is often claimed)
because e.g. of tail effects from the two populations, from the noise,
etc. Furthermore, the existence of known concurrent 
processes (due e.g. to end-range alphas, 
neutrons, fission fragments or in some case also the so--called surface electrons),
whose contribution cannot be estimated and subtracted in any
reliable manner at the needed level of precision, 
excludes that an unambiguous result on WIMP
presence can be obtained following a similar approach. 

Moreover, when using similar procedures, the real reached sensitivity  is 
based e.g. on the proper estimate of the systematic errors,
on the accuracy of all the involved procedures and
on the proper accounting of all the
related efficiencies, on the proper knowledge of the energy scale and energy threshold (see also \S \ref{sc:qf}) 
and on the verified stability of the running conditions. Consider
e.g. the difficulty to manage the efficiency due to 
the coincidence  of the few keV  heat/ionization or heat/scintillation signals or, in case of
the two-phases LXe detectors, the triggering of the
primary and secondary scintillations.
We note also that sometimes in literature some methodologically uncorrect methods are also considered 
which allow to claim for a larger sensitivity than the correct one.

In conclusion, the possibility
to achieve a control of the systematic error in rejection procedures at level of 
$\simeq 10^{-4}$, as it has recently been claimed (see \S \ref{compdir}), appears unlikely whatever
rejection approach would be considered.

Finally, it is worth to note that rejection strategies cannot safely be applied 
to the data when a model independent signature based on the correlation of the measured experimental
rate with the Earth galactic motion is pursued (see later); in fact, 
the effect searched for (which is typically at level of few \%) 
would be largely affected by the uncertainties
associated to the -- always statistical -- rejection procedure. On the other hand the
signature itself acts as an effective background rejection as pointed out e.g. for the WIMP
annual modulation signature since ref. \cite{Freese}.

\subsection{An unambiguous signature for WIMPs in the galactic halo is needed}
\label{sc:sign}

To obtain a reliable signature for WIMPs is necessary
to follow a suitable model independent approach. 
In principle, three main possibilities exist;
they are based on the correlation between the distribution of the events, detected 
in a suitable underground set-up, with the galactic motion of the Earth.

\begin{figure}[!ht]
\centering
\includegraphics[height=6.cm]{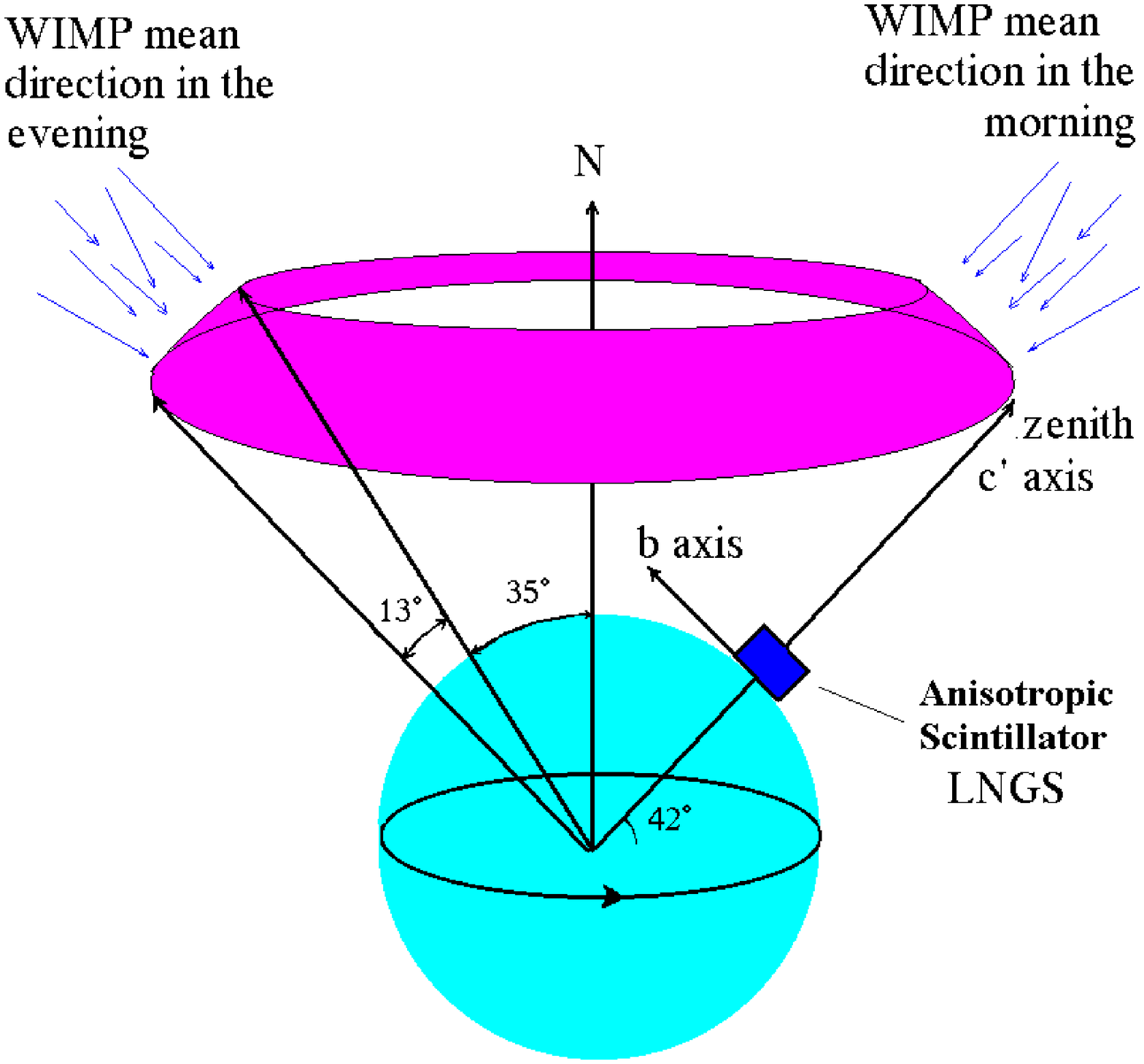}
\includegraphics[height=6.cm]{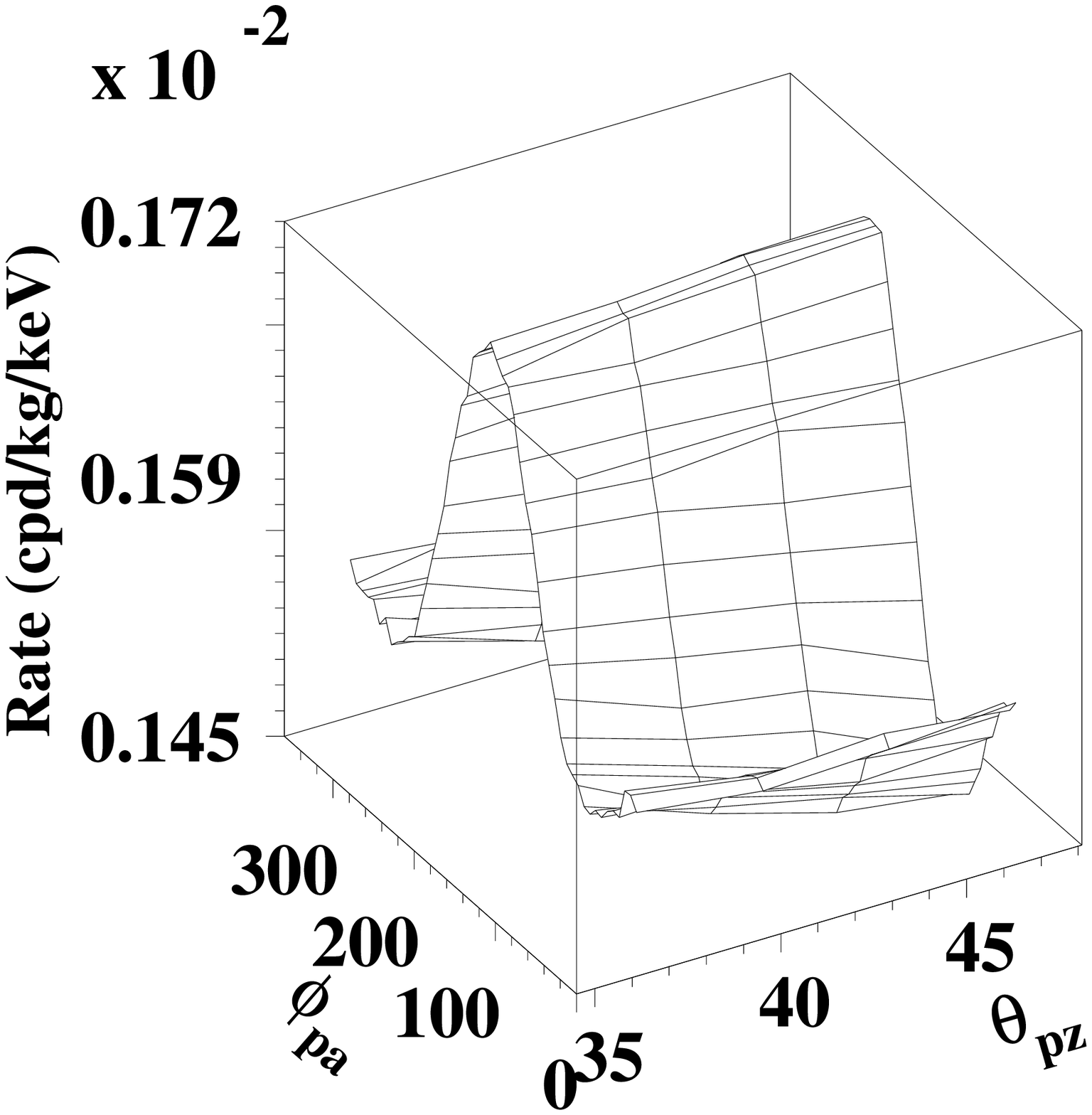}
\caption{Left: schematic representation of the experimental approach considered in ref.
\cite{anisotr} to investigate the correlation between 
the recoil direction and the Earth velocity direction by using anisotropic scintillators.
The anisotropic scintillator is placed ideally at LNGS with $c'$ axis in
the vertical direction and $b$ axis pointing to the North. The area
in the sky from which the WIMPs are preferentially expected is highlighted. Right:
expected rate, in the  3-4 keV energy
window, versus the detector (or Earth) possible velocity directions. This example refers
to the particular assumptions of a WIMP mass equal to 50
GeV,
a WIMP-proton cross section equal to
$3\cdot 10^{-6}$ pb and to the model framework of ref. \cite{anisotr}.
The dependence on the ``polar-azimuth" angle ($\phi_{pa}$) induces a diurnal variation of
the rate.}
\label{fg:direz1}
\end{figure}

The first one correlates the recoil direction with that of the Earth velocity,
but it is practically discarded mainly because of the technical difficulties in reliably and
efficiently
detecting the short recoil track. Few R\&D attempts have been carried out so far such as e.g. 
\cite{besida,Ukdir}, while a suggestion -- based on the use of
anisotropic scintillators -- was originally proposed by DAMA collaborators in
ref. \cite{Ncim} 
and recently revisited in ref. \cite{anisotr}.
As an example,
Fig. \ref{fg:direz1} (left) shows
a schematic representation of the experimental approach studied in ref. \cite{anisotr};
an example of the dependence of the expected rate on the WIMP arrival
direction, with respect to the crystal axes, for the considered experimental case
is given in Fig.\ref{fg:direz1} (right).

\begin{figure}[!ht]
\centering
\includegraphics[height=6.cm]{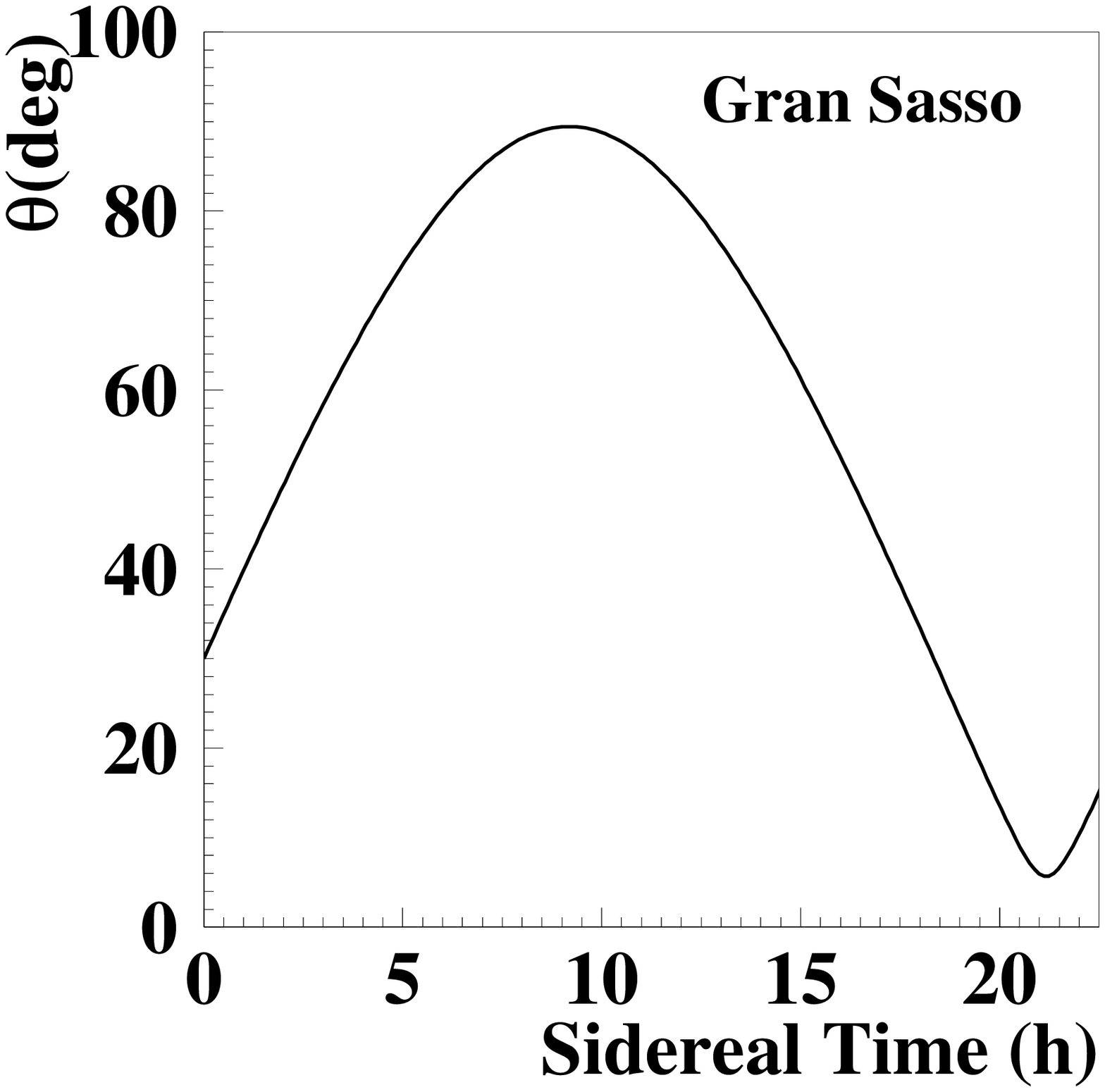}
\includegraphics[height=6.cm]{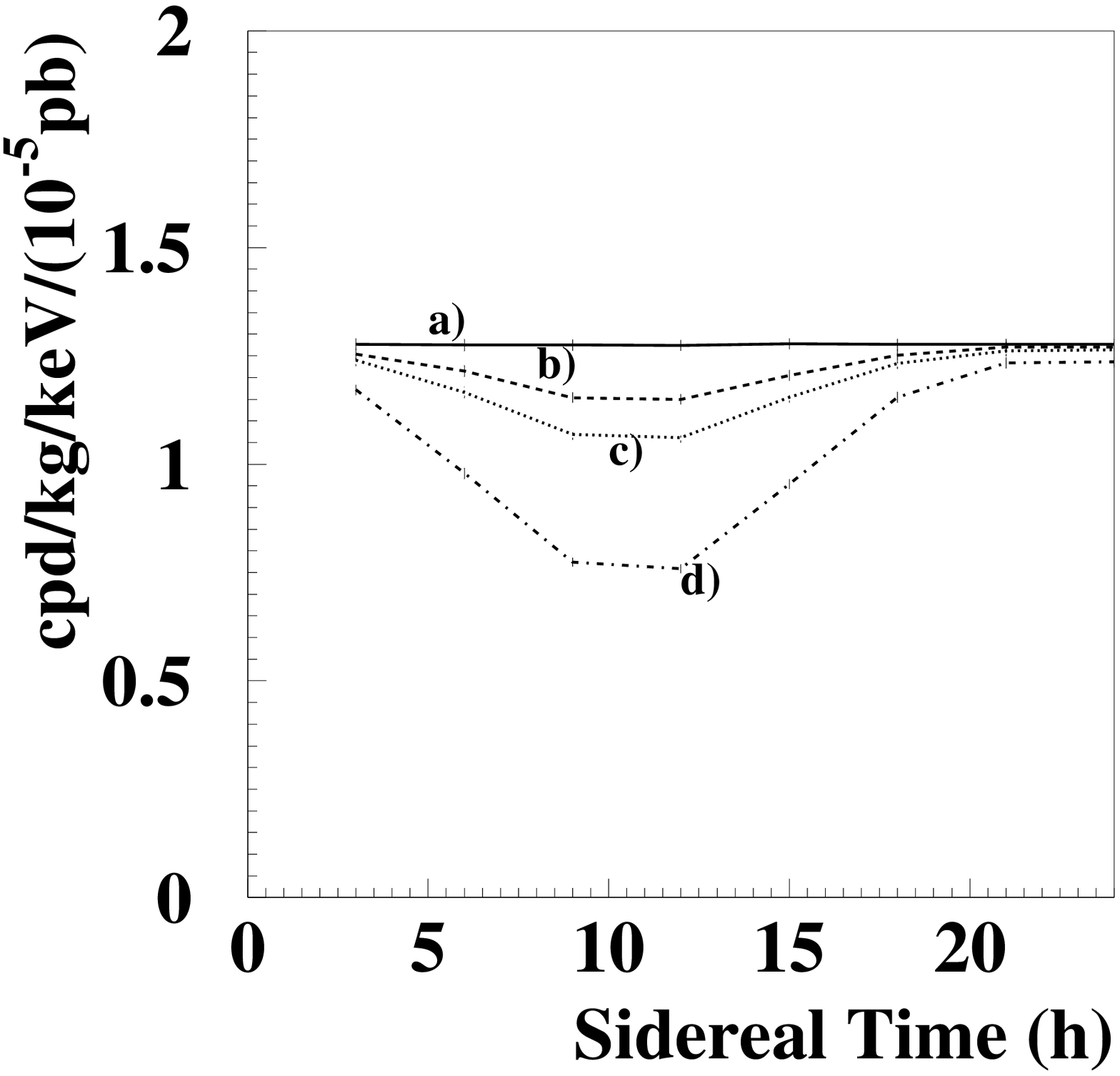}
\caption{Schematic description of the 
approach which correlates the
time occurrence of each event with the diurnal rotation of the Earth.
Left: the $\theta$ angle (defined by the Earth velocity in the Galactic frame 
with the vector joining the center of the Earth to the position of the
laboratory) as a
function of the sidereal time;
here the case for the Gran Sasso National Laboratory of the I.N.F.N.
is considered.
Right: signal rate expected in the 2--6 keV energy interval when assuming 
a 60 GeV WIMP mass, a WIMP-proton cross section equal to:
a) $7.0\cdot 10^{-6}$ pb, b) $5\cdot 10^{-2}$ pb, c) $10^{-1}$ pb, d) $1.0$ pb,
and the model framework of ref. \cite{Diu99}.}
\label{fg:diurna1}
\end{figure}

The second approach correlates the
time occurrence of each event with the diurnal rotation of the Earth.
In fact, a diurnal variation of the low energy rate in WIMP direct searches 
can be expected during the sidereal day since the Earth shields a given detector
with a variable thickness, eclipsing the WIMP ``wind'' \cite{diuav}.
However, this effect can be appreciable 
only for relatively high cross section candidates and,
therefore, it can only test a limited range of Cold Dark Matter halo density.
For a recent experimental result see e.g. ref. \cite{Diu99}, where a statistics of 14962
kg$\cdot$day
collected by DAMA/NaI has been investigated in the light of this signature. As an example
the dependence of $\theta$ (the angle defined by the Earth velocity in the Galactic frame
with the vector joining the center of the Earth to the position of the laboratory) on
the sidereal time, is shown in Fig. \ref{fg:diurna1}(left) in case of the Gran Sasso National
Laboratory location.
The expected signal rate, in case of the experimental set-up and assumptions quoted in
ref. \cite{Diu99},
is given in Fig. \ref{fg:diurna1} (right).

\begin{figure}[!ht]
\centering
\includegraphics[height=6.cm]{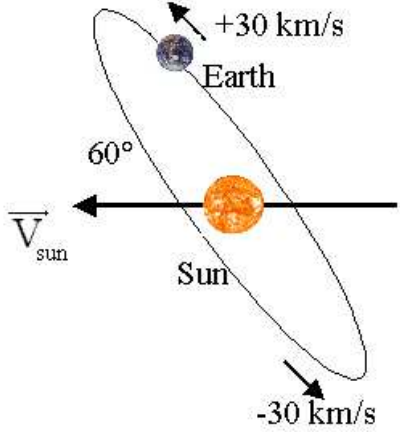}
\caption{Schematic view of the Earth motion around the Sun.}
\label{fg:annmod1}
\end{figure}

The third possibility, feasible and able to test a large interval of 
cross sections and of WIMP halo densities, is the so-called annual modulation 
signature \cite{Freese}. This is the main signature exploited by DAMA/NaI 
\cite{Mod1,Mod2,Ext,Mod3,Sist,Sisd,Inel,Hep}.
The annual modulation of the signal rate is induced by the Earth revolution
around the Sun; as a consequence, the Earth is 
crossed by a larger WIMP flux in June (when its rotational velocity is summed 
to the one of the solar system with respect to the Galaxy)
and by a smaller one in December (when the two velocities are subtracted) (see
Fig.\ref{fg:annmod1}).

In particular, the expected differential rate
as a function of the recoil energy, $dR/dE_R$ (see \S\ref{sc:rate} for detailed 
discussion), depends on the WIMP velocity distribution 
and on the Earth's velocity in the galactic frame, $\vec{v}_e(t)$.
Projecting $\vec{v}_e(t)$ on the galactic plane, one can write:
\begin{equation}
v_e(t) = v_{\odot} + v_{\oplus} cos\gamma cos\omega(t-t_0)
\end{equation}
here $v_{\odot}$ is the Sun's velocity with respect 
to the galactic halo ($v_{\odot} \simeq v_0 + 12$ km/s and $v_0$ is 
the local velocity whose value is in the range 170-270 km/s \cite{Ext,loc}); 
$v_{\oplus}$ = 30 km/s is the Earth's orbital
 velocity around the Sun on a plane with inclination
 $\gamma$ = 60$^o$ respect to the galactic plane; furthermore, 
$\omega$= 2$\pi$/T with T=1 year and roughly t$_0$ $\simeq$ 2$^{nd}$ June 
(when the Earth's speed is at maximum). 
The Earth's velocity can be 
conveniently expressed in unit of
 $v_0$: $\eta(t) = v_e(t)/v_0 = \eta_0 + 
 \Delta\eta cos\omega(t-t_0)$, 
where -- depending on the assumed value of the 
local velocity -- $\eta_0$=1.04-1.07 is the yearly average of $\eta$ and 
$\Delta\eta$ = 0.05-0.09. Since $\Delta\eta\ll\eta_0$, the expected counting 
rate can be expressed by the first order Taylor approximation:
\begin{equation}
\frac{dR}{dE_R}[\eta(t)] = \frac{dR}{dE_R}[\eta_0] +
\frac{\partial}{\partial \eta} \left( \frac{dR}{dE_R} \right)_{\eta =
\eta_0} \Delta \eta \cos\omega(t - t_0) .
\end{equation}
Averaging this expression in a $k$-th energy interval one obtains:
\begin{equation}
	S_k\lbrack\eta(t)\rbrack = S_k\lbrack\eta_0\rbrack
  + \lbrack\frac{\partial  S_k}{\partial \eta}\rbrack_{\eta_0}
\Delta\eta cos\omega(t-t_0) =S_{0,k} + S_{m,k}cos\omega(t-t_0),
\label{eq:sm}
\end{equation}
with the contribution from the highest order terms less than 0.1$\%$.
The first time-independent term is: 
\begin{equation}
S_{0,k} = \frac{1}{\Delta E_k}
\int_{\Delta E_k} \frac{dR}{dE_R}[\eta_0] dE_R,
\end{equation}
while the second term is the modulation amplitude given by:
\begin{equation}
S_{m,k} = \frac{1}{\Delta E_k} \int_{\Delta E_k} \frac{\partial}{\partial \eta}
\left( \frac{dR}{dE_R} \right)_{\eta = \eta_0} \Delta \eta
dE_R \simeq \frac{S_k[\eta_{max}] - S_k[\eta_{min}]}{2} ,
\label{eq:svlpsm}
\end{equation}
with $\eta_{max}$ = $\eta_0 + \Delta \eta$ and $\eta_{min}$
= $\eta_0 - \Delta \eta$.
The $S_{0,k}$ and $S_{m,k}$ are functions of the parameters associated with the WIMP
interacting particle (such as e.g. mass and  
interaction cross sections), of the experimental response of the detector, of the
considered model framework and of the
related parameters (see later).

It is worth to note
that the S$_{m,k}$ values can be not only positive, but also negative or zero,
due to the expected energy distribution 
profiles in June and in December within a finite energy 
window \cite{Bel96}. Therefore, the highest sensitivity can be obtained when 
considering the smallest energy bins allowed by the available
statistics in the energy region of interest.

Although the modulation effect 
is expected to be relatively small (the fractional difference between 
the maximum and the minimum of the rate is of order of $\simeq$ 7\%), 
a suitable large-mass, low-radioactive set-up with an efficient 
control of the running conditions -- such as DAMA/NaI \cite{Nim98} -- 
would point out its presence. In fact, 
a suitable correlation analysis can allow to extract 
even a small periodic component, superimposed with a time 
independent signal and a background \cite{Freese}.
With the 
present technology, the annual modulation remains the 
main signature of a WIMP signal.

In addition, the annual modulation signature is very distinctive
since a WIMP-induced seasonal effect must simultaneously satisfy
all the following requirements: the rate must contain a component
modulated according to a cosine function (1) with one year period (2)
and a phase that peaks roughly around $\simeq$ 2$^{nd}$ June (3);
this modulation must only be found
in a well-defined low energy range, where WIMP induced recoils
can be present  (4); it must apply to those events in
which just one detector of many actually "fires", since
the WIMP multi-scattering probability is negligible (5); the modulation
amplitude in the region of maximal sensitivity must be $\lsim$7$\%$ (6).
Only systematic effects able to fulfil these 6 requirements could mimic
this signature and -- as far as we know --  
no other effect investigated so far in the field of rare processes offers 
a so stringent and unambiguous signature.

Of course, the amount of the measured effect depends e.g. on the sensitivity of the 
experiment to the coupling of the WIMP candidate,
on the WIMP particle physics features, on
the nuclear features of the used tar\-get-nuc\-leus 
and on the quality of the running conditions.

\section{The DAMA experiment}

The DAMA experiment has been worked and works as an observatory for rare processes
(such as WIMP direct detection,
$\beta\beta$ decay processes, charge-non-conserving processes,
Pauli exclusion principle violating
processes, nucleon instability, solar
axions and exotics
\cite{Dax,Besimp,qballs,Bea96,Lxe20,Nim98,Psd96,Xe98,Diu99,Mod1,Mod2,Ext,Mod3,Sist,Sisd,Inel,Bel96,tutti,Xeapp,Ela99,Qf01,Hep})
by developing and using low radioactive scintillators.
It is installed deep underground in the Gran Sasso National Laboratory of I.N.F.N..

The main developed and used
experimental set-ups are: the 
$\simeq$ 100 kg NaI(Tl) set-up (DAMA/NaI) \cite{Nim98} (which has completed its data taking in July
2002), 
the $\simeq$ 6.5 kg liquid Xenon 
set-up (DAMA/LXe) \cite{Xeapp}, the so-called ``R\&D'' apparatus (DAMA/R\&D) and the new  LIBRA (Large sodium
Iodide Bulk for RAre processes; $\simeq$ 250 kg of ultra-radiopure NaI(Tl)) 
set-up (DAMA/LIBRA) whose installation has been started in fall 2002 and which is presently in test
run.
Moreover, an underground low-background germanium detector allows 
to select materials for radiopurity.

In the following the final model independent result of DAMA/NaI 
on the investigation of the WIMP annual modulation signature is discussed.
We remind that DAMA/NaI is the largest mass, highest sensitivity experiment, built before the new DAMA/LIBRA, 
having as main aim the investigation of WIMPs in the galactic halo. It was a pioneer experiment 
proposed in 1990 \cite{prop}, which has opened 
for other experiments and approaches in the field; moreover,
its results on the investigation of WIMPs in the galactic halo by the annual modulation signature 
have - by the fact - motivated the wide interest in the field arisen in recent years.

\subsection{DAMA/NaI}
\label{app}
 
The DAMA/NaI set-up \cite{Nim98}
can effectively exploit the WIMP annual modulation signature because of its
well known technology, of its high intrinsic radiopurity, of its mass,
of its suitable control of all the operational parameters and
of the deep underground experimental site.

The detailed description of the $\simeq$ 100 kg NaI(Tl) DAMA set-up,
of its radiopurity, of its performance, of the used
hardware procedures, of the determination of the experimental quantities 
and of the data reduction has been given in refs. \cite{Nim98,Mod3,Sist}.

Here we only recall that the detectors used in the annual 
modulation studies are nine 9.70 kg highly radiopure 
NaI(Tl) scintillators especially 
built for this purpose in a joint effort with Crismatec company.
The bare NaI(Tl) crystals are 
encapsulated in suitably radiopure Cu housings;
10 cm long Tetrasil-B light guides act as optical windows on the two 
end faces of the crystals and are 
coupled to specially developed 
EMI9265-B53/FL photomultipliers (PMT), which are supplied by
positive voltage with ground cathode. The two PMTs of a detector
work in coincidence and their threshold is set at the single 
photoelectron level; the measured light response is 5.5 -- 7.5 photoelectrons/keV
depending on the detector \cite{Nim98}. The software energy 
threshold has been cautiously taken at 2 keV 
\cite{Nim98,Psd96,Sist}.
The detectors are inside a low radioactivity sealed copper
box installed in the center of a low 
radioactivity Cu/Pb/Cd-foils/polyethylene/paraffin
shield.  Moreover, 
about 1 m concrete (made from the Gran Sasso
rock material) almost fully surrounds (outside the barrack)
this passive shield, acting as a further neutron moderator.
The copper box is maintained in a high purity (HP) Nitrogen 
atmosphere in slightly overpressure with respect to the external environment.
Furthermore, also the whole shield is sealed and maintained in the HP
Nitrogen atmosphere. The whole installation is air-conditioned and the temperature is monitored.
On the top of the shield a glove-box (also maintained in the HP Nitrogen
atmosphere) is directly connected to the 
inner Cu box, housing the detectors, through Cu pipes. The pipes 
are filled with low 
radioactivity Cu bars (covered by 10 cm of low radioactive Cu and 15 cm of low
radioactive Pb) which can be removed to allow the insertion
of radioactive sources for calibrating the detectors in the same 
running condition, without any contact with external air \cite{Nim98}.

An hardware/software system to monitor 
the running conditions has been operative;
in particular, several probes have been read out by the data acquisition
system and stored with the production data. Moreover, self-controlled
computer processes are operational to automatically control
several parameters and to manage alarms \cite{Nim98,Sist}.

The electronic chain and the data acquisition system used during the DAMA/NaI-0
to DAMA/NaI-5 running periods has been described in ref. \cite{Nim98}. At completion of the
DAMA/NaI-5 data taking (summer 2000) the whole electronics and DAQ have been 
completely
substituted; they are briefly summarized in the following. 
This new system has been operative during DAMA/NaI-6 to -7 running
periods, that
is up to the end of the DAMA/NaI data taking. 

The new DAQ system
has been based 
on a Digital Alpha Workstation with Digital Unix operating system interfaced with 
the VXI
and CAMAC components of the electronic chain via a GPIB bus; the acquisition 
program has been developed on the basis
on the system discussed in ref. \cite{acq} and on a specific applicative software. 
In the new configuration the HV power supply for the PMTs has been given by a CAEN 
multichannel
voltage supply with voltage stability of 0.1\%.

\begin{figure}[!ht]
\centering
\epsfig{file=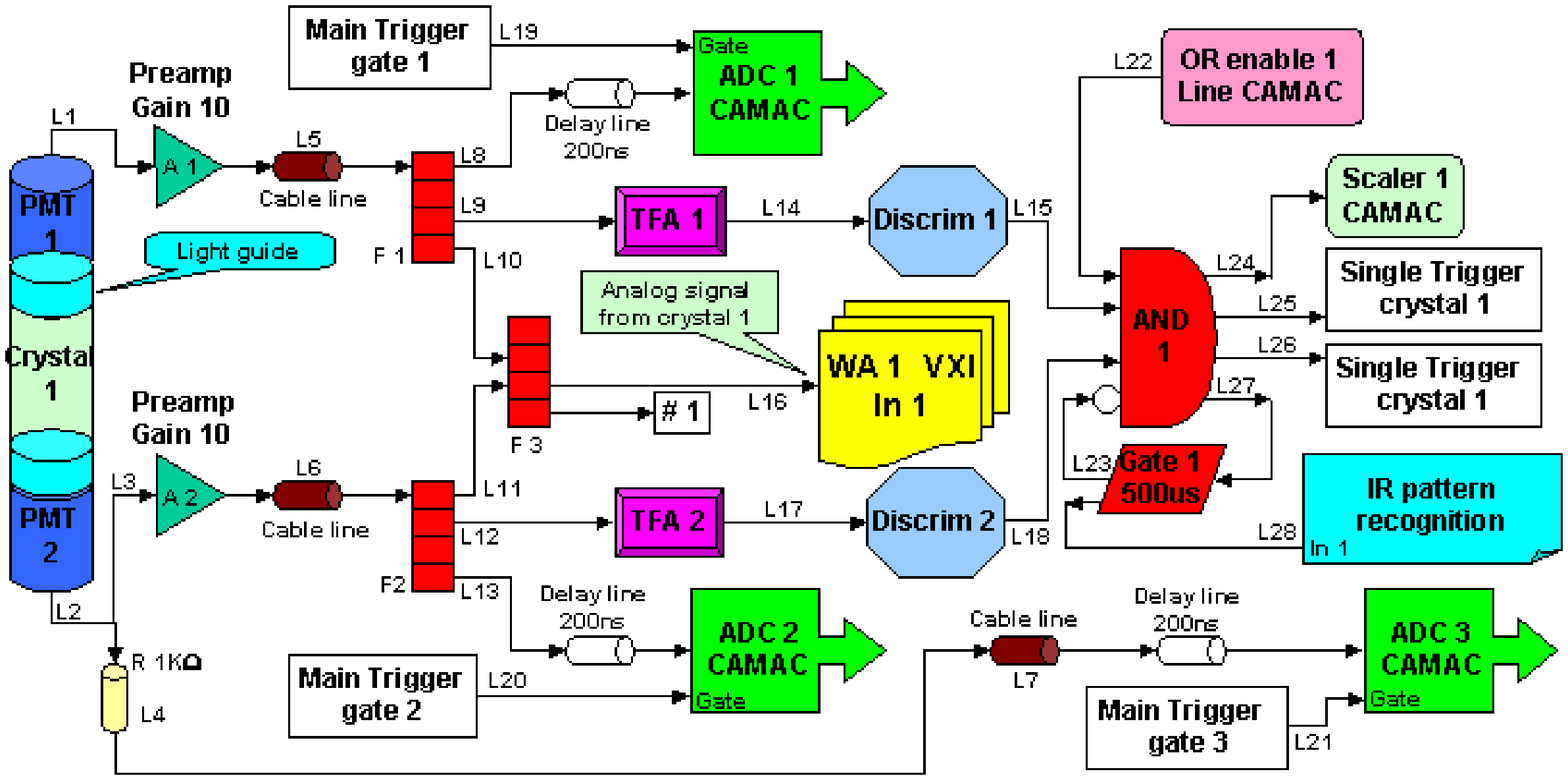,height=6.2cm}  
\epsfig{file=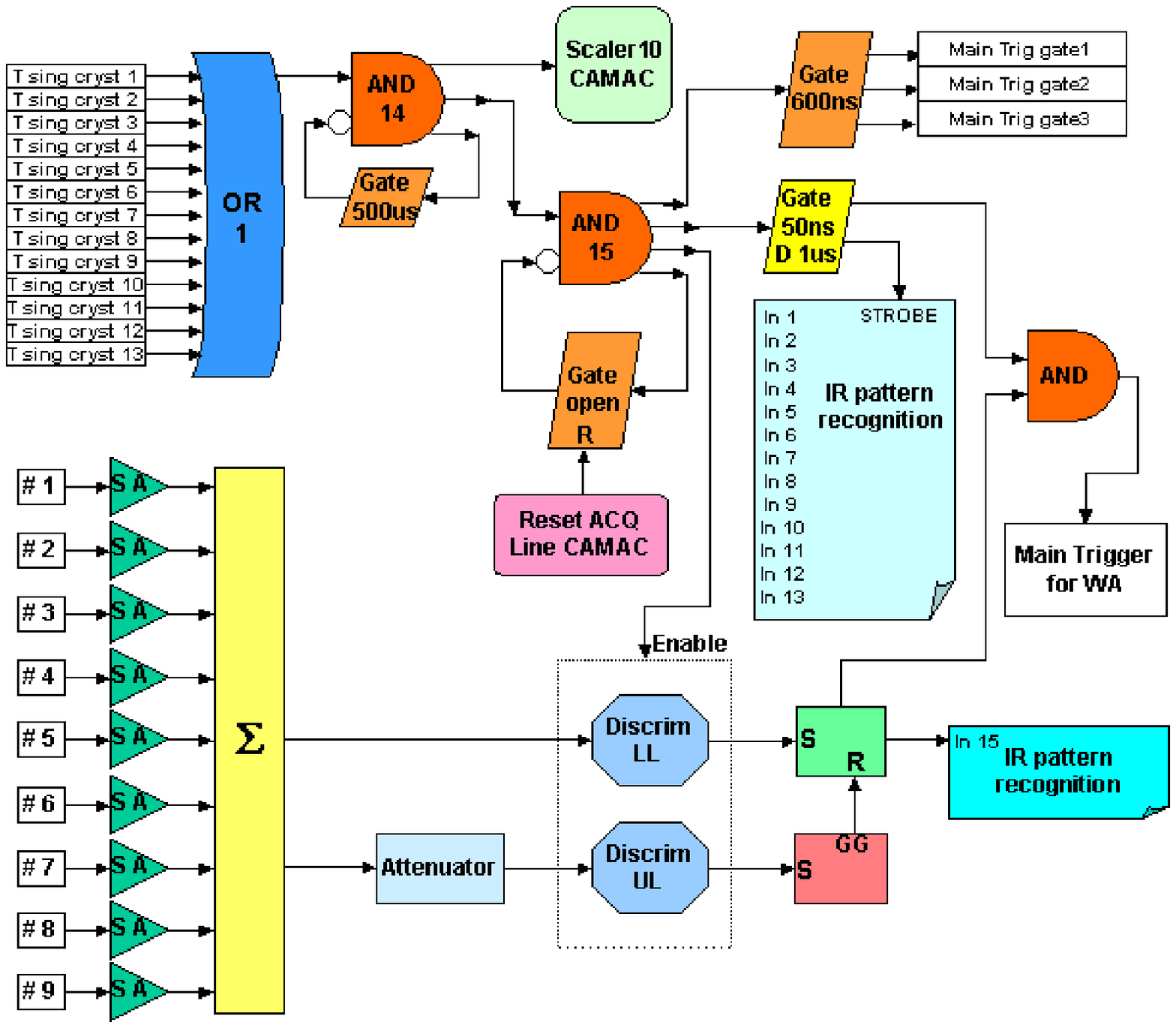,height=5.cm}
\hspace{1.cm}
\epsfig{file=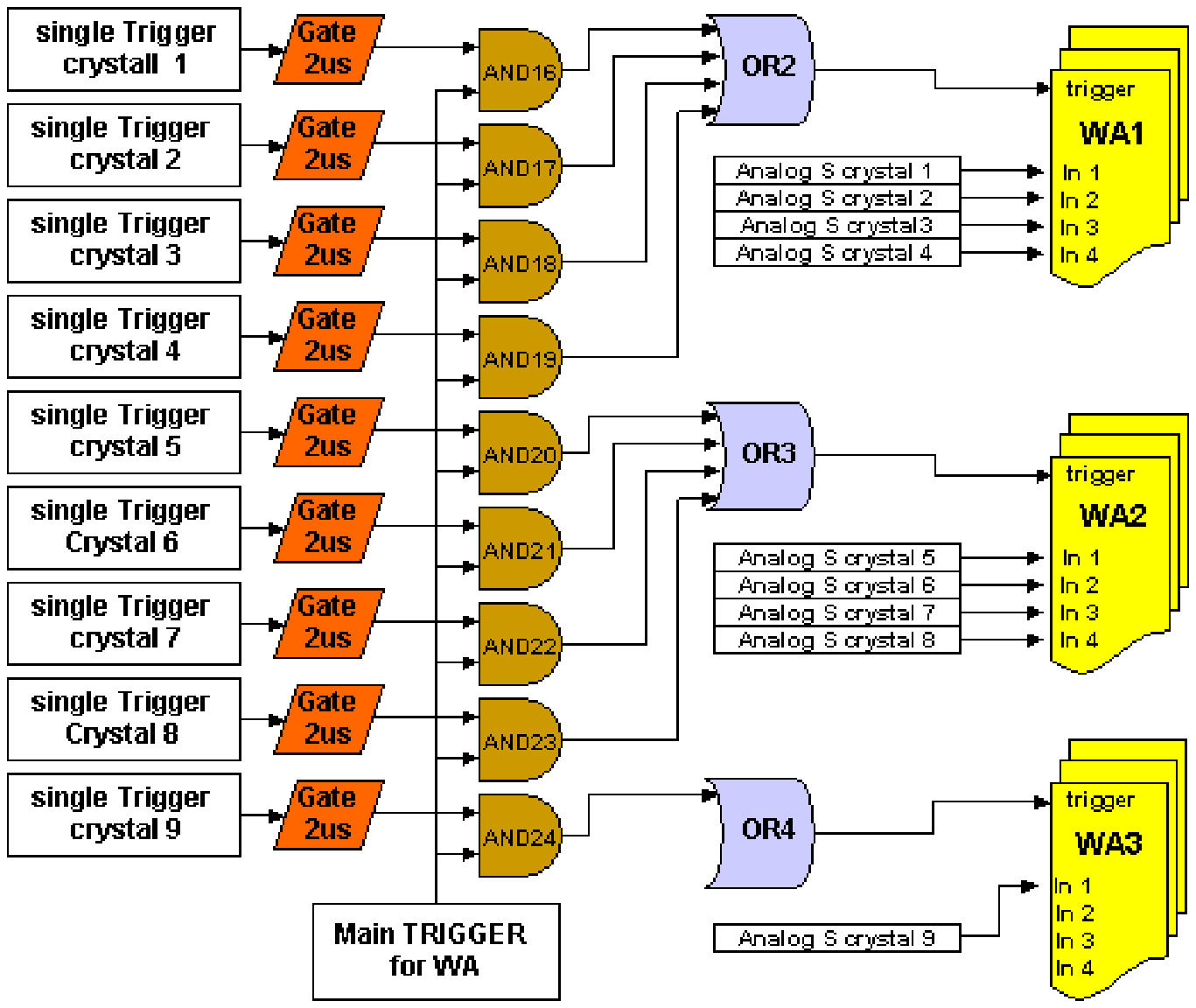,height=5.cm}
\vspace{-0.1cm}
\caption{The new electronic chain installed after the completion of the DAMA/NaI-5 
running period. Top: schema of the
electronic chain of a single detector with its trigger in the new electronic chain.
Bottom left: schema of the main trigger of the new acquisition system.
Bottom right: schema of the trigger of the new Waveform Analyzers ($WA$).}
\label{fg:st}
\vspace{-0.4cm}
\end{figure}

In Fig. \ref{fg:st} the analogic part of this new electronic chain for one
detector and its trigger are shown as well as 
the main trigger of the acquisition system and 
the trigger system of the new waveform analyzer.
We note that the analogic part and the trigger for single detector are similar to
the ones in the previous electronic chain, but the multiplexer system --
which was used in the past (having there at disposal 
only one single channel Transient Digitizer
LeCroy 8828D with 200 MSample/s sampling frequency) -- has been removed since now 
each detector has a devoted waveform analyzer ($WA$) channel. 
This is accomplished using fast VXI Tektronix 
four-channel TVS641A digitizers with a sampling frequency of 1 GSample/s 
and 250 MHz bandwidth. 
The digitizers provide a VXI word-serial protocol
for commands, while a specific Fast Data Channel (FDC) driver 
developed by \cite{dangelo} has been used to speed up the data transfer 
through the GPIB bus.
The main trigger part of the electronic schema and the high energy data 
acquisition are also similar to the ones in the previously used 
electronic chain \cite{Nim98}. 

For completeness and for template purpose, let us briefly describe this new electronic chain.
The signal $L1$ (see Fig. \ref{fg:st} for the definition of the symbols) 
from the first PMT is sent to the $A1$ preamplifier 
having 0-250 MHz bandwidth, a factor 10 gain and
a voltage integral linearity $\pm 0.2\%$. The signal 
$L2$ from the second PMT is divided in two branches: 
19/20 of the signal is sent to the input of the $A2$ preamplifier, while 
the remaining 1/20 -- suitably delayed -- feeds  
a charge ADC ($ADC3$) channel. This last part processes the 
pulses with amplitude such to saturate the remaining part of the
electronics  (they correspond to high energy events).
The preamplified signals -- through linear Fan-in/Fan-out devices --
provide the inputs for the charge ADCs (left signals and right signals)
and for the VXI waveform analyzer ($WA1$ in Fig. \ref{fg:st}) (which processes the signal 
in a 2 $\mu$s time window)\footnote{For completeness we note that the 4 
detectors named SIMP in ref. \cite{Nim98} which have been used only as
additional anticoincidence and -- sometimes -- in special triggers,
 have not been equipped in the new electronic chain with waveform analyzers.}.

The electronic devices, that
provide the trigger of a single detector, are shown in Fig. \ref{fg:st}.
In particular, 
the copies of the PMT signals are the inputs of the 
Timing Filter Amplifiers ($TFA1$ and $TFA2$) which amplify and integrate 
the signal (integration time 50 ns); their outputs are discriminated 
($Discrim1$ and $Discrim2$) with single photoelectron  
trigger level. The coincidence ($AND1$) between the two logical NIM outputs 
provides the $single$ trigger of the detector. 
The other inputs of $AND1$ are:
the signal $L23$ given by a Gate Generator ($GG1$) which allows to reject 
afterglow and Bi-Po events in a 500$\mu$s time window after the occurrence of the
event (introducing
a systematic error on the measured rate of about 10$^{-4}$) and 
the signal $L22$ given by a CAMAC I/O Register.
The latter  permits to enable or disable the $single$ detector trigger
during the calibrations.
The outputs of $AND1$ provide: i) the signal for 
a CAMAC scaler ($Scaler1$) to count the events for 
each detector; ii) the $L25$ and $L26$ used in the 
main trigger (see later); iii) $L27$ giving the start to the Gate
Generator ($GG1$) which - in addition to the veto of the coincidence --
gives the signal $L28$ issued to a 16-bit CAMAC I/R Pattern Recognition 
which allows to identify the detector or the detectors which have generated
the trigger.  

The general trigger of the acquisition -- see Fig. \ref{fg:st} -- is provided
by the logic $OR1$ of all the crystals.
The output of $AND14$ is issued to a 
Scaler, which counts the number of main triggers sent to the coincidence
$AND15$. The coincidence $AND15$ generates triggers only when
the acquisition is ready to manage them. 
Therefore, the dead time of the acquisition is properly accounted in the 
estimate of the running time by using the information from the scaler 
after $AND14$. 
When a general trigger occurs, the following logic signals are issued to: 
i) the Gate Generator ($GG2$) generating the 600 ns gates the charge ADCs;
ii) the Delay Gate Generator which gives the strobe signal to 
the I/R Pattern Recognition and generates the LAM (and, therefore, 
the interrupt to the CPU of the acquisition computer) in the CAMAC system;
iii) the Delay Gate Generator which gives the signal to the 
trigger of the waveform analyzers.
This last condition is verified only if the total energy 
deposited in the detectors is in an energy window suitably chosen
(1 to 90 keV). For this purpose, each line feds a Spectroscopy Amplifier 
whose gain is equalized in order to have the same response for 
each detector. Therefore, a Single Channel Analyzer 
made by the two discriminators, {\it Discrim LL} and {\it Discrim UL}, 
allows to select only events in the chosen energy window. 

A devoted electronic circuit \cite{Tesi}, shown in Fig. \ref{fg:st},
allows to trigger only the 
$WA$'s which correspond to fired detectors;
it gives a trigger to each $WA$ when: i) at least one of its corresponding lines
has a trigger; ii) the {\it main trigger} is present; iii) 
the total energy of the events is in the chosen energy window.
Let us remind that for the events with energy outside this energy window
(e.g. high energy events) the ADC values are acquired in any case. 

\vspace{0.4cm}

As regards other aspects, we recall that
the linearity and the energy resolution of the detectors have been 
investigated using several sources \cite{Nim98,Sist} such as,
for the low energy region, $^{55}$Fe (5.9 keV X-rays), 
$^{109}$Cd (22 keV X-rays and 88 keV $\gamma$ line) and 
$^{241}$Am (59.5 keV $\gamma$ line) sources.
In particular, in the production runs, the knowledge of 
the energy scale is assured by periodical calibrations with 
$^{241}$Am source and by monitoring (in the production data themselves
summed every $\simeq$ 7 days) the position and energy resolution 
of the 46.5 keV $\gamma$ line of the $^{210}$Pb 
\cite{Nim98,Psd96,Mod1,Mod2,Mod3,Sist}. The
latter peak is 
present -- at level of few counts per day per kg (cpd/kg) -- in the 
measured energy distributions mainly
because of a contamination (by environmental Radon) of the external surface of the crystals'
Cu housings, occurred 
during the first period of the underground storage of the detectors.
The calibration sources are introduced in
the proximity of the detectors by means of the pipes connected -- as already described --
with the upper "glove-box", which is also continuously maintained in the HP
Nitrogen atmosphere.

As in every experiment in the field, 
obvious noise events (whose number sharply decreases when increasing
the number of available photoelectrons) have to be removed; the used 
procedure has been described e.g. in refs. 
\cite{Nim98,Sist}\footnote{This procedure assures also the rejection 
of any possible contribution either from afterglows 
(when not already excluded by the dedicated 500 $\mu$s veto time; 
see above) induced by high energy events 
or from any possible $\check{C}$erenkov pulse in the
light guide or in the PMTs; in fact, they also have time decay of order of tens ns as the noise events.}. 
We remind that the noise in this experiment is given by 
PMT fast single photoelectrons with decay times of the order of tens ns,
while the "physical" (scintillation) pulses have decay times of order 
of hundreds ns. Thus, the large difference in decay times and the relatively
large number of available photoelectrons response 
assure an effective noise rejection \cite{Nim98,Mod3,Sist}. 
Several variables
can be built by using the pulse information
recorded by the waveform analyzer \cite{Nim98,Sist}).
In particular, for each energy bin, we plot
the $Y$ = $ Area(from~0~ns~to~50~ns) \over {Area(from~0~ns~to~100~ns)}$ value
versus
the $X$ = $ Area(from~100~ns~to~600~ns) \over {Area(from~0~ns~to~600~ns)}$
value calculated for every event. In the $X,Y$ plane the slow scintillation
pulses are grouped roughly
around ($X \simeq 0.7$, $Y \simeq 0.5$) well separated
from the noise population which is grouped around small $X$
and high $Y$ values (see e.g. ref. \cite{Nim98}). 
The scintillation
pulses are selected by applying an acceptance window in $X,Y$.
Since the statistical spread of the two populations
in the $X,Y$ plane becomes larger when the
number of available photoelectrons and the signal/noise ratio
decrease, windows with smaller acceptance become
necessary to maintain the same noise rejection power.
In the DAMA/NaI experiment they are kept enough stringent
to assure also the absence of any possible residual noise tail
in the scintillation data to be analysed \cite{Nim98}.
According to standard procedures, the acceptance of
the considered window for scintillation pulses in the $X,Y$ plane
is determined by applying the same procedure to the scintillation data
induced -- in the same energy intervals -- by calibration sources \cite{Nim98,Mod3,Sist}.
In particular, for this purpose, about 10$^4$ - 10$^5$ events per keV are typically
collected in the low energy region just above the 2 keV software energy
threshold during routine calibration runs \cite{Nim98,Sist}. 
All the periodical long calibration procedures \cite{Nim98,Psd96}
and the time specifically allocated for maintenance and/or for improvements 
are the main components affecting the duty cycle of the experiment.
Moreover, in the DAMA/NaI-1 running period the data have been taken only in the two extreme 
conditions for the annual modulation signature (see Table \ref{tab:modep}).

The energy threshold, the PMT gain, the electronic line stability are
continuously verified and monitored during the data taking by the 
routine calibrations, by the position and energy resolution of the $^{210}$Pb line (see
above) and by the study of the hardware rate behaviours with time.

In particular, the measured low energy distributions of interest for the WIMP investigation have been given 
in refs. \cite{Diu99,Mod3,Sist,Ela99}, where  
the corrections for 
efficiencies and acquisition dead time have already been applied.
We note that usually in DAMA/NaI the low energy distributions refer to 
those events where only one detector of many actually fires
(that is, each detector has all the 
others in the same installation
as veto; this assures a background 
reduction, which is of course impossible when a single detector is used).

\section{The first DAMA/NaI results on the annual
modulation signature} 

The presence of a model independent effect has been firstly pointed out since 
the TAUP conference in 1997 \cite{taup97} and corollary model dependent quests for a candidate
particle have been analysed in some of the many possible model
frameworks, improving the quest with time (see Table \ref{tab:modep}). 
Cumulatively during
four annual cycles a model independent effect
(exposures up to 57986 kg $\cdot$ day; see Table \ref{tab:modep}) has been pointed out
\cite{Mod3,Sist}. 
No systematics or side reactions able to mimic the annual modulation 
signature has been found \cite{Sist};
this can be well understood when considering the particularly stringent
and numerous specific requirements which identify the WIMP annual modulation signature
itself (see \S \ref{sc:sign}). 
\begin{table}[!ht]
\vspace{-0.2cm}
\caption{Summary of the first running periods 
which have already cumulatively shown a 4 $\sigma$ C.L. model
independent effect. The related references are given in the third column.
In the last column the improvements with time
in the model-dependent quest for the candidate are summarized.}
\vspace{-0.1cm}
\begin{center}
\scriptsize
\begin{tabular}{|c|c|c|c|}
\hline \hline
 & &  & Considered scenarios \\
Periods  & Statistics (kg $\times$ day) & Ref. & in corollary quests for the candidate \\
 & & & (prior on $m_W$ from\\ 
 & & & accelerator expts included)\\
\hline \hline
           &                  &               & WIMP pure SI, \\
           &                  &               & Isothermal spherical halo, \\
DAMA/NaI-1 & 3363.8 (winter)  & \cite{Mod1}   & $v_0 = 220$ km/s, \\
           & + 1185.2 (summer) &              & Helm Form Factor, \\
           &                  &               & All the parameters fixed \\
           &                  &               & to their central value \\
\hline
           &                  &               &      as DAMA/NaI-1          \\
DAMA/NaI-2 & 14962            & \cite{Mod2},     &   + halo (co-)rotation   \\
 & (Nov. $\rightarrow$ end of July) & \cite{Ext} &   + uncertainty on $v_0$ \\
\hline
DAMA/NaI-3 & 22455            & \cite{Mod3}   &   as DAMA/NaI-2 \\ 
           & (middle Aug. $\rightarrow$ end of Sept.) &     &   + prior from DAMA/NaI-0 \\
\hline
DAMA/NaI-4 & 16020            & \cite{Mod3}   & as DAMA/NaI-3   \\ 
           & (middle Oct. $\rightarrow$ middle Aug.) &   &     \\
\hline
           &                  & \cite{Mod3},  & as DAMA/NaI-3 and DAMA/NaI-4 \\
           &                  & \cite{Sist},  & + SI\&SD, ``inelastic'' \\
TOTAL      & 57986            & \cite{Sisd},  &  + Other consistent halo models,\\ 
           &                  & \cite{Inel},  &  SD form factor from \cite{res97} \\
           &                  & \cite{Hep}    &  Uncertainties on some parameters\\
\hline
           &                  &               & Limits on recoils measured \\
+ DAMA/NaI-0 & 4123.2         & \cite{Psd96}  & by pulse shape  \\
           &                  &               & discrimination  \\
\hline\hline
\end{tabular}
\end{center}
\label{tab:modep}
\vspace{-0.4cm}
\end{table}
No other experiment has at present suitable sensitivity, mass and control of the
running
conditions to effectively exploit the WIMP annual modulation signature as DAMA/NaI.

As mentioned, the implications of the observed model independent effect have
been in addition studied also under some -- of the many  possible -- 
different model--dependent frameworks. In fact, some scenarios for purely 
spin-independent (SI), purely spin-dependent (SD) \footnote{For the sake of completeness, 
we comment that JHEP 0107 (2001) 044 is not at all in conflict with a possible SD solution
since it considered only two particular purely SD couplings (of the many possible) in a strongly
model dependent context and using modulation amplitudes valid instead only in a particular
purely SI case.},
mixed SI and SD coupled WIMPs and also WIMPs with {\em preferred inelastic} scattering 
\cite{Mod1,Mod2,Ext,Mod3,Sist,Sisd,Inel,Hep} have been considered, including in the data 
analyses the lower bound 
on the mass of the supersymmetric candidate
as derived from the LEP data in the usually adopted supersymmetric
schemes based on GUT assumptions as e.g. in ref. \cite{Dpp0}.  
This corollary investigation on the quest for a candidate particle 
has been improved with time in several aspects as summarized in 
Table \ref{tab:modep}. 
Theoretical implications of these results in terms
of a neutralino with dominant SI interaction
have been discussed e.g. in ref. \cite{Botdm,Ar02} for some theoretical model frameworks
and in terms of an heavy neutrino of the fourth family
in ref. \cite{Far}.

\subsection{Comparison with some model dependent results}

\subsubsection{... from direct searches}
\label{compdir}

As mentioned above no other experiment directly comparable with the model independent 
DAMA/NaI result on WIMPs in the galactic halo is available at present.

Only few experiments \cite{CDMS,Edel,UKXe},
which use different target nuclei and
different methodological approaches,
have released extremely poor statistics
following the so-called model dependent "traditional'' approach (see 
section \ref{trad}). We have reported several 
times (see e.g. ref. \cite{Conf}) some specific arguments;
here we only summarize in Table \ref{tb:modepcomp1} some main
items.

\begin{table}[!ht]
\caption{Features of the first DAMA/NaI results on the WIMP annual modulation signature 
(57986 kg $\times$ day exposure) \cite{Mod1,Mod2,Ext,Mod3,Sist,Sisd,Inel,Hep}
with those of refs. \cite{CDMS,Edel,UKXe}. See text. Here (as well as in the text) keV always means 
keV electron equivalent if not otherwise mentioned.}
\vspace{-0.4cm}
\begin{center}
\scriptsize
\begin{tabular}{|c|c|c|c|c|}
\hline \hline
  &  &  & & \\
  & DAMA/NaI  & CDMS-I & Edelweiss-I & Zeplin-I \\
  &  &  & & \\
\hline\hline
  &  &  & &\\
Signature    & annual  & None  & None & None   \\
  & modulation &  & & \\
\hline
  &  &  &   & \\
Target-nuclei & $^{23}$Na, $^{127}$I & $^{nat}$Ge &  $^{nat}$Ge &  $^{nat}$Xe \\
\hline
  &  &  &   & \\
Technique    & well known &  poorly & poorly & critical optical  \\
  &  & experienced  & experienced & liquid/gas interface \\
  &  &  & &  in this realization\\
\hline
  &  &  & & \\
Target mass &  $\simeq 100$ kg & 0.5 kg & 0.32 kg & $\simeq 3$ kg\\
\hline
  &  &  & & \\
Exposure     & 57986 kg $\times$ day & 15.8 kg $\times$ day  
                  & 8.2 kg $\times$ day  & 280 kg $\times$ day\\
\hline
  &  &  & & \\
Depth of the  & 1400 m  & 10 m  & 1700 m & 1100 m \\
experimental site &  &  & & \\
\hline
  &  &  & & \\
Software energy & 2 keV  & 10 keV  & 20 keV & 2 keV  \\ 
threshold & (5.5 -- 7.5 p.e./keV) &  & & (but: $\sigma/E =100$\%   \\
 &  &  & & mostly   \\
 &  &  & & 1 p.e./keV; \cite{UKXe})\\
 &  &  & & (2.5 p.e./keV \\
 &  &  & & for 16 days; \cite{ZepI}) \\
\hline
  &  &  & & \\
Quenching       & Measured & Assumed  = 1 & Assumed = 1 & Measured  \\
factor &  &  & &  \\
\hline
  &  &  & & \\
Measured event & $\simeq 1$ cpd/kg/keV  & $\simeq 60$ cpd/kg/keV & 
2500 events & $\simeq 100$ cpd/kg/keV \\
rate in low &   &(10$^5$ events)  & total &  \\
energy range & &  & & \\
\hline
  &  &  & & \\
Claimed events  & & 23 in Ge, 4 in Si, & 0 &  $\simeq$ 20-50 cpd/kg/keV \\
after rejection & &  4 multiple evts in Ge &  &  after rejection and \\
procedures      & & + MonteCarlo on    & &  ?? after standard PSD \\
                     & &  neutron flux  & & \cite{UKXe,ZepI} \\
\hline
  &  &  & & \\
Events satisfying & modulation           & & & \\ 
the signature     & amplitude      & & & \\ 
in DAMA/NaI       & integrated over the  & & & \\ 
                  & given exposure & & & \\ 
                  & $\simeq 2000$ events & & & \\ 
\hline
                & & from few down        & from few down        & \\ 
Expected number & & to zero depending    & to zero depending    & depends on   \\
of events from  & & on the models        & on the models        & the models \\ 
DAMA/NaI effect & & (and on quenching    & (and on quenching    &(even zero) \\
                & & factor)              & factor)              & \\
\hline\hline
\end{tabular}
\end{center}
\label{tb:modepcomp1}
\end{table}

\normalsize

In particular, these experiments exploit a huge data selection releasing
typically extremely poor exposures with    
respect to generally long data taking and, in some cases, to several used detectors.
Their counting rate is very high and few/zero events are claimed after applying
several strong and hardly safe rejection procedures 
(involving several orders of magnitude; see Table \ref{tb:modepcomp1}).
These rejection procedures are also poorly described and, often,
not completely quantified.  
Moreover, most efficiencies and physical quantities entering in the interpretation of
the 
claimed selected events (see Table \ref{tb:modepcomp1}) have never been discussed in the needed 
details; as an example, we mention the case 
of the quenching factor of the recoil target nuclei in
the whole bulk material for the bolometer cases, which is arbitrarily assumed to be
1 (see \S \ref{sc:qf}), implying a substantially arbitrarily assumed
energy scale and energy threshold. The reproducibility of the results over different
running periods has also not been proved as well as the values of the effective sensitive
volumes for read-outs of the two signals (when applied) and
the overall efficiencies.  
Further uncertainties are present when, as in
ref. \cite{CDMS}, a neutron background modeling and 
subtraction is pursued in addition.

As regards in particular the Zeplin-I result of ref. \cite{UKXe,ZepI},
a very low energy threshold is claimed (2 keV), although the 
light response is  very poor: between $\simeq$ 1 ph.e./keV \cite{UKXe}
(for most of the time)
and $\simeq$ 2.5 ph.e./keV (claimed for 16 days) \cite{ZepI} \footnote{ 
For comparison we remind that the data of the DAMA/LXe
set-up, which has a similar light response, are analysed by using the much more realistic
and safer software energy threshold of 13 keV \cite{Xe98}.}.
Moreover, a strong data filtering is applied to
the high level of measured counting rate ($\simeq$ 100 cpd/kg/keV at low energy, 
which is nearly two orders of magnitude larger that the DAMA NaI(Tl) background in the 
same energy region)
by hardware vetoes, by fiducial volume cuts and,
largely, by applying down to few keV a standard pulse shape discrimination procedure,
although the LXe scintillation pulse
profiles (pulse decay time $<$ 30 ns) 
are quite similar even to noise events in the lower energy bins and in spite of the
poor light response. Quantitative information on experimental quantities related to the used
procedures has not yet been given \cite{UKXe,ZepI}

In addition to the experimental aspects, these experiments generally perform an
uncorrect quotation of the DAMA/NaI first quests for a 
purely SI coupled candidate in some given model frameworks and ignore the 
published interpretation of the DAMA/NaI model independent effect in 
terms of candidates with other kind of couplings. Anyhow, 
intrinsically no reliable result can be achieved in a comparison 
of the exclusion plots quoted by these experiments with regions
allowed by DAMA/NaI
in corollary quests for a candidate. In fact, any exclusion plot always refers to a particular model
framework where, in addition, all the involved nuclear cross sections are scaled to cross section(s) on nucleon
(see \S \ref{sc:rate}); thus, it
has no "universal" validity and -- even 
within the same general assumptions for a model (e.g. purely  SI coupling) --  
the proper accounting for
parameters uncertainties, scaling laws uncertainties, 
form factors uncertainties, halo model uncertainties, etc. (see \S \ref{sc:halo}) 
significantly vary 
the result of any comparison (even when assuming as correct the evaluation of the selected number 
of events and the energy scale and energy threshold determinations given in 
refs. \cite{CDMS,Edel,UKXe,ZepI}). 
Moreover, there exist scenarios (see e.g. later in \S \ref{sc:halo}) 
to which Na and I are sensitive and 
other nuclei, such as e.g. $^{nat}$Ge, $^{nat}$Si and $^{nat}$Xe, are not. Just as an example, a possible 
WIMP with a SI cross section of few 10$^{-7}$ pb and with 
SD cross section of few 10$^{-1}$ pb would produce a sizeable signal
in DAMA/NaI, but almost nothing in the Ge and Si experiments of refs. \cite{CDMS,Edel}
as well as in the Xenon target of ref. \cite{UKXe,ZepI}
if the SD component
would have $\theta \simeq 0$ or $\theta \simeq \pi$ (see \S \ref{sc:rate1}). 

In conclusion: 
\begin{enumerate}
\item  
no other experiment, whose result can be directly comparable in a model independent
   way with that of DAMA/NaI, is available so far.
\item
as regards in particular CDMS-I, EDELWEISS-I and Zeplin-I, e.g.:

   i)   they are insensitive to the model independent WIMP annual modulation
        signature exploited by DAMA/NaI; 
   ii)  they use different methodological approaches, which do not allow any 
        model independent comparison and they have different sensitivities to 
        WIMPs; in particular, the number of counts they could expect on the 
        basis of the model independent DAMA/NaI result varies from few to zero 
        events depending on the models, on the assumptions and on the 
        theoretical/experimental parameters' values adopted in the calculations;  
   iii)	they do not make neither correct nor complete comparisons with the DAMA/NaI experimental
        result;   
   iv)  they use extremely poor statistics; 
   v)   they reduce their huge measured counting rate 
        of orders of magnitude by various rejection procedures
        claiming for very optimistic rejection powers;
   vi)  their energy scale determination and/or
        energy threshold appear questionable (in the first two cases 
        because of the quenching factors values
        and in the second because of the poor number of photoelectrons/keV);
    etc. 

\end{enumerate}

Finally, in addition, these experiments \footnote{Recent updates of some results, appeared during the publication of this
paper, leave the arguments discussed in this section unchanched in the essence.} intrinsically
could never reliably
claim for a signal because of the used approach, as mentioned in \S \ref{sc:rej}.

\subsubsection{... from indirect searches}

It has been suggested that Dark Matter particles 
could loose their velocity down to 
a value lower than the escape velocity of a celestial body (Earth, Sun)
scattering off nuclei and, therefore, 
remaining trapped in its gravitational field. 
Subsequently, via their annihilation in the celestial bodies 
or in the Galactic halo they could give rise to high
energy neutrinos, positrons, antiprotons and gamma's.
In principle, the Sun could capture WIMPs more effectively than the Earth 
because of the higher escape velocity, but the smaller distance detector --
center of the Earth and the "resonant" scattering on the heavy nuclei in the 
Earth (mostly on iron) could compensate this effect. 

A possible signature of WIMP annihilation in celestial bodies is given by
the produced $\nu_\mu$, whose
interactions in the rock below a
detector would give rise to "upgoing" muons in the detector itself. 
The expected $\mu$ flux depends on the WIMP 
annihilation rate in the celestial body and on the neutrino energy
spectrum produced 
in the annihilations. However, several sources of uncertainties are present
in similar estimates (and, therefore, in the obtained results) such as for example 
the assumption that a "steady state" has been reached in the considered celestial body
and the significant uncertainty which arises from the estimate and subtraction
of the existing competing process offered by the 
atmospheric neutrinos.

Anyhow, when a model and the related parameters' values are assumed,
it is possible to estimate the differential flux expected for the secondary neutrinos.
According to ref. \cite{Bot99} (where the neutralino 
in the MSSM model has been considered), this flux can be written as 
\begin{equation}
  \frac{dN_{\nu}}{dE_{\nu}}=\frac{\Gamma_{A}}{4\pi d^2}
  \sum_{F,f}B_{\chi f}^{(F)}\frac{dN_{f\nu}}{dE_{\nu}} ,
\label{eq:fnuind}
\end{equation}
where $\Gamma_A$ is the annihilation rate, $d$ is the distance between
the detector and the source (e.g. the Earth center or the Sun center),
$F$ is the final state of the annihilation process, 
$B_{\chi f}^{(F)}$ are the branching ratios of the heavy quarks decays;
the $dN_{f\nu}/dE_{\nu}$ term 
represents the differential distribution of neutrinos produced by 
$\tau$ and by the quarks and gluons hadronization and of the subsequent 
semileptonic decays of the produced hadrons.

Considering, in particular, the $\nu_{\mu}$ and 
$\overline{\nu}_{\mu}$, an estimate of the produced neutrino flux 
can be obtained by measuring
the up-going muons given by the $\nu_{\mu}$ and 
$\overline{\nu}_{\mu}$ interactions with the rock below  
the detector.
Their energy distribution can be written as:
\begin{equation}
\frac{dN_{\mu}}{dE_{\mu}}=N_A
\int_{E_{\mu}^{th}}^{\infty}dE_{\nu}\int_{0}^{\infty}dX
\int_{E_{\mu}}^{E_{\nu}} dE'_{\nu}P_{surv}(E_{\mu},E'_{\mu};X)
\frac{d\sigma(E_{\mu},E'_{\mu})}{dE'_{\mu}}\frac{dN_{\nu}}{dE_{\nu}} ,
\end{equation}
where  $X$ is the muon range in the rock,
$d\sigma(E_{\mu},E'_{\mu})/dE'_{\mu}$ is the charge current cross section for
muon production of energy  $E'_{\mu}$ from a 
neutrino of energy $E_{\nu}$ and 
$P_{surv}(E_{\mu},E'_{\mu};X)$ is the survival probability
of a muon with $E'_{\mu}$  initial energy and 
$E_{\mu}$ final energy after crossing a thickness 
$X$ of rock;
$E_{\mu}^{th}$ is, finally, the energy threshold of the detector.
The function $P_{surv}(E_{\mu},E'_{\mu};X)$ obviously account for the muon energy loss in the
rock.

As mentioned, the up-going muons produced by atmospheric neutrinos 
are side reactions for the process searched
for, however -- in principle -- they are expected to have 
a flat angular distribution while those induced by WIMPs have a preferred 
impinging direction (e.g. the Sun--laboratory direction or Earth center--laboratory
direction).

Model dependent analyses with a similar approach have been carried out 
by large experiments deep underground such as e.g. MACRO and Superkamiokande.
It is worth to remark that no quantitative comparison
can be directly performed between the results obtained in direct 
and indirect searches because it strongly depends on the assumptions 
and on the considered model frameworks.
In particular, a comparison 
would always require the calculation and the consideration of 
all the possible WIMP configurations in the
given particle model (e.g. for $\chi$: in the allowed 
parameters space), since it does not exist
a biunivocal correspondence between the observables in the
two kinds of experiments: WIMP-nucleus elastic 
scattering cross section (direct detection case) and flux of 
muons from neutrinos (indirect detection case). In fact, 
the counting rate in direct search depends on 
the spin-dependent (SD) and on the spin-independent (SI) cross 
sections of the elastic processes, while the muon flux
is connected not only to them, but also
to the WIMP annihilation cross section.
In principle, the three cross sections can be correlated,
but only when a specific model is adopted and by 
non directly proportional relations.
\begin{figure}[!h]
\centerline{\hbox{ \psfig{figure=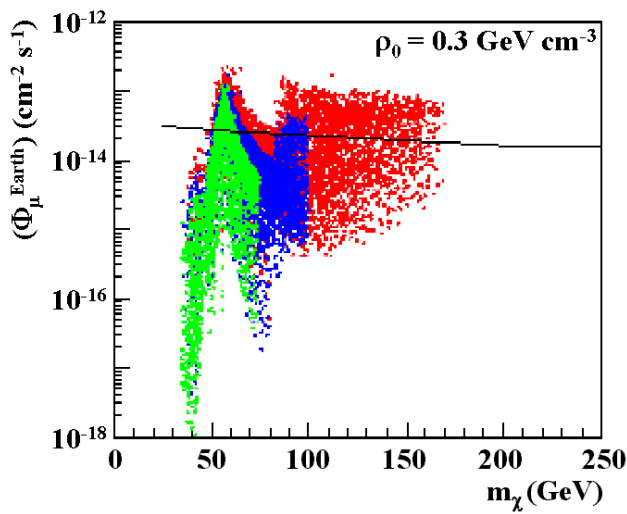,height=7.0cm} }}
\vspace{-0.4cm}
\caption{Scatter plot for the up-going muon flux from the center of the Earth 
vs neutralino mass.
The configurations (MSSM) -- here the model dependent 
constraints from the LEP data of 2000 have been used -- 
have been selected by the DAMA annual modulation region obtained for the model 
framework considered in ref. \cite{Mod3}.
For details see \cite{Botdm}.
The solid line is the
model dependent upper bound derived from MACRO experiment; the one from Superkamiokande is only 
marginally more stringent.}
\label{fg:figbot}
\end{figure}
As an example, we report in Fig. \ref{fg:figbot} the scatter plot for 
the up-going muon flux from the center of the Earth for
a standard Maxwellian distribution versus $\chi$ mass in MSSM \cite{Botdm};
here the configurations have been selected by the DAMA annual 
modulation region for
the particular purely SI model framework considered in ref. \cite{Mod3}.
As it is evident, the  up-going muon flux spans several orders of magnitude
although the cross section of the DAMA region, allowed in the model framework considered there, 
spans almost one. 
The solid line in this figure is the
model dependent upper bound derived from MACRO experiment\cite{Macro}; 
the one from Superkamiokande is only
marginally more stringent, thus it is still compatible with the DAMA result even when -- 
as in the quoted ref. \cite{Mod3} -- the
uncertainties on several assumptions and parameters are not yet included (see e.g. \cite{Hep}).

\begin{figure}[!hb]
\centering
\vspace{-0.2cm}
\epsfig{file=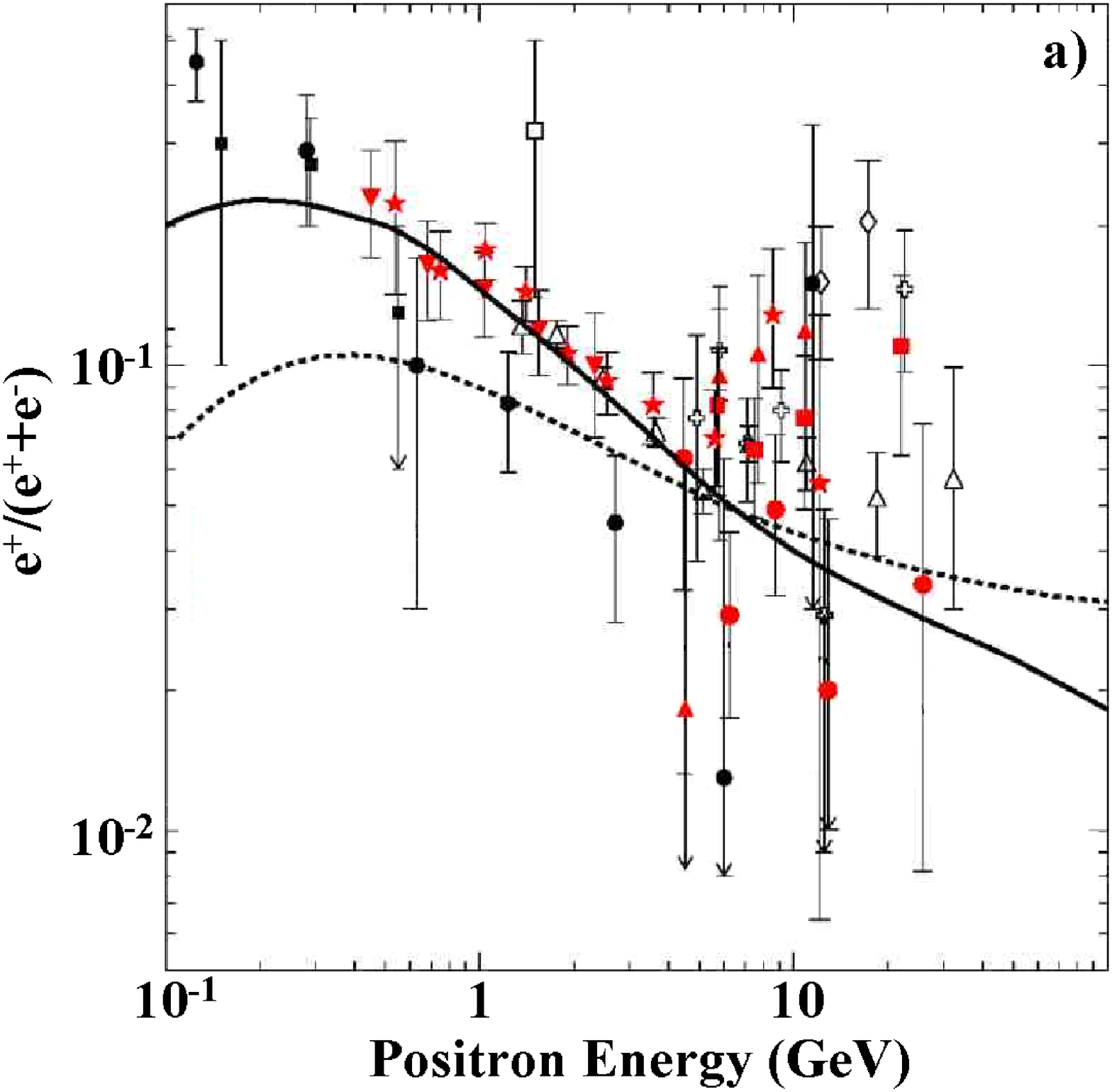,height=5.3cm}
\epsfig{file=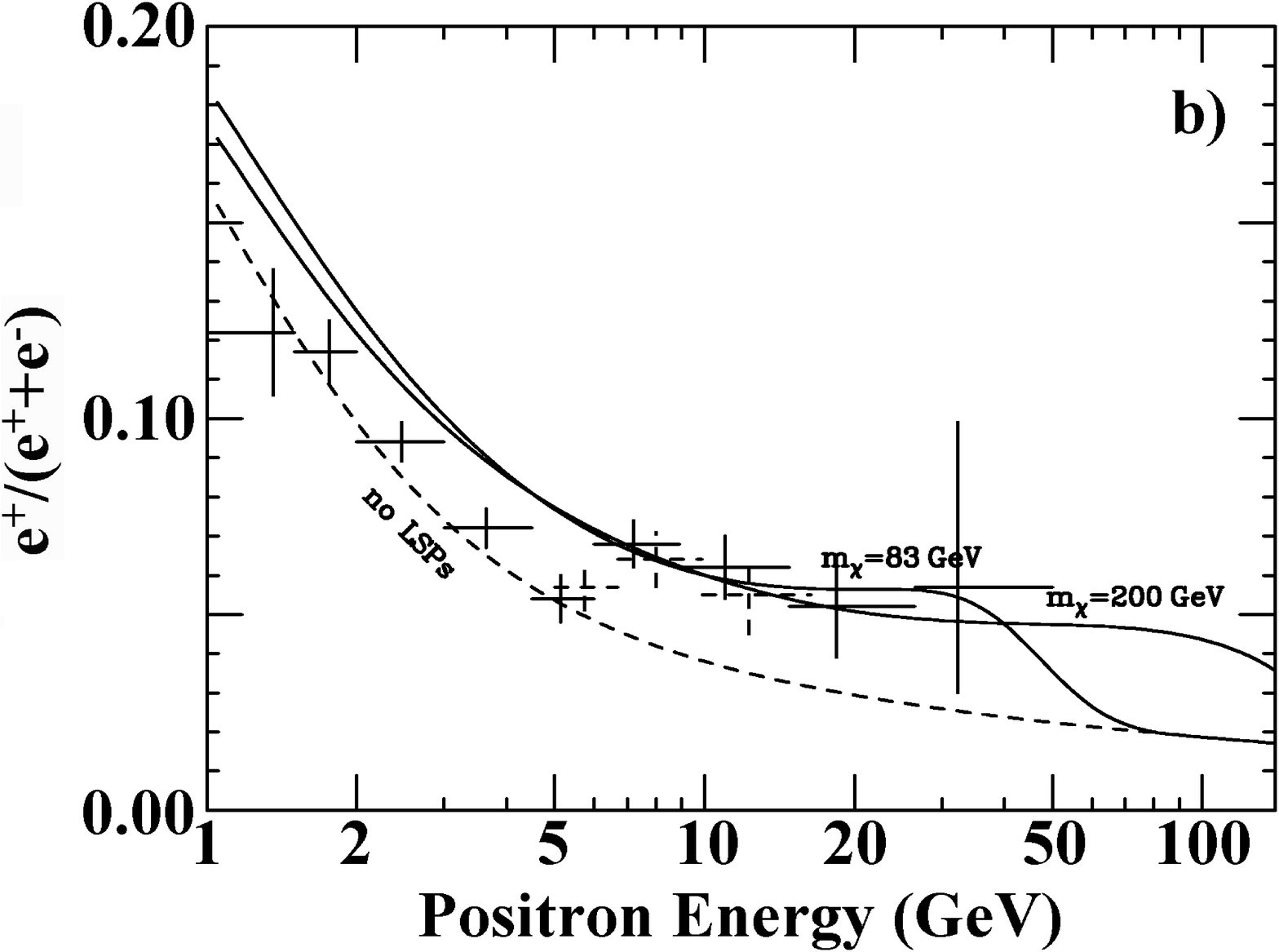,height=5.cm}
\epsfig{file=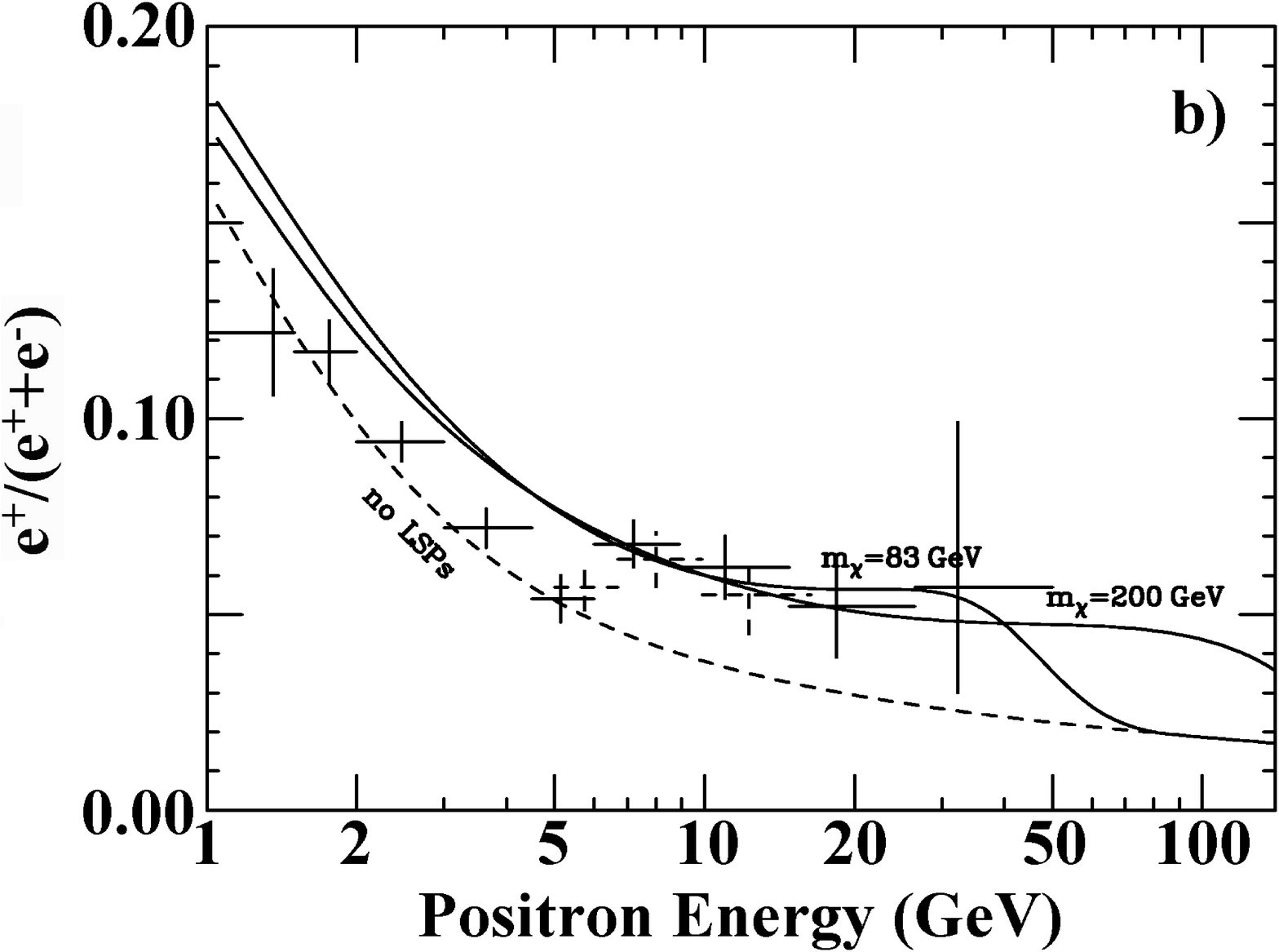,height=5.2cm}
\vspace{-0.3cm}
\caption{Experimental results and theoretical predictions for the indirect searches: 
a) Positron fraction measured by several experiments (Caprice98, AMS, Caprice94,
HEAT, Clem et al., TS93, MASS89, Golden et al., Muller and Tang, Fanselow et al.; see 
ref. \cite{morse1} for details).
The two lines are two models of background from secondary production;
an excess of positrons over the
background modelling is present in both cases.
b) Positron fraction measured by HEAT in the interpretation of ref. \cite{heat},
where the excess of positrons with respect to the background modelling is explained in
terms of neutralino annihilation in the galactic
halo.
c) Gamma ray energy spectrum of the inner Galaxy as measured by EGRET 
compared with the background modelling (lower line). The large excess of 
gamma's is explained in ref. \cite{morse2} by a neutralino annihilation in the
galactic
halo (the upper curve is the total contribution). These preliminary 
indications are not in
conflict with the DAMA/NaI model independent results previously published.}
\label{fg:indirect}
\vspace{-0.5cm}
\end{figure}
As we mentioned at the beginning, the annihilation of the Dark Matter 
particles in the halo could also produce antimatter particles and gamma's. 
The antimatter searches have to be carried out outside the 
atmosphere, i.e. on balloons or satellites. In particular, the WIMP 
annihilation would result in an excess of antiprotons 
or of positrons up to the WIMP mass with respect to the
background arising from other possible sources. Again the estimate and subtraction
of such a background together with the influence of the Earth and of the galactic magnetic field
on these particles plays a crucial role on the possibility of a reliable extraction of 
a signal. However, 
at present some interesting results have been reported in the analysis of the HEAT 
balloon-borne experiment and in some others, as reported in Fig. \ref{fg:indirect}a) and b).
In fact, an excess of positrons  with energy $\simeq 5 - 20$ GeV 
has been found; it -- interpreted in
terms of WIMP annihilation 
\cite{heat} -- gives a result not in conflict 
with the effect observed by DAMA/NaI.
Further results can be expected in future by experiments operating in space \cite{Space}.

As regards the possibility to detect $\gamma$'s from WIMP annihilation
in the galactic halo, experiments in space are planned. 
However, at present it is difficult to estimate their
possibilities considering 
e.g. the background level, the uncertainties in its reliable estimate 
and subtraction as well as the smallness of the expected signal (even more, 
if a subdominant component would be present) when 
properly calculated with rescaling procedure.
However, we mention the analysis of ref. \cite{morse} 
which already suggests the presence of a $\gamma$ excess from the center of the Galaxy 
in the EGRET data  \cite{egret} as reported in Fig. \ref{fg:indirect}c). 
This excess match with a possible 
WIMP annihilation in the galactic halo \cite{morse} and 
is not in 
conflict with the DAMA/NaI model independent result previously published.
Other activities are in preparation and will further clarify the situation \cite{glast}.

We stress, however, that the specific parameters of a WIMP candidate 
(mass and cross sections), which 
can be derived from the indirect searches, critically depend
on several assumptions used in the calculations such as the estimation 
of the background, the halo model, the amount of WIMP in the galactic dark halo, 
the annihilation channels, the transport of charged particle to the Earth, etc.; thus,
they have the same relative meaning as those obtained in the quest for 
a candidate in direct search approach as described later.

\subsection{Conclusions}

 In conclusion, no model independent comparison with the DAMA/NaI effect is available. 
Only few model dependent approaches have been used in the direct search approach 
to claim for a particular model
dependent comparison, which appears in addition -- as discussed above --  
neither based on solid procedures nor fully correct nor complete. 
On the other hand, the indirect search approaches, which also can offer  
only model dependent comparisons, are either not in contradiction or 
in substantial agreement with the DAMA/NaI observed effect.

Thus, the interest in the further available DAMA/NaI data is increased; 
the model independent result on the WIMP annual modulation signature
from the data of seven annual cycles as well as some (of the many possible) model
dependent quests 
for a candidate are discussed in the following, reviewing as well
the general aspects related to the WIMP direct detection.  

\vspace{-0.3cm}
\section{The DAMA/NaI model independent result on the WIMP annual modulation signature
from the data of seven annual cycles}
\vspace{-0.2cm}

As mentioned, the main goal of the DAMA/NaI experiment 
is the investigation on a WIMP component in the galactic halo 
by exploiting the WIMP annual modulation signature introduced in 
\S \ref{sc:sign}.
The experiment has collected data during seven annual cycles.

The data of each annual cycle have been taken in the same
experimental conditions; in particular,
the Copper box housing the detectors has always been closed and
sealed and the detectors have always been in contact only with an
atmosphere of HP Nitrogen \cite{Nim98,Sist}.
Moreover, the data taking of each annual cycle has been started before the
expected minimum
of the signal rate (which is roughly around $\simeq$ 2$^{nd}$
December) and concluded after the expected maximum (which is
roughly around $\simeq$ 2$^{nd}$ June).

As mentioned, several operational parameters have been regularly acquired with the
production data, such as the operating temperature, the HP Nitrogen flux into the inner
Cu box housing the detectors, the pressure of
the HP Nitrogen atmosphere in the
inner Cu box, 
the environmental Radon from which however the detectors are excluded (see above and later)
and
the hardware rate (including the noise) above single photoelectron threshold. Computer
controlled processes immediately inform the operator during production runs in case one of the
parameters
goes outside the stringent allowed interval of stability values. In addition, the
recorded parameters
values allow a deep analysis and control of possible systematics as performed e.g. in ref.
\cite{Sist}
and discussed in the following.

\subsection{The evidence}
\label{sc:evi}

A model independent investigation of the annual modulation
signature has been realized by exploiting the time behaviour of the
residual rates of the single hit events in the lowest energy regions 
over the seven annual cycles (total exposure: 107731 kg $\cdot$ day), 
as previously performed in refs. \cite{Mod3,Sist}. 
\begin{figure}[!hb]
\begin{center}
\vspace{-1.0cm}
\epsfig{file=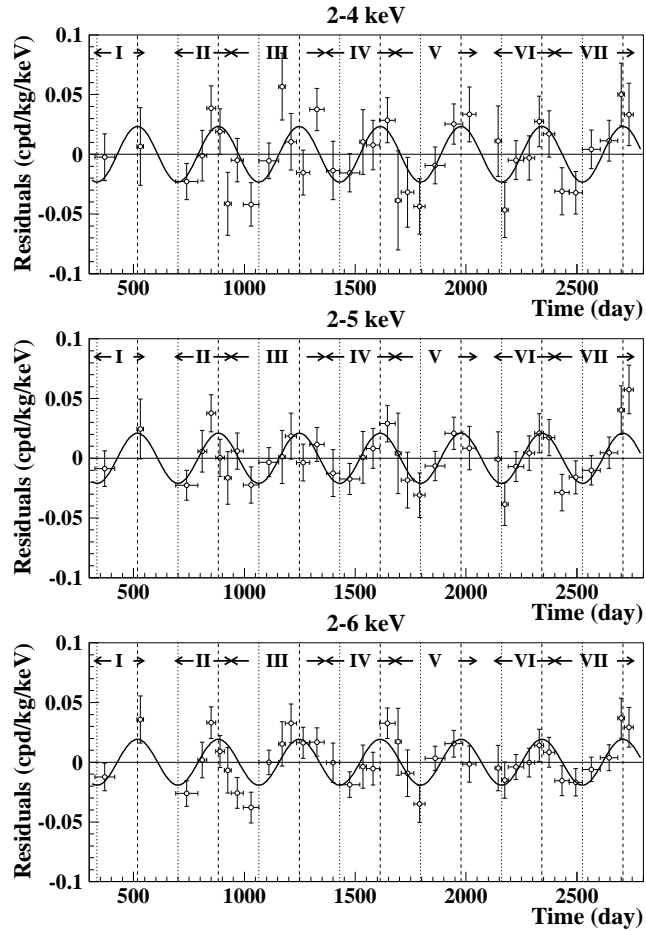,height=13cm}
\end{center}
\vspace{-1.0cm}
\caption{Model independent residual rate for single hit events, in the (2--4), (2--5) 
and (2--6) keV energy intervals as a function of the time elapsed since
January 1-st of the first year of data taking. The
experimental points present the errors as vertical
bars and the associated time bin width as horizontal bars. The 
superimposed curves represent the cosinusoidal functions 
behaviours expected for a WIMP signal 
with a period equal to 1 year and phase at $2^{nd}$ June; the modulation
amplitudes have been obtained by best fit. See text.
The total exposure is 107731 kg $\cdot$ day.}
\label{fig1}
\vspace{-0.2cm}
\end{figure}
These
residual rates are calculated  from the measured event rate after subtracting
the constant part (the weighted mean of the residuals must obviously be zero
over each period): $<r_{ijk} - flat_{jk}>_{jk}$. There $r_{ijk}$ is the rate
in the considered $i$-th time interval for the $j$-th detector in
the $k$-th considered energy bin,
while $flat_{jk}$ is the rate of the $j$-th detector in the $k$-th energy bin averaged
over the cycles. The average is made on all the detectors ($j$ index) and on
all the energy bins in the considered energy interval.


This model independent approach on the data of the seven annual cycles 
offers an immediate evidence of the presence of 
an annual modulation of the rate of the single hit events
in the lowest energy region  
as shown in Fig. \ref{fig1}, where the time behaviours of the (2--4), (2--5) and (2--6) keV
single hit residual rates are
depicted. They refer to
4549, 14962, 22455, 16020, 15911, 16608, 17226 kg $\cdot$ day 
exposures, respectively for the DAMA/NaI-1 to -7 running periods
\footnote{In particular, the DAMA/NaI-5 data have been collected 
from August 1999 to end of July 2000
(statistics of 15911 kg $\cdot$ day); then, the DAQ and
the electronics have been fully substituted (see \S \ref{app}).
Afterwards, the DAMA/NaI-6 data have been collected from
November 2000 to end of July 2001 (statistics of 16608 kg $\cdot$ day), while
the DAMA/NaI-7 data have been collected from August 2001 to July 2002
(statistics of 17226 kg $\cdot$ day), when the data taking with this set-up has 
been concluded.}.


In fact, the data favour the presence of a modulated cosine-like behaviour 
($A \cdot$ cos$\omega (t-t_0)$) at 6.3 $\sigma$ C.L. 
\footnote{It is worth to note that the confidence level given 
in ref. \cite{Mod3} was instead referred to the particular model framework
considered there in the quest for a candidate.
Here the confidence level refers to the
model independent effect itself and is calculated on the basis of the 
residual rate in the (2--6) keV energy interval. 
Applying the same procedure to the residuals given in ref. \cite{Mod3},
one gets 4.6 $\sigma$ C.L. which is in agreement with the presently quoted value 
once scaling it by the square root of the ratio of the relative exposures.}
and their fit for the (2--6) keV 
larger statistics energy interval offers modulation amplitude equal to
$(0.0200 \pm
0.0032)$ cpd/kg/keV,
$t_0 = (140 \pm 22)$ days and
$T = \frac{2\pi}{\omega} = (1.00 \pm 0.01)$ year, all parameters kept free in the fit.
The fitting function has been derived from eq. (\ref{eq:sm}) integrated over each time
bin. The period and phase agree with those expected in the case of a
WIMP induced effect ($T$ = 1 year and $t_0$ roughly at $\simeq$ 152.5-th day of the year). 
The $\chi^2$ test on the (2--6) keV residual rate in Fig. \ref{fig1}
disfavours the hypothesis of unmodulated behaviour giving a probability of $7 \cdot
10^{-4}$ ($\chi^2/d.o.f.$ = 71/37).
We note that, for simplicity, in Fig. \ref{fig1} 
the same time binning already considered in ref. \cite{Mod3,Sist} has been
used. The result of this approach is similar by choosing other time binnings; moreover,
the results given in the following are not dependent on time binning at all.


The residuals given in Fig. \ref{fig1} have also been fitted, according to the previous procedure,
fixing 
the period at 1 year and the phase at $2^{nd}$ June; the best fitted modulation
amplitudes
are: $(0.0233 \pm 0.0047)$ cpd/kg/keV for the (2--4) keV energy interval, $(0.0210 \pm 0.0038)$
cpd/kg/keV for the (2--5) keV energy interval,
$(0.0192 \pm 0.0031)$ cpd/kg/keV for the (2--6) keV energy interval,
respectively.

\begin{figure}[!ht]
\begin{center}
\vspace{-0.7cm}
\epsfig{file=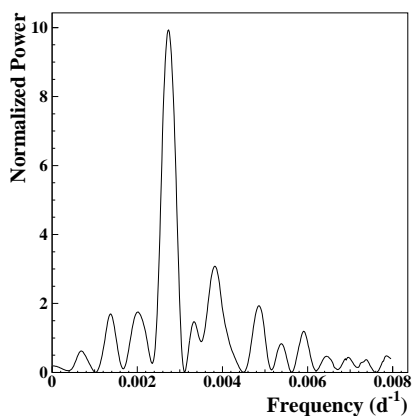,height=5.5cm}
\end{center}
\vspace{-0.7cm}
\caption{Power spectrum of the measured (2--6) keV single
hit residuals calculated according to ref. 
\cite{Lomb}, including also the treatment of the experimental errors and of the time binning.
As it can be seen, the principal mode corresponds to a frequency of $2.737 \cdot 10^{-3}$ d$^{-1}$,
that is to a period of $\simeq$ 1 year.}
\label{fig2}
\normalsize
\end{figure}

The same data have also been investigated by a Fourier analysis (performed according to ref.
\cite{Lomb} including also the treatment of the experimental errors and of the time binning),
obtaining
the result shown in Fig.~\ref{fig2}, 
where a clear peak for a period of 1 year is evident.

\begin{figure}[!ht]
\vspace{-0.7cm}
\centering
\epsfig{file=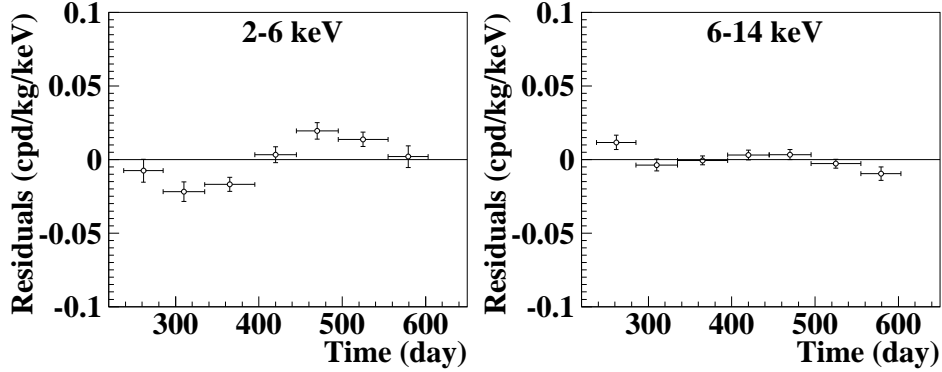,height=6.cm}
\vspace{-1.cm}
\caption{Single hit residual rate in a single annual cycle from the total exposure of
107731 kg $\cdot$ day. The
experimental points present the errors as vertical
bars and the associated time bin width as horizontal bars.
The initial time is taken at August 7$^{th}$.
Fitting the data with a cosinusoidal function with period of 1 year and phase at 152.5 days,
the following amplitudes are obtained: $(0.0195 \pm 0.0031)$ cpd/kg/keV and $-(0.0009 \pm 0.0019)$
cpd/kg/keV, respectively. Thus, a clear modulation is present in the lowest energy region, while it
is absent just above.}
\label{fig3}  
\vspace{-0.3cm}
\end{figure}

In Fig. \ref{fig3} the single hit residual rate in a single annual cycle from the total exposure of
107731 kg
$\cdot$ day is presented for two different energy intervals; as it
can be seen the modulation is clearly present in the (2--6) keV energy region, while 
it is absent just above.

\begin{figure}[!h]
\begin{center}
\vspace{-1cm}
\mbox{\epsfig{file=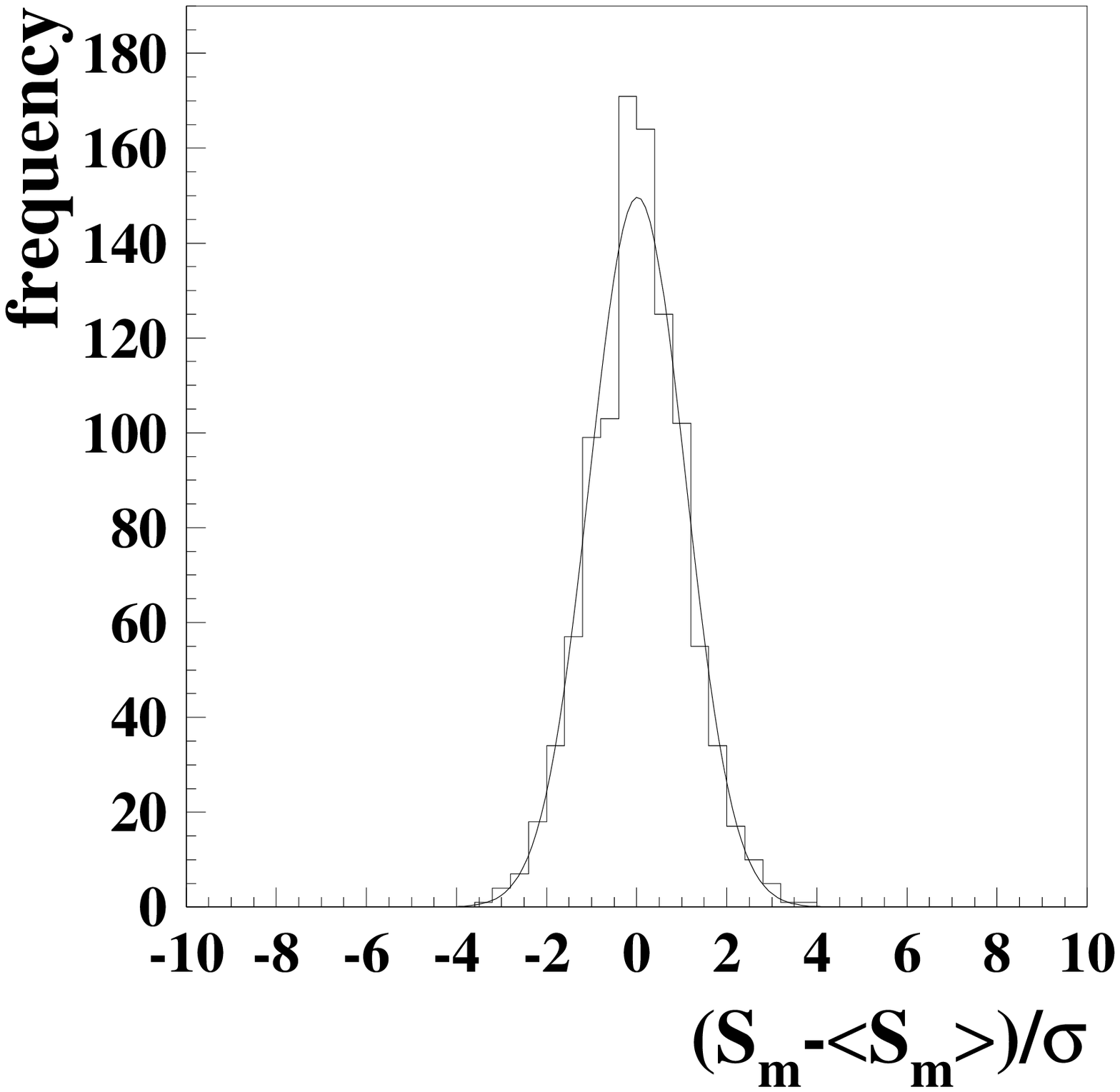,height=6.0cm}}
\mbox{\epsfig{file=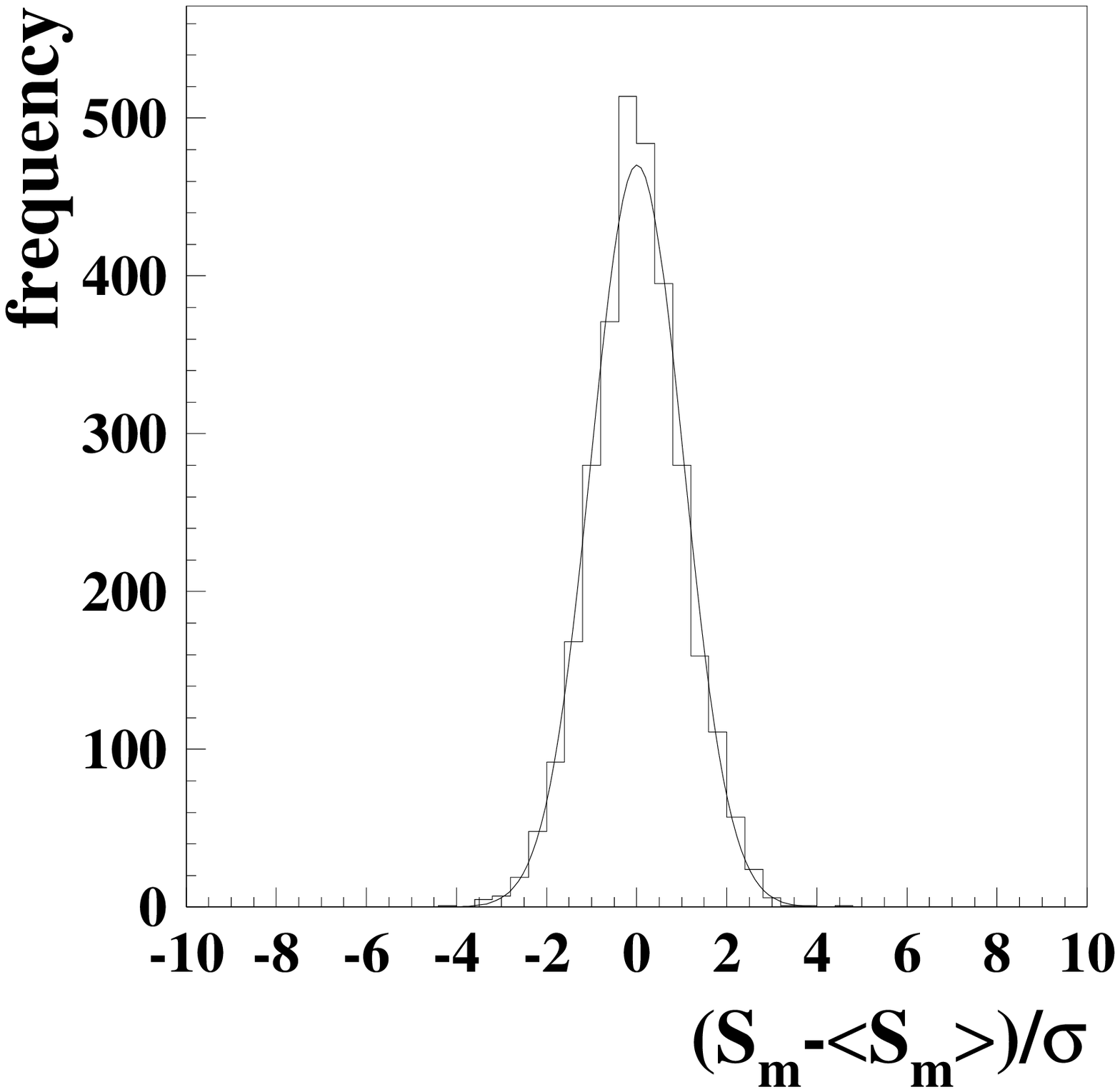,height=6.0cm}}
\end{center}
\vspace{-1cm}
\caption[]{
Distributions of the variable $\frac {S_m-<S_m>}{\sigma}$ (where $\sigma$ is the error
associated to the $S_m$) evaluated for each detector,   
for each annual cycle and each considered energy bin: i) in the region of interest for   
the observed modulation, 2--6
keV (left panel);
ii) including also the energy region just above, 2--14 keV (right panel). See text.}
\label{gaus1}
\end{figure}

\vspace{0.2cm}

Finally, Fig. \ref{gaus1} shows the 
distributions of the variable $\frac {S_m-<S_m>}{\sigma}$, where $S_m$ are the
modulation amplitudes evaluated by a maximum likelihood method \cite{Mod1} for each detector,
each annual cycle and each considered energy bin (taken
there as an example equal to 0.25 keV) and $\sigma$ are their
errors. The $<S_m>$ represent the mean values of the modulation amplitudes over
the detectors and the annual cycles for each energy bin.
The left panel of Fig. \ref{gaus1} shows the distribution referred to the
region of interest for the observed modulation: 2--6
keV, while the right panel includes also the energy region just above: 2--14 keV. 
These distributions allow one to conclude that the individual $S_m$
values follow a
normal
distribution, since the variable $\frac {S_m -<S_m>}{\sigma}$ is distributed as a 
gaussian with an unitary standard deviation. In particular, this demonstrates that the
modulation amplitudes are statistically well distributed in all the crystals,
in all the data taking periods and  
considered energy bins.


\subsection{The investigation of possible systematic effects and side 
reactions}

\vspace{0.4cm}

As previously mentioned -- to mimic the annual modulation
signature a systematic effect or side reaction should not only be quantitatively
significant, but also able to satisfy the
six requirements as for a WIMP induced effect (see \S \ref{sc:sign});
no effect able to mimic the signature has been found.
\begin{table}[!hbt]
\caption{Modulation amplitudes obtained by fitting the time behaviours of the main
running parameters including a WIMP-like cosine
modulation. These running parameters, acquired with the production data, are: 
i) the operating temperature of the detectors;
ii) the HP Nitrogen flux in the inner Cu box 
housing the detectors; iii) the pressure of the  HP Nitrogen atmosphere of the
inner Cu box
housing the detectors; iv) the environmental Radon in the inner part of the 
barrack 
from which the detectors
are however excluded (see text); v) the hardware rate above single photoelectron 
threshold. See also the discussion in the whole section.}
\vspace{0.4cm}
\centering
\footnotesize
\begin{tabular}{|c||c||c||c||} \hline
  & & & \\
  & DAMA/NaI-5 & DAMA/NaI-6 & DAMA/NaI-7 \\ 
  & & & \\
\hline
  & & & \\
Temperature & $-(0.033 \pm 0.050) ^{\circ}$C &
              $(0.021 \pm 0.055) ^{\circ}$C  & 
              $-(0.030 \pm 0.056) ^{\circ}$C  \\
  & & & \\
Flux  & $(0.03 \pm 0.08)$ l/h &
        $(0.05 \pm 0.14)$ l/h & 
        $(0.07 \pm 0.14)$ l/h \\
  & & & \\
Pressure   & $-(0.6 \pm 1.7) 10^{-3}$ mbar &
              $(0.5 \pm 2.5) 10^{-3}$ mbar &  
              $(0.2 \pm 2.8) 10^{-3}$ mbar \\
  & & & \\
Radon       & $-(0.09 \pm 0.17)$ Bq/m$^{3}$ &
              $(0.06 \pm 0.14)$ Bq/m$^{3}$ & 
              $-(0.02 \pm 0.03)$ Bq/m$^{3}$ \\
  & & & \\
Hardware rate & $(0.10 \pm 0.17) 10^{-2}$ Hz &
              $-(0.09 \pm 0.19) 10^{-2}$ Hz &  
              $-(0.22 \pm 0.19) 10^{-2}$ Hz \\
  & & & \\
\hline\hline
\end{tabular}
\label{tb:par567}
\end{table}
A careful investigation of all the known possible sources of systematics and side
reactions has been regularly carried out by DAMA/NaI and presented at time of each data 
release \cite{Mod1,Mod2,Mod3,Sist}.
In particular, in ref. \cite{Sist} a detailed discussion has been carried out
considering in the quantitative evaluations the data of DAMA/NaI-3 and -4. 
The same analysis is presented in the following considering in the
quantitative evaluations the data
of the DAMA/NaI-5 to -7 running periods; it offers a general description of this approach 
as well.

\vspace{0.4cm}

First of all the time behaviours 
of the main running parameters acquired with the production data
have been investigated. In particular, the modulation amplitudes given in  Table
\ref{tb:par567} (see also next subsections)
have been obtained for each  
annual cycle when fitting the
behaviours
including a WIMP-like cosine modulation.
As it can be seen, all the measured amplitudes are
compatible with zero.

\subsubsection{The Radon}

As already discussed elsewhere \cite{Nim98,Sist},
the detectors are excluded from the environmental air which
contains traces of radioactive Radon gas ($^{222}$Rn
-- T$_{1/2}$ = 3.82 days -- and of
$^{220}$Rn -- T$_{1/2}$ = 55 s -- isotopes, which belong to the
$^{238}$U and $^{232}$Th chains, respectively), whose 
daughters attach themselves to
surfaces by various processes. 
In fact: i) the walls, the floor and the top of the installation are insulated by 
Supronyl (permeability: 2 $\cdot$ 10$^{-11}$
cm$^2$/s \cite{woj}) and a large flux of HP Nitrogen is released
in the closed space of that inner barrack;
an Oxygen level alarm informs the operator
before entering it, when necessary; ii) the whole passive shield is sealed in a 
plexiglas
box and maintained continuously in HP Nitrogen atmosphere in slight overpressure
with respect to the environment as well as 
the upper glove box for calibrating the detectors; iii)
the detectors are housed in an inner sealed 
Cu box also maintained continuously in HP Nitrogen atmosphere in slight 
overpressure
with respect to the environment; the Cu box can enter in contact only with 
the upper glove box -- during calibrations -- which is 
also continuously maintained in HP Nitrogen
atmosphere in slightly overpressure with respect to the external environment. 

\begin{figure}[!htb]
\centering
\epsfig{file=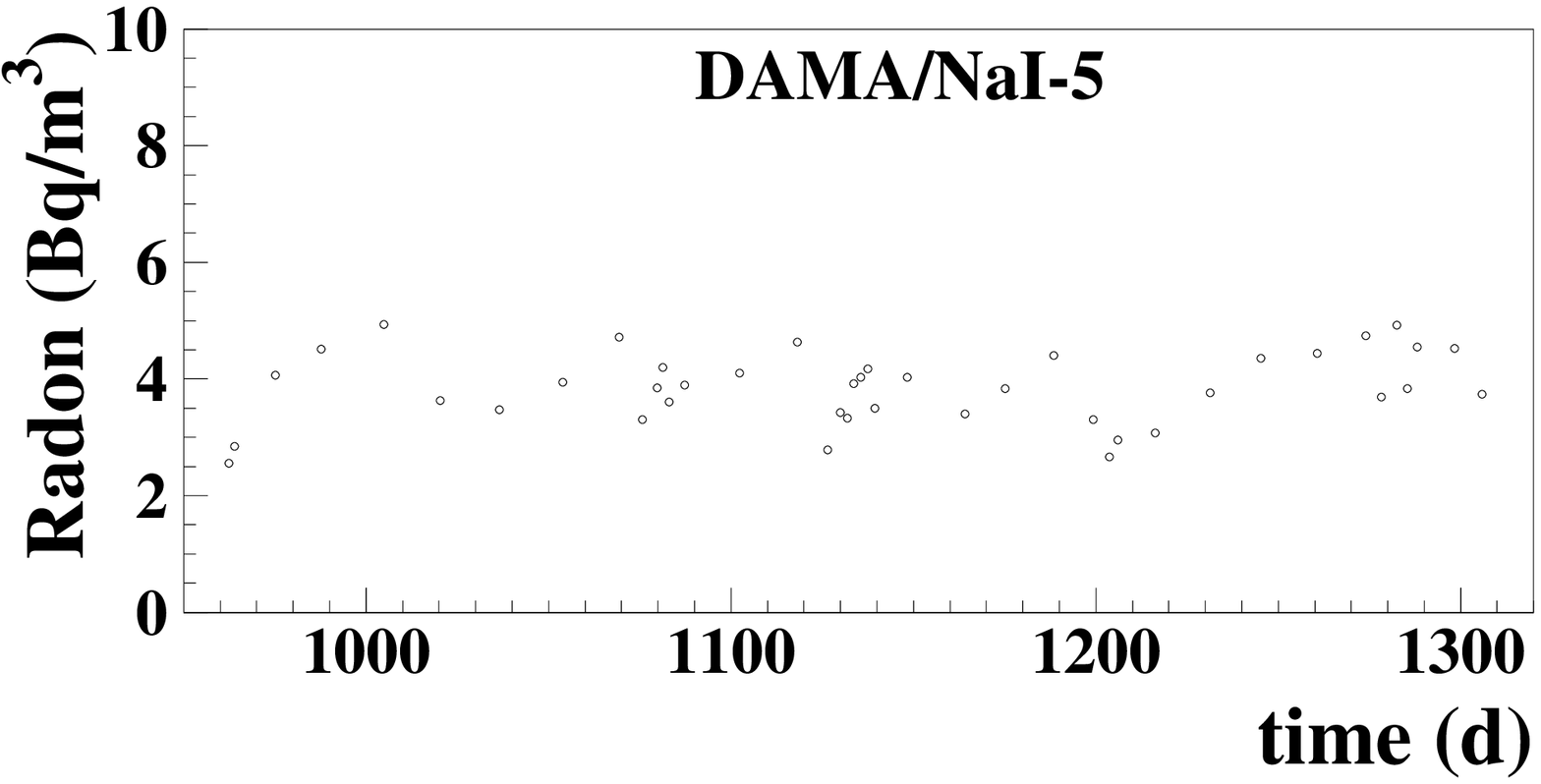,height=3.2cm}
\hspace{-0.2cm}
\epsfig{file=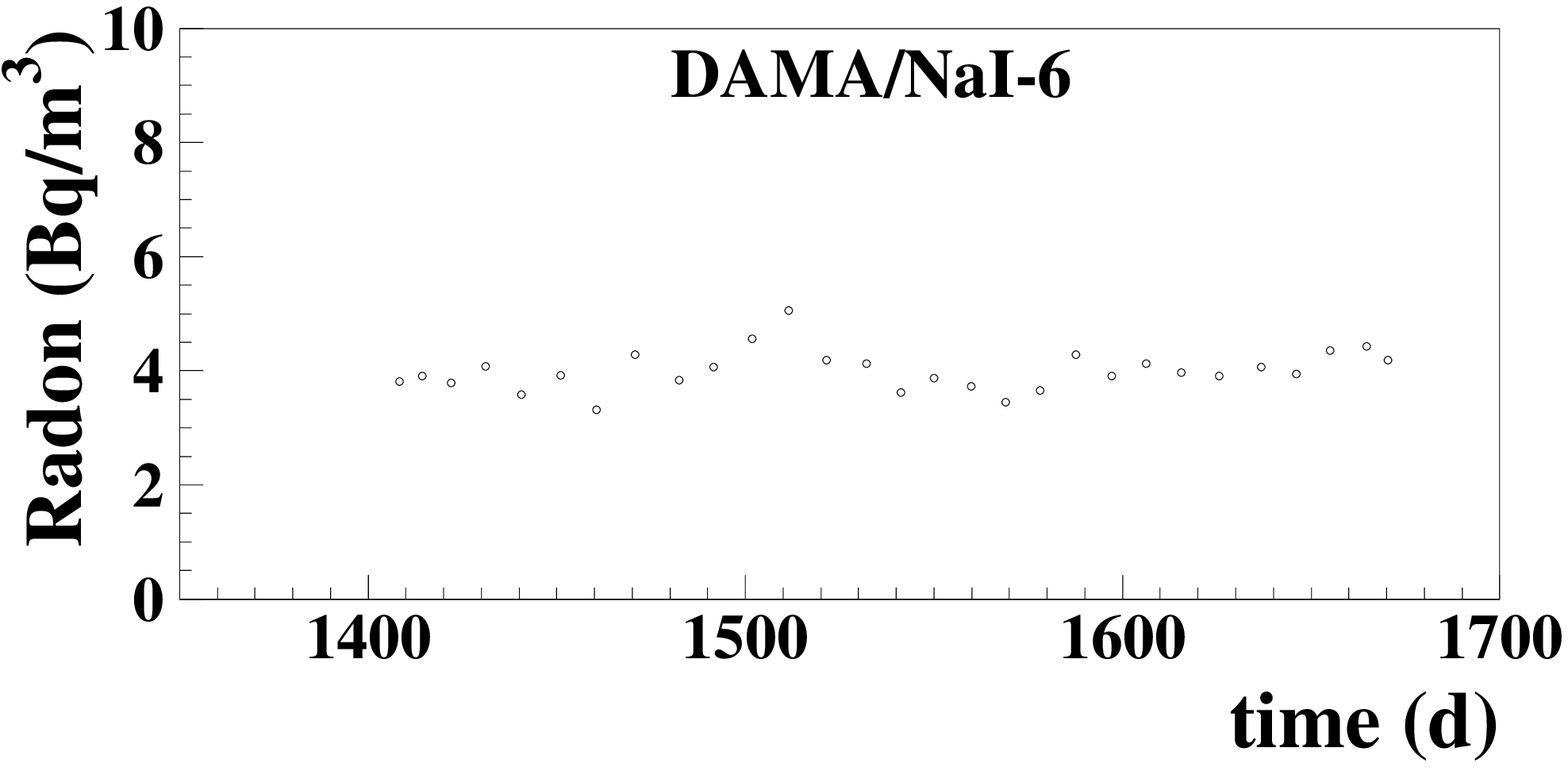,height=3.2cm}
\hspace{8.cm}
\epsfig{file=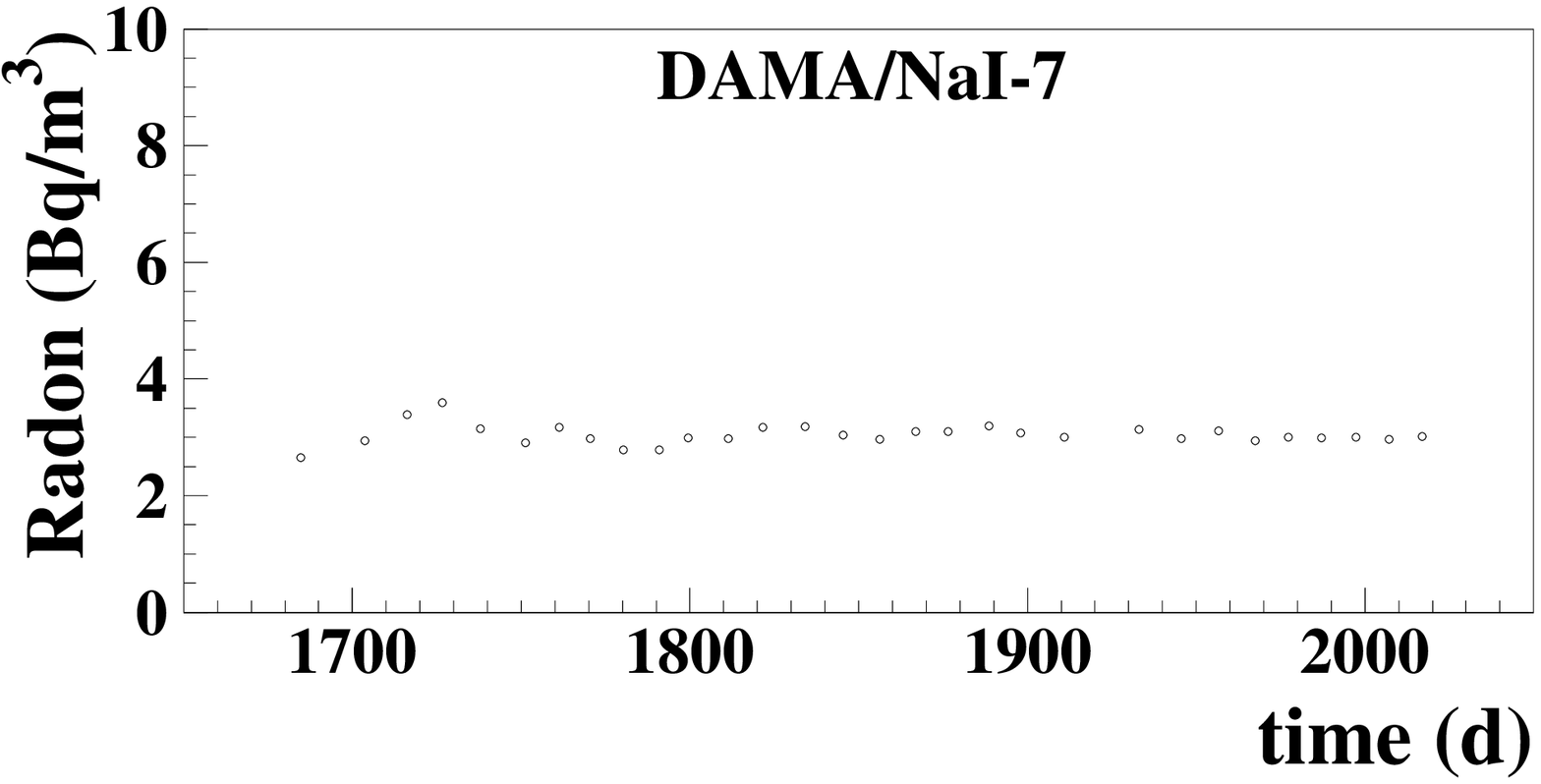,height=3.2cm}
\caption[]{Time behaviours of the environmental Radon in the inner part of the barrack (from which the detectors 
are however excluded; see text) during the DAMA/NaI-5 to -7
running periods, respectively. The measured values are at the level of sensitivity 
of the used radonmeter.}
\label{fig_radon}
\end{figure}

Notwithstanding the above considerations, the Radon in the
installation outside the plexiglas box, containing the passive shield,
is continuously monitored; it is at level of sensitivity
of the used Radonmeter (see Fig. \ref{fig_radon}). No modulation of external Radon, from which 
the detectors are anyhow excluded,
is observed as quantitatively reported in Table \ref{tb:par567}.

\begin{figure}[!htb]
\begin{center}
\vspace{-0.4cm}
\epsfig{file=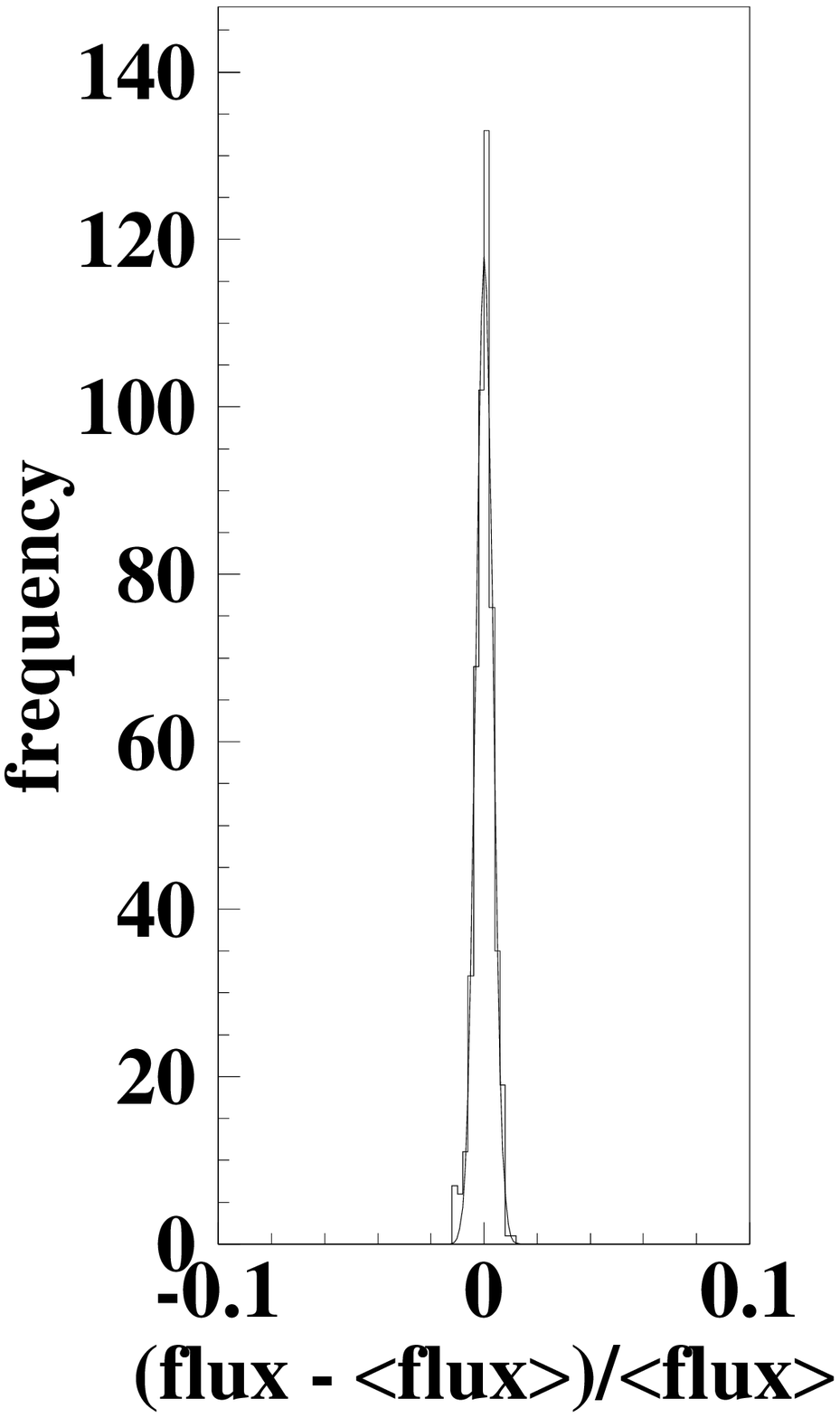,height=7cm}
\epsfig{file=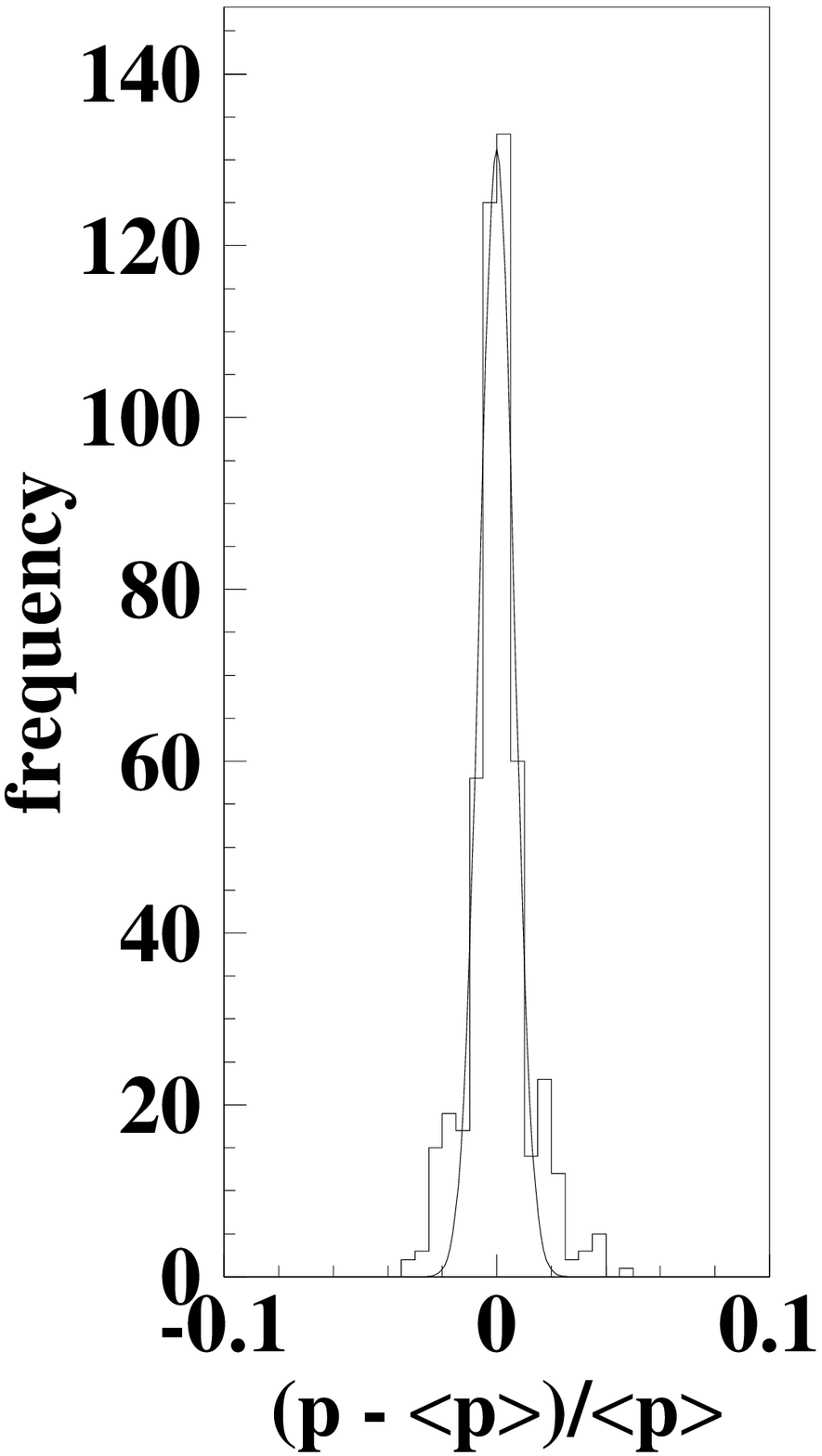,height=7cm}
\vspace{-.6cm}
\end{center}      
\caption[]{Distribution of
the relative variations of the HP Nitrogen flux in the inner Cu box housing the detectors
and of its pressure as measured during the
DAMA/NaI-5 to -7 running periods.}
\label{fig_pres}
\end{figure}

In Fig. \ref{fig_pres} the distribution of
the relative variations of the HP Nitrogen flux in the inner Cu box housing the detectors
and of its pressure as measured during the 
DAMA/NaI-5 to -7 running periods are shown (the typical flux mean value for each annual
cycle is of order of $\simeq$ 260 l/h and the typical overpressure mean value is of order of 2 mbar).

As already reported in ref. \cite{Sist}, also 
possible Radon trace in the HP Nitrogen atmosphere inside the Cu box 
has been estimated by searching for the double coincidences of
the gamma-rays (609 and 1120 keV) from $^{214}$Bi
Radon daughter, obtaining an upper 
limit on 
the possible Radon concentration in the Cu box HP Nitrogen atmosphere:
$< 4.5 \cdot 10^{-2}$ Bq/m$^3$  (90\% C.L.); thus, 
roughly $<$ 4 $\cdot 10^{-4}$ cpd/kg/keV can be expected from this source in the lowest energy 
bins of interest from the obtained result on the WIMP annual modulation signature \cite{Sist}.
This has allowed us to show that even an hypothetical, e.g. 10\%, modulation of possible Radon in 
the HP Nitrogen Cu box atmosphere would correspond to 
$< 0.2\%$ of the observed modulation amplitude.

Finally, it is worth to note that, while the possible presence of a sizeable quantity 
of Radon nearby a detector would forbid the investigation of the WIMP annual 
modulation 
signature (since every Radon variation would induce variation in the measured 
background
and the continuous pollution of the exposed surfaces by the non-volatile 
daughters), it cannot mimic the WIMP 
annual modulation signature in experiments such as DAMA/NaI
which record the whole energy distribution. In fact, possible presence of Radon 
variation
can be easily identified in this case, since it would induce rate 
variation also in 
other energy regions than the one of interest for the WIMP search,
that is some of the six requirements 
of the WIMP annual modulation signature would fail.

In conclusion, no significant effect is possible from the Radon.

\subsubsection{The temperature}

To avoid any significant temperature variation and, in particular,
to maintain suitably stable the temperature of the electronic devices
the installation, where the $\simeq$ 100 kg NaI(Tl)
set-up is operating, is air-conditioned.
Moreover, the operating temperature of the detectors 
in the Cu box (stored with the
production data) is read out by a probe located inside the multi-tons passive shield, 
whose huge heat capacity assures further a relevant stability of the detectors' 
operating temperature \cite{Nim98,Mod1,Mod2,Sist,3k99}.

\begin{figure}[!htb]
\begin{center}
\vspace{-0.4cm}
\epsfig{file=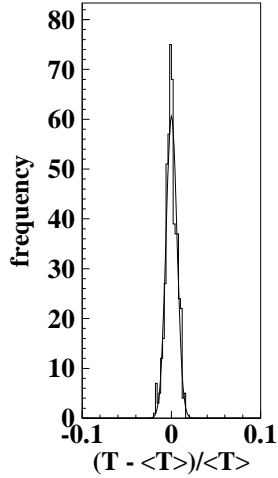,height=6.5cm}
\vspace{-.6cm}
\end{center}
\caption[]{Distribution of the relative variations of the operating
temperature
measured during the DAMA/NaI-5 to -7 running periods.}
\label{fig_temp}
\end{figure}

\begin{figure}[!ht]
\begin{center}
\vspace{-1.2cm}
\epsfig{file=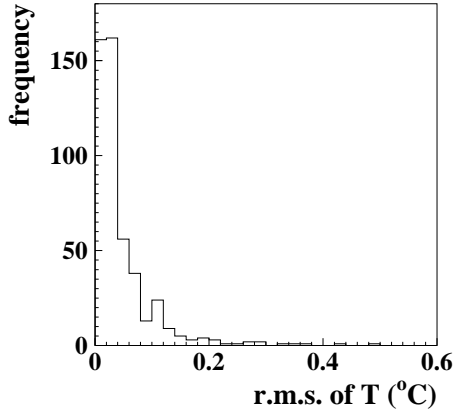,height=6.5cm}
\vspace{-.6cm}
\end{center}
\caption[]{Distribution of the root mean square (r.m.s.) 
detectors' operating temperature
variation within periods
with the same calibration factors (typically $\simeq$ 7 days)
during the DAMA/NaI-5 to -7 running periods.
The mean value is 0.05 $^o$C.}
\label{fig_rms_T}
\vspace{-0.4cm}
\end{figure}

Information for the new DAMA/NaI-5 to 7 running periods
can be derived from Fig. \ref{fig_temp}; moreover, no evidence of any 
operating
temperature modulation has been observed as quantitatively reported in Table \ref{tb:par567}.
Notwithstanding, to properly evaluate the real effect of possible variations of the
detectors' operating temperature on the light output,
we consider -- according to the procedure given in ref. \cite{Sist} --
the distribution of the root mean square temperature 
variation within periods 
with the same calibration factors (typically $\simeq$ 7 days); 
this is given in Fig.~\ref{fig_rms_T} cumulatively for the three data sets.

Considering the obtained mean value of the root mean square detectors' operating 
temperature variation: $\simeq$ 0.05 $^o$C,
and the known value of the slope of the light output around its value:
$\lsim$ -0.2\%/$^o$C, the 
relative light output variation is $\lsim$ 10$^{-4}$, which corresponds
to $\lsim$ 0.5\% of the modulation amplitude observed in the lowest energy 
region of the production data, $S_m^{obs}$, since 
the counting rate is $\simeq 1.0$ cpd/kg/keV in the region of interest 
\cite{Sist,Diu99} 
and $S_m^{obs}$ is  $\simeq 0.02$ cpd/kg/keV (see previous section).

Thus, any significant effect from the detectors' operating temperature is further
excluded.
As in ref \cite{Sist}, for the sake of completeness, we comment that 
sizeable temperature variations could also induce variations in
the electronic noise, in the Radon release from the rocks and, therefore, in
the environmental background;
these specific topics
will be further  analysed in the following, where cautious upper limits on their 
possible effect are given.

Finally, it is worth to remark that any hypothetical effect induced by 
temperature variations would fail at least some 
of the six requirements needed to mimic the annual modulation 
signature (such as e.g. the $4^{th}$ and the $5^{th}$).

In conclusion, all the arguments given above
exclude any role of possible effects on the observed rate modulation 
correlated with temperature.

\subsubsection{The noise}

Despite the stringent used noise rejection procedure (see refs.
\cite{Nim98,Mod2,Mod3,Sist} and the
brief mention in \S \ref{app}), the role 
of possible noise tail in the data after the noise rejection has been 
quantitatively investigated
\cite{Sist}.

\begin{figure}[!h]
\vspace{-2.2cm}
\epsfig{file=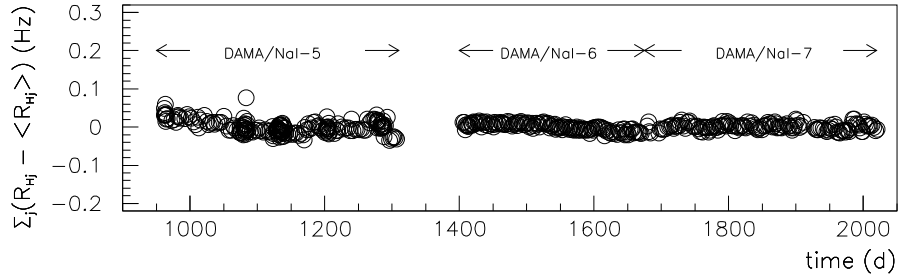,height=14cm}  
\vspace{-9.0cm}
\caption[]{Time behaviour of the variable $R_H = \Sigma_j (R_{Hj} - <R_{Hj}>)$, where
$R_{Hj}$ is the hardware rate of each one of the nine
detectors above single photoelectron threshold (that is including the noise), 
$j$ identifies the detector and $<R_{Hj}>$ is the mean value of $R_{Hj}$ in the
corresponding running period.}
\label{fig_stab_rh}
\vspace{-0.1cm}
\end{figure}

In particular, the hardware rate of each one of the nine
detectors above a single photoelectron, $R_{Hj}$ ($j$ 
identifies the detector), can be considered; 
in fact, it is significantly determined 
by the noise. For this purpose the variable
$R_H = \Sigma_j (R_{Hj} - <R_{Hj}>)$ can be built (where in our case
$<R_{Hj}> \lsim 0.25$ Hz \cite{Nim98}); its time behaviour 
during the DAMA/NaI-5 to -7 running periods 
is shown in Fig. \ref{fig_stab_rh}.

\begin{figure}[!ht]
\begin{center}
\vspace{-0.9cm}
\mbox{\epsfig{file=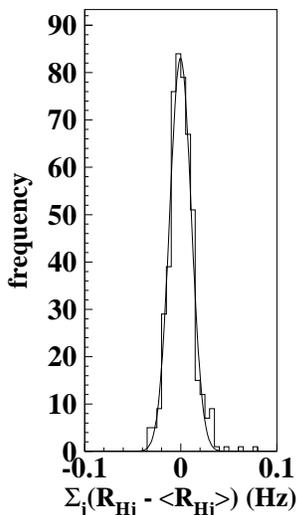,height=7cm}}  
\vspace{-0.4cm}
\caption[]{Distributions of R$_H$ during the DAMA/NaI-5 to -7 running periods; 
 see text.}
\label{fg:fig_rh}
\end{center}
\vspace{-0.6cm}
\end{figure}

As it can be seen in Fig. \ref{fg:fig_rh},
the cumulative distribution of $R_H$ for 
the  DAMA/NaI-5 to -7 running periods 
shows a gaussian behaviour
with $\sigma$ = 0.5\%, value well in agreement with that expected 
on the basis of simple statistical arguments.

Moreover, by fitting the time behaviour of R$_H$ in the three data taking periods -- including a
WIMP-like modulated term --
a modulation amplitude compatible with zero:
$-(0.06 \pm 0.11) \cdot 10^{-2}$ Hz, is obtained. From this value 
the upper limit at 90\% C.L. on the modulation amplitude 
can be derived: $<$ 1.3 $\cdot$ 10$^{-3}$ Hz.
Since the typical noise contribution to the hardware rate 
of each one of the 9 detectors is
$\simeq$ 0.10 Hz, the upper limit on the noise relative 
modulation amplitude is given by:
$ \frac{1.3 \cdot 10^{-3} Hz} {9 \times 0.10 Hz} \simeq 1.4 \cdot 10^{-3}$  (90\% C.L.).
Therefore, even in the worst hypothetical case of a 
10\% contamination of the residual noise -- after rejection -- in the 
counting rate, the noise contribution to the modulation   
amplitude in the lowest energy bins would be 
$<$ 1.4 $\cdot$ 10$^{-4}$ of the total counting rate.
This means that an hypothetical noise modulation could account at maximum
for absolute amplitudes of the order of few 10$^{-4}$ cpd/kg/keV, that is
$<$1\% of the observed annual modulation amplitude \cite{Mod3}.

In conclusion, there is no evidence for any role of an hypothetical tail of
residual noise after rejection.

\subsubsection{The efficiencies}

The behaviour of the used efficiencies during the 
whole data taking periods has even been investigated. 
\begin{figure}[!hb]
\begin{center}
\vspace{-0.8cm}
\hspace{2.cm}
\mbox{\epsfig{file=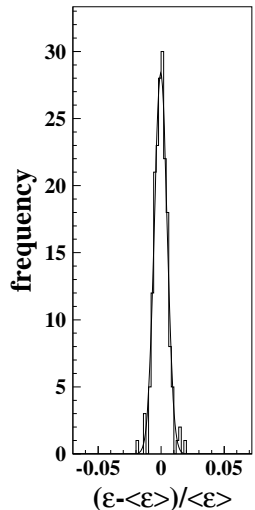,height=7cm}}
\end{center}
\vspace{-0.8cm}
\caption[]{Distribution of the percentage variations
of the efficiency values with the respect to their mean values during the DAMA/NaI-5 to 
-7 running periods; see text.}
\label{fig_eff}
\vspace{-0.2cm}
\end{figure}   
Their possible 
time variation depends essentially on the stability
of the efficiencies related to the  previously mentioned
acceptance windows, which are regularly measured by dedicated 
calibrations (see e.g. ref. \cite{Nim98,Mod2,Mod3}).

In particular, we show in Fig.~\ref{fig_eff}
the percentage variations of the efficiency values
in the (2-8) keV energy interval considering 2 keV bins. They
show a gaussian distribution with $\sigma$ = 0.5\%
for DAMA/NaI-5 to -7, cumulatively. Moreover, we have verified
that the time behaviour of these percentage variations
does not show any modulation with period
and phase expected for a possible WIMP signal.
In Table \ref{tb:eff567}
the modulation amplitudes of the efficiencies in each energy
bin between 2 and 10 keV are reported, showing that they are all consistent with zero.
In particular, modulation amplitudes -- considering the three periods 
together -- equal to $(0.7 \pm 1.0) \cdot 10^{-3}$ and
$(0.1 \pm 0.8) \cdot 10^{-3}$ are found in the (2-4) keV and
(4-6) keV energy bins, respectively; both consistent with zero. 

\begin{table}[!ht]
\caption{Modulation amplitudes 
obtained by fitting the time behaviour of the
efficiencies including a WIMP-like cosine
modulation for 
the DAMA/NaI-5 to -7 running periods.}
\vspace{0.4cm}
\centering
\begin{tabular}{|c|c|c|c|} \hline
  & \multicolumn{3}{c|}{Amplitude ($\times 10^{-3}$)}  \\
  & & & \\
Energy  & DAMA/NaI-5 & DAMA/NaI-6 & DAMA/NaI-7\\
  & & & \\
\hline
  & &  & \\
2-4 keV   & $(1.0 \pm 3.3)$ & $(1.8  \pm 1.5)$  & $ -(0.4 \pm 1.5)$ \\
4-6 keV   & $(1.6 \pm 2.3)$ & $(0.7  \pm 1.3)$  & $ -(0.9 \pm 1.2)$ \\
6-8 keV   & $(1.0 \pm 1.8)$ & $-(0.1 \pm 1.0)$  & $  (0.3 \pm 1.0)$ \\
8-10 keV  & $(0.7 \pm 1.3)$ & $(0.3  \pm 0.8)$  & $  (1.5 \pm 1.0)$ \\
 & & &  \\
\hline\hline
\end{tabular}
\label{tb:eff567}
\end{table}

Thus, also the unlikely idea of a possible role played by the 
efficiency values in the observed effect is ruled out.

\subsubsection{The calibration factor}

In long term running conditions, the
knowledge of the energy scale
is assured by periodical calibration with $^{241}$Am source and
by continuously monitoring within the same production data
(grouping them each $\simeq$ 7 days)
the position and resolution of the $^{210}$Pb peak
(46.5 keV), mentioned in \S \ref{app} \cite{Nim98,Mod1,Mod2,Mod3,Sist}.
Although it is highly unlikely that a variation of the calibration factor
(proportionality factor between the area of the recorded pulse  
and the energy), $tdcal$,
could play any role, 
according to e.g. ref. \cite{Sist}
a quantitative
investigation on that point has been carried out.

\begin{figure}[!ht]
\begin{center}
\vspace{-0.3cm}
\hspace{1.1cm}
\mbox{\epsfig{file=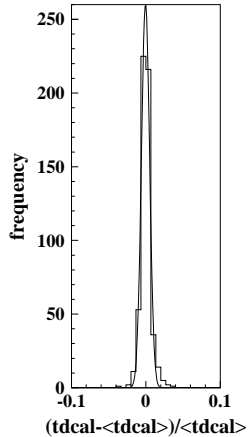,height=6cm}}   
\end{center}
\vspace{-0.5cm}
\caption[]{Distribution of the percentage variations
of the energy scale factors ($tdcal$) in the DAMA/NaI-5 to -7 running periods 
without applying any correction; see text.
The standard deviation is 0.5\%.}
\label{fig_tdcal}
\vspace{-0.2cm}
\end{figure}

For this purpose, the distribution of the relative variations of 
$tdcal$ -- without applying any correction --
estimated from the position of the $^{210}$Pb peak
for all the 9 detectors during the DAMA/NaI-5 to -7 
running periods is given 
in Fig.~\ref{fig_tdcal}. This distribution
shows a gaussian behaviour with
$\sigma \lsim 0.5\%$. Since the results of the routine calibrations
are obviously properly taken into account in the data analysis,  
such a result allows to conclude that
the energy calibration factor for each detector
is known with an uncertainty $\ll 1\%$ within every 7 days interval.

As discussed also in ref. \cite{Sist},
the variation of the calibration factor for each detector,
within each interval of $\simeq$ 7 days, would give rise to an additional energy
spread ($\sigma_{cal}$) besides the detector energy resolution
($\sigma_{res}$). The total
energy spread can be, therefore, written as: $\sigma = \sqrt{\sigma^2_{res} +
\sigma^2_{cal}} \simeq \sigma_{res} \cdot
[1+\frac{1}{2} \cdot (\frac{\sigma_{cal}}{\sigma_{res}})^2]$; 
clearly the contribution due to the calibration factor variation
is negligible since
$\frac{1}{2} \cdot (\frac{\sigma_{cal}/E}{\sigma_{res}/E})^2 \lsim   
7.5 \cdot 10^{-4} \frac{E}{20 keV} $ (where the adimensional ratio
$\frac{E}{20 keV}$ accounts for the energy dependence of this limit value).
This order of magnitude is confirmed by a MonteCarlo calculation,
which credits -- as already reported in ref. \cite{Sist} 
a maximum value of the effect of similar variations of $tdcal$ on the
modulation amplitude equal to $1.6 \cdot 10^{-4}$, giving
an upper limit $< 1\%$ of the modulation amplitude measured at   
very low energy.

Thus, also the unlikely idea that the calibration factor could 
play a role can be safely ruled out.

\subsubsection{The background}

Similarly as done for the previous data sets (see e.g. ref. \cite{Sist}),
in order to verify the absence of any significant background modulation, 
the energy distribution measured during the data taking periods
in energy regions not of interest 
for the WIMP-nucleus elastic scattering 
can be investigated in order to verify if 
the modulation detected in the lowest energy 
region could be ascribed to a background modulation.
In fact, the background in the lowest energy region can be 
essentially due to "Compton" electrons, X-rays and/or Auger
electrons, muon induced events, etc., which are strictly correlated
with the events in the higher energy part of the spectrum.
Thus, if a modulation with time detected 
in the lowest energy region would be due to
a modulation of the background (instead of the possible signal) with time,
an equal or larger (sometimes much larger)
modulation in the higher energy regions should be present.
For this purpose, we have investigated the rate 
integrated above 90 keV,  R$_{90}$, as a function of the time.

\begin{figure}[!ht]
\begin{center}
\vspace{-0.3cm}
\hspace{1.1cm}
\mbox{\epsfig{file=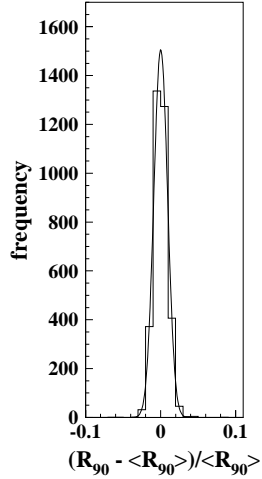,height=6.5cm}}   
\end{center}
\vspace{-0.3cm}
\caption[]{Distributions of the percentage variations
of R$_{90}$ with respect to the mean values for all the detectors in the DAMA/NaI-5 to -7 
running periods; see text.}
\label{fig_r90}
\vspace{-0.2cm}
\end{figure}

In Fig.~\ref{fig_r90} the distribution of the percentage variations
of R$_{90}$ with respect to the mean values for all the nine detectors
during the DAMA/NaI-5 to -7 running periods is given. 
They show cumulative gaussian behaviours
with $\sigma$ $\simeq$ 0.9\%, well accounted by the statistical 
spread expected from the used sampling time.
This result excludes any significant background variation.

Moreover, fitting
the time behaviour of R$_{90}$, a WIMP-like modulation amplitude 
compatible with zero is found in each running period:
($0.09 \pm 0.32$) cpd/kg, 
($0.06 \pm 0.33$) cpd/kg and  -($0.03 \pm 0.32$) cpd/kg for DAMA/NaI-5, DAMA/NaI-6
and DAMA/NaI-7, respectively.
This excludes the presence of a background
modulation in the whole energy spectrum at a level much
lower than the effect found
in the lowest energy region;
in fact, otherwise -- considering the R$_{90}$ mean values --
a modulation amplitude of order of tens cpd/kg, that is
$\simeq$ 100 $\sigma$ 
far away from the measured value, would be present.

A similar analysis performed on the data collected just above the energy
region, where
the modulation
is observed, that is in (6 -- 10) keV, gives: -(0.0076 $\pm$ 0.0065)
cpd/kg/keV,
(0.0012 $\pm$ 0.0059) cpd/kg/keV and 
(0.0035 $\pm$ 0.0058) cpd/kg/keV for the three periods, respectively.

The results of this subsection also demonstrate
that the production data satisfy the $4^{th}$ requirement 
of the WIMP annual modulation signature.

Notwithstanding the results given above already account
also for the background component due to the neutrons, 
a further additional 
independent and cautious analysis to estimate their possible contribution
has been given in ref. \cite{Sist}.
In particular, 
the effect of the about 1 m concrete (made from the Gran Sasso
rock material) which, as known, almost fully surrounds (outside the barrack) 
the DAMA/NaI passive shield -- acting as a further neutron moderator -- 
has not been cautiously included
in these estimates, which are recalled in the following.

As regards the thermal neutrons, the neutron capture reactions 
$^{23}Na(n,\gamma)^{24}Na$ and
$^{23}Na(n,\gamma)^{24m}Na$ (cross section to thermal neutrons
equal to 0.10 and 0.43 barn, respectively \cite{Led}) have been investigated.
The capture rate is: $\simeq 0.2$ captures/day/kg since 
the thermal neutron flux has been measured to 
be 1.08 $\cdot 10^{-6}$ neutrons $\cdot$ cm$^{-2} \cdot$ s$^{-1}$ 
\cite{Neu89} \footnote{
Consistent upper limit on the thermal neutron flux 
have been obtained with the $\simeq$ 100 kg DAMA NaI(Tl) set-up
considering 
these same capture reactions \cite{Nim98}.}.
Assuming cautiously a 10\% modulation of the thermal neutrons flux, 
the corresponding modulation amplitude in the lowest energy region
has been calculated by MonteCarlo 
program to be $< 10^{-5}$ cpd/kg/keV, that is
$<0.05\%$ of the modulation amplitude we found in the lowest energy interval of
the production data. In addition, a similar contribution cannot anyhow 
mimic the annual modulation signature since it would fail some of 
the six requirements quoted in \S \ref{sc:sign} (such as e.g. the $4^{th}$ 
and the $5^{th}$).

A similar analysis has also been carried out for the fast neutrons case \cite{Sist}.
From the fast neutron flux 
measured at the Gran Sasso underground 
laboratory,
$0.9 \cdot 10^{-7}$ neutrons $\cdot$ cm$^{-2} \cdot$ s$^{-1}$ \cite{Cri},
the differential counting rate above 2 keV has been estimated
by MonteCarlo to be 
$\simeq 10^{-3}$ cpd/kg/keV. Therefore, assuming -- also in this case --
cautiously a 10\% modulation of the fast neutron flux,
the corresponding modulation amplitude is 
$<0.5\%$ of the modulation amplitude found in the lowest energy interval. 
Moreover, also in this case
some of the six requirements mentioned above would fail.

Finally, possible side reactions have been also carefully searched for. 
The only process which has been found 
as an hypothetical possibility is the muon flux modulation reported 
by the MACRO experiment \cite{Mac97}. In fact, MACRO has observed 
that the muon flux shows a nearly sinusoidal time behaviour 
with one year period and maximum in the summer with amplitude
of $\simeq$ 2 \%; this muon flux modulation is correlated
with the temperature of the atmosphere. 
A simple calculation to estimate the 
modulation amplitude expected 
from this process in our set-up has been introduced in ref. \cite{Sist} and is recalled in the following.
In fact, the muon flux ($\Phi_\mu$) and the
yield of neutrons produced by muons measured at the underground
Gran Sasso National Laboratory ($Y$) 
are:
$\Phi_\mu \simeq 20 $ muons m$^{-2}$d$^{-1}$ \cite{Mac97}
and 
$Y \simeq (1$ -- $7) \cdot 10^{-4}$ neutrons per muon per g/cm$^2$ \cite{Agl},
respectively. Thus,
the fast neutron rate produced by muons is given by: 
$R_n = \Phi_\mu \cdot Y \cdot M_{eff}$,
where $M_{eff}$ is the effective mass 
where muon interactions can give rise to events detected in the DAMA 
set-up.
Consequently, the annual modulation amplitude 
in the lowest energy region induced in our experiment by a muon flux 
modulation as measured by MACRO \cite{Mac97} can be estimated as:
$S_m^{(\mu)} = R_n \cdot g \cdot \epsilon \cdot f_{\Delta E} \cdot 
f_{single} \cdot 2\% / (M_{set-up} \cdot \Delta E)$,
where $g$ is a geometrical factor, $\epsilon$ is the detection efficiency 
for elastic scattering interactions, $f_{\Delta E}$ is the acceptance
of the considered 
energy window (E $\ge$ 2 keV),
$f_{single}$ is the "single hit" efficiency and 2\% is the MACRO 
measured effect. Since 
$M_{set-up} \simeq$ 100 kg and $\Delta E \simeq$  4 keV, assuming the very
cautious values $g \simeq \epsilon \simeq f_{\Delta E} \simeq f_{single} 
\simeq 0.5$ 
and $M_{eff}$ = 15 t, one obtains:
$S_m^{(\mu)} < (1$ -- $7) \cdot 10^{-5}$ cpd/kg/keV, that is 
$<0.3\%$ of the 
modulation amplitude we observe \cite{Mod3}. We stress that -- in addition --
the latter value has been 
overestimated of orders of magnitude both because of the extremely cautious 
values assumed in the
calculation and, as mentioned, of the omission of the effect of the
$\simeq$ 1 m concrete neutron moderation. Finally,
we remark that not only the modulation of the muon flux observed by MACRO 
would give rise in our set-up to a quantitatively 
negligible effect, but -- in addition -- 
some of the six requirements necessary to mimic 
the annual modulation signature (such as e.g. the $4^{th}$ and 
the $5^{th}$) would fail. 
Therefore, it can be safely ignored.

Just for the sake of completeness, we remind that the contribution of solar neutrinos,
whose flux is also expected to be modulated, is many orders of magnitude 
lower than the measured rate \cite{Naisol}.

In conclusion, the results presented in this section demonstrate
that the production data satisfy the $4^{th}$ requirement of a WIMP induced effect 
and -- at the same time -- exclude that
the annual modulation observed
in the lowest energy region could be ascribed to modulation of any kind of possible background.

\subsection{Conclusions on the DAMA/NaI model independent result}

No modulation has been
found in any of the considered 
possible source of systematics; thus, upper limits (90\% C.L.)  
on the possible contributions to a modulated amplitude
have been calculated and are summarized in Table \ref{tb:sist}.
\begin{table}[h]
\vspace{-0.3cm}
\caption{Summary of the results obtained by investigating possible sources of 
systematics
or of side reactions
in the data of the DAMA/NaI-5 to 7 running periods.
None able to give a modulation amplitude different from zero has been found; 
thus cautious upper limits (90\% C.L.)
on the possible contributions to a modulation amplitude
have been calculated and are shown here
in terms of the measured model independent modulation amplitude, $S_m^{obs}$, we have
observed (see \S \ref{sc:evi}).
As it can be seen none (nor
their cumulative) effect could account for the 
measured modulation; moreover, as discussed in details already in
ref.\cite{Sist},
none of them could mimic the signature.}
\vspace{-0.2cm}
\begin{center}
\begin{tabular}{|c|c|c|}
\hline \hline
 Source    & Main comment &  Cautious upper limit \\
& & (90\%C.L.) \\
\hline\hline
Radon    & Sealed Cu Box in  &  $<0.2\% S_m^{obs}$\\
         & HP Nitrogen atmosphere &  \\
\hline
Temperature   & Air conditioning  & $<0.5\% S_m^{obs}$\\
\hline
Noise     & Efficient rejection & $<1\% S_m^{obs}$\\
\hline
Energy scale & Routine  & $<1\% S_m^{obs}$\\
  & + intrinsic calibrations & \\
\hline
Efficiencies    & Mainly routine measurements & $<1\% S_m^{obs}$\\
\hline
Background      &  No modulation observed   & $<0.5 \% S_m^{obs}$ \\
                &  above 6 keV; this limit & \\ 
                &  includes possible effect & \\
                &  of thermal and fast neutrons &  \\
\hline
Side reactions & From muon flux variation& $<0.3\% S_m^{obs}$ \\
               & measured by MACRO  & \\
\hline
\multicolumn{3}{|c|} {In addition: no effect can mimic the signature} \\
\hline \hline
\end{tabular}
\end{center}
\label{tb:sist}
\vspace{-0.2cm}
\end{table}
In particular, they cannot account for the
measured modulation because quantitatively not relevant and, as discussed 
in details already in ref. \cite{Sist},
none of them is able to mimic the observed effect; in fact, 
none can satisfy all the above mentioned peculiarities of the signature.

The quantitative investigations discussed above offer a complete 
analysis of known sources of possible systematic effects. We can 
conclude that a relative systematic error, affecting the energy 
spectrum, of order of $\lsim 10^{-3}$ is credited by these 
investigations, while the results on the analysis of R$_{90}$
exclude the presence of a possible background modulation even 
at more stringent level. Furthermore, no systematic effect 
or side reaction able to mimic a WIMP induced effect, that is to be not only
quantitatively significant, but also able to 
satisfy all the many requirements of the signature (see \S \ref{sc:sign}), 
has been found.

\vspace{0.3cm}

In conclusion, the presence of an annual modulation in the residual rate 
in the lowest energy interval (2 -- 6) keV 
with all the features expected for a WIMP component in the galactic halo
is supported by the data at 6.3 $\sigma$ C.L.

\section{Corollary results: quests for a candidate particle in some model
frameworks}

On the basis of the previous main result, a corollary 
investigation can also be
pursued on the nature and coupling of a WIMP candidate.
This latter investigation is instead model
dependent and -- considering the large uncertainties which exist on the 
astrophysical, nuclear and particle physics
assumptions and parameters needed in the calculations -- has no general
meaning (as it is also the case of exclusion plots and of the WIMP parameters evaluated in indirect
search experiments). Thus, it should be
handled in the most general way as we have preliminarily pointed out with time in the past
\cite{Mod1,Mod2,Ext,Mod3,Sist,Sisd,Inel,Hep} and we will again show 
in the following sections. The results we will discuss are, of course, not exhaustive
and many other different allowed regions can be obtained by varying the 
assumptions within the many possible model frameworks, which at present level of knowledge 
cannot be disentangled (e.g. open questions: i) which is the right nature for the
WIMP particle;
ii) which is its right couplings with ordinary matter;  iii) which are the right form factors and
related parameters for each target nucleus; iv) 
which is the right spin factor for each target nucleus; v) which are the right scaling laws; vi) which
is the
right halo model and
related parameters; vii) which are the right values of the experimental parameters which we can
determine
only with associated uncertainties;  etc.).
The situation is analogous for results presented in form of exclusion plots, which also have 
not an "universal" meaning, as well as for the results of WIMP specific parameters
inferred by indirect searches.

\subsection{Estimates of WIMP direct detection rates}
\label{sc:rate}

In the following the main bases necessary to perform model 
dependent analyses in this field are given.
The approximations and intrinsic uncertainties can be inferred straight-forward.

\subsubsection{WIMP-nucleus elastic scattering}
\label{sc:rate1}

The studied process is the WIMP-nucleus elastic scattering and the
measured quantity is the recoil energy.
In the general form
the differential energy distribution of the recoil nuclei
can be calculated \cite{Psd96,Boat}
by means of the differential cross section of the WIMP-nucleus elastic 
processes
\begin{eqnarray}
\frac{d\sigma}{dE_R}(v,E_R) = \left( \frac{d \sigma}{dE_{R}} \right)_{SI}+
\left( \frac{d \sigma}{dE_{R}} \right)_{SD} = 
\nonumber \\
= \frac{2 G_F^2 m_N}{\pi v^2} 
\textrm{\{} \left[ Zg_p + (A-Z)g_n \right]^2 F^2_{SI}(E_R) + 
8\Lambda^2 J(J+1) F_{SD}^2(E_R) \textrm{\}} ,
\label{eq:prosezdiff}
\end{eqnarray}
where: $G_F$ is the Fermi coupling constant; $m_N$ is the nucleus mass;
$v$ is the WIMP velocity in the laboratory frame; 
$E_R = m^2_{WN} v^2 (1-cos \theta^*)/m_N$  
(with $m_{WN}$ WIMP-nucleus reduced mass and $\theta^*$ scattering angle 
in the WIMP-nucleus c.m. frame) is the recoil energy; $Z$ is the 
nuclear charge and  $A$ is the atomic number;
$g_{p,n}$ are the effective WIMP-nucleon coupling strengths for SI interactions;
$\Lambda^2 J(J+1)$ is a spin factor. Moreover, 
$F^2_{SI}(E_R)$
is the SI form factor (see later), while 
$F_{SD}^2(E_R)$  is the SD form factor (see later)
for which an universal formulation is not possible since 
in this case 
the internal degrees of the WIMP particle model (e.g. supersymmetry
in case of neutralino) cannot be completely decoupled 
from the nuclear ones.
It is worth to note that this 
adds significant uncertainty in the model dependent 
results.

Furthermore, it can be demonstrated 
that $\Lambda = \frac{a_p < S_p > + a_n < S_n >}{J}$ with $J$ nuclear spin,
with $a_{p,n}$ effective WIMP-nucleon coupling strengths for SD interaction and with $<S_{p,n}>$
mean values of the nucleon spins in the nucleus.
Therefore, the differential cross section and,
consequently, the expected energy distribution depend on the WIMP mass and on 
four unknown parameters of the theory: $g_{p,n}$ and $a_{p,n}$. 

The total cross section for WIMP-nucleus elastic scattering
can be obtained by integrating equation (\ref{eq:prosezdiff}) over 
$E_R$ up to $E_{R,max} = \frac {2 m^2_{WN} v^2} {m_N}$:
\begin{eqnarray}
 \sigma(v) & = & \int^{E_{R,max}}_0
\frac{d\sigma}{dE_{R}}(v, E_R) dE_R =
 \frac {4} {\pi} G_F^2 m^2_{WN} 
\textrm{\{} \left[ Zg_p + (A-Z)g_n \right]^2 G_{SI}(v) + 
\nonumber \\
& & + 8 \frac {J+1} {J} \left[ a_p <S_p> + a_n <S_n> \right]^2 G_{SD}(v)
\textrm{\}}.
\label{eq:cs1}
\end{eqnarray}
\noindent Here $G_{SI}(v) = \frac {1} {E_{R,max}} \int^{E_{R,max}}_0    
F^2_{SI}(E_R) dE_R$; $G_{SD}(v)$ can be derived straightforward.

The standard point-like cross section 
can be evaluated in the limit $v \rightarrow 0$ (that is 
in the limit $G_{SI}(v)$ and $G_{SD}(v)$ $\rightarrow 1$).
Knowing that $<S_{p,n}> = J = 1/2$ for single nucleon, 
the SI and SD point-like cross sections on proton and on neutron
can be written as:
\begin{equation}  
\sigma^{SI}_{p,n}=\frac{4}{\pi} G_F^2 m_{W(p,n)}^2 g_{p,n}^2 \\
\sigma^{SD}_{p,n}=\frac{32}{\pi}\frac{3}{4} G_F^2 m_{W(p,n)}^2a_{p,n}^2,
\end{equation}
where $m_{Wp} \simeq m_{Wn}$ are the WIMP-nucleon reduced masses.

As far as regards the SI case, the first term within squared 
brackets in eq. (\ref{eq:cs1}) can be arranged in the form 
\begin{equation} 
\left[ Zg_p + (A-Z)g_n \right]^2 = \left( \frac {g_p + g_n} {2} \right)^2 \left[
1 - \frac {g_p - g_n} {g_p + g_n} \left(1 - \frac {2 Z} {A}\right) \right]^2
A^2  = g^2 \cdot A^2 .
\end{equation} 
Considering $\frac {Z} {A}$ nearly constant for the nuclei typically
used in direct searches for Dark Matter particles, the coupling term $g$
is generally assumed -- in a first approximation -- as independent on the 
used target nucleus. Under this assumption, the nuclear parameters can be decoupled from the
particle parameters and 
a generalized SI WIMP-nucleon cross section: 
$\sigma_{SI} = \frac{4}{\pi} G_F^2 m_{Wp}^2 g^2$, can be conveniently introduced.

As far as regards the SD couplings, let us now introduce the useful notations \cite{Sisd}
\begin{equation}
\bar{a}=\sqrt{a_p^2+a_n^2}, \\
tg \theta = \frac{a_n}{a_p}, \\
\sigma_{SD} =\frac{32}{\pi}\frac{3}{4} G_F^2 m_{Wp}^2\bar{a}^2,
\end{equation}
where $\sigma_{SD}$ is a suitable SD
WIMP-nucleon cross section. 
The SD cross sections on proton and neutron can be, then, written
as: 
\begin{equation}  
\sigma^{SD}_{p}=\sigma_{SD} \cdot cos^2 \theta \\
\sigma^{SD}_{n}=\sigma_{SD} \cdot sin^2 \theta.
\end{equation}

In conclusion, equation (\ref{eq:prosezdiff}) can be re-written in terms of 
$\sigma_{SI}$, $\sigma_{SD}$ and $\theta$ as:
\begin{eqnarray}
\frac{d\sigma}{dE_R}(v,E_R) = \frac{m_N}{2m_{Wp}^2v^2} \cdot
\Sigma(E_R) ,
\label{eq:prosezdiff4}
\end{eqnarray} 
with 
\begin{eqnarray}
\Sigma(E_R) & = &
\textrm{\{} A^2\sigma_{SI} F_{SI}^2(E_R)+ \nonumber \\
& & + \frac{4}{3} \frac{(J+1)}{J} \sigma_{SD} \left[ <S_p>\cos \theta + 
<S_n>\sin \theta \right]^2  F_{SD}^2(E_R) \textrm{\}}.
\label{eq:prosezdiff41}
\end{eqnarray}
The mixing angle $\theta$ is defined in the $\left[ 0, \pi \right)$ interval;
in particular, $\theta$ values in the second sector 
account for $a_p$
and $a_n$ with different signs.
As it can be noted from its definition \cite{Boat}, 
$F_{SD}^2(E_R)$ depends on $a_p$ and $a_n$ only 
through their ratio and, consequently, depends on $\theta$,
but it does not depend on $\bar{a}$.

Finally, setting the local WIMP density, $\rho_W = \xi \rho_0$,
where $\rho_0$ is the local halo density and $\xi$ \footnote{Pay attention that in ref. 
\cite{Mod2,Mod3,Sisd,Inel} the same symbol indicates instead a different quantity:
$\xi = \rho_W / (0.3 GeV cm^{-3})$.}
($\xi\leq 1$)
is the fractional amount of local WIMP density,
and the WIMP mass, $m_W$,
one can write the energy distribution of the recoil rate ($R$) in the form
\begin{eqnarray}
\frac{dR}{dE_R} & = & N_{T}\frac{\rho_{W}}{m_W}\int^{v_{max}}_{v_{min}(E_{R})} 
\frac{d\sigma}{dE_{R}}(v,E_{R}) v f(v) dv = \nonumber \\
& & N_{T} \frac{\rho_0 \cdot m_N}{2 m_W \cdot m_{Wp}^2} 
\xi \Sigma(E_R) I(E_R),
\label{labelt}
\end{eqnarray}
where: $N_T$ is the number of target nuclei and 
$I(E_{R})=\int^{v_{max}}_{v_{min}(E_{R})}dv\frac{f(v)}{v}$
with $f(v)$ WIMP velocity distribution in the Earth frame; 
$v_{min} = \sqrt{\frac {m_N \cdot E_R}{2 m^2_{WN}}}$ 
is the minimal WIMP velocity providing $E_R$ recoil energy; 
$v_{max}$ is the maximal WIMP velocity in the halo
evaluated in the Earth frame. 

The differential distribution of the detected
energy, $E_{det}$, for a mul\-ti\-ple-nuc\-lei de\-tec\-tor (as e.g. the NaI(Tl))
can be easily derived:

 \begin{eqnarray}
\frac{dR}{dE_{det}}(E_{det}) & = &
\int K(E_{det},E') \cdot
\sum_{x=nucleus}
\frac{dR_x}{dE_R}\left(E_R=\frac{E'}{q_x}\right) 
\cdot dE',
\label{labelmul}
\end{eqnarray}

\noindent where $q_x$ is the quenching factor for the $x$ recoiling nucleus and
$K(E_{det},E')$ takes into account the response and energy resolution of the
detector; generally it has a gaussian behaviour.

 It is worth to remark, as it can
be inferred by eq. (\ref{eq:prosezdiff}),
that only nuclei with spin different from zero
are sensitive to WIMPs with both SI and SD couplings. 
This is the case of the $^{23}$Na and $^{127}$I nuclei,
odd-nuclei with an unpaired proton, constituents of the DAMA/NaI 
detectors. Thus, the purely SI coupling scenario widely
considered in this field represents only 
a particular case of the more general framework of a WIMP candidate 
with both mixed SI and SD couplings.
Therefore, in the following analyses, we will consider some of the possible scenarios for the
mixed SI and SD couplings and, then, also the sub-cases 
of pure SI and pure SD couplings.

\subsubsection{WIMPs with {\it preferred inelastic} scattering}
\label{sc:inel}

It has been suggested \cite{Wei01} also the possibility that the
annual modulation of the low energy rate observed by DAMA/NaI 
could be induced by 
WIMPs with {\it preferred inelastic} scattering: relic particles that cannot scatter elastically 
off nuclei. As discussed in ref. \cite{Wei01}, the inelastic Dark Matter 
could arise from a massive complex scalar split into two approximately 
degenerate real scalars or from a Dirac fermion split into two 
approximately degenerate Majorana fermions, namely $\chi_+$ and $\chi_-$,
with a $\delta$ mass splitting. In particular, a specific
model featuring a real component of the sneutrino,
in which the mass splitting naturally arises, has been given in ref.
\cite{Wei01} and mentioned here in \S \ref{sc:part}.
The detailed discussion of the theoretical arguments 
on such inelastic Dark Matter can be found in ref. \cite{Wei01}. 
In particular, there has been shown that for the $\chi_-$ inelastic scattering 
on target nuclei a kinematical constraint exists which favours
heavy nuclei (such as $^{127}$I) with respect to 
lighter ones (such as e.g. $^{nat}$Ge) as target-detectors media.
In fact, $\chi_{-}$ can only inelastically scatter
by transitioning to $\chi_{+}$ (slightly heavier state than $\chi_{-}$)
and this process can occur
only if the $\chi_{-}$ velocity, $v$, is larger than:
\begin{equation}
v_{thr} = \sqrt{\frac{2\delta}{m_{WN}}}.
\label{eq:constraint}
\end{equation}
This kinematical constraint becomes increasingly severe 
as the nucleus mass, $m_N$, is decreased \cite{Wei01}. 
For example, if $\delta \gsim$ 100 keV, a signal rate 
measured e.g. in Iodine will be a factor about 10 or more 
higher than that measured in Germanium \cite{Wei01}.
Moreover, this model scenario implies some peculiar features when 
exploiting the WIMP annual modulation signature 
\cite{Freese}; in fact -- with respect to the
case of WIMP elastically scattering -- it would give rise to an enhanced 
modulated component, $S_m$, with respect to the unmodulated one, $S_0$,
and to largely different behaviours with energy for
both $S_0$ and $S_m$ (both show a higher mean value) \cite{Wei01}.

The {\em preferred inelastic} Dark Matter scenario \cite{Wei01}
offers further possible model frameworks and has also the merit to naturally 
recover the sneutrino as a WIMP candidate (see e.g. \S \ref{sc:inel}).

The differential energy distribution of the recoil nuclei
in the case of inelastic processes can be calculated 
by means of the differential cross section of the WIMP-nucleus inelastic 
processes:
\begin{eqnarray}
{d\sigma \over d\Omega^*} = 
\frac {G_F^2 m_{WN}^2} {\pi^2}
\left[ Zg_p + (A-Z)g_n \right]^2 F^2_{SI}(q^2) 
\cdot \sqrt{1 - \frac {v_{thr}^2} {v^2}} , 
\label{eq:rate1}
\end{eqnarray}
where $d\Omega^*$ is the 
differential solid angle in the WIMP-nucleus c.m. frame; 
$q^{2}$ is the squared three-momentum transfer. 

In the inelastic process the recoil energy depends on the 
scatter angle, $\theta^*$, in the c.m. frame according to:
\begin{equation}
E_R = \frac {2 m_{WN}^2 v^2} {m_N} \cdot \frac {1-\frac{v_{thr}^2} {2 v^2}-
\sqrt{1- \frac {v_{thr}^2} {v^2}} \cdot cos \theta^*}{2}.
\label{eq:erec1}
\end{equation}
Thus, we can write:
\begin{equation}
dE_R = \frac {2 m_{WN}^2 v^2} {m_N} \cdot \sqrt{1- \frac {v_{thr}^2} {v^2}} 
\cdot \frac {d\Omega^*}{4\pi}.
\label{eq:erec2}
\end{equation}

From eq. (\ref{eq:rate1}) and (\ref{eq:erec2}) 
we derive the differential cross section as a
function of the recoil energy, $E_R$, and the WIMP velocity, $v$:
\begin{eqnarray}
\frac{d\sigma}{dE_R}(v,E_R) = 
\frac{2 G_F^2 m_N}{\pi v^2} 
\left[ Zg_p + (A-Z)g_n \right]^2 F^2_{SI}(E_R).
\label{eq:prosezdiff2}
\end{eqnarray}
Here we apply the relation $q^2 = 2 m_N E_R$.

The minimal WIMP velocity, $v_{min}(E_R)$,  providing $E_R$ recoil 
energy in the inelastic process is:
\begin{equation} 
v_{min}(E_R) = 
\sqrt{\frac{m_N E_R}{2 m_{WN}^2}} \cdot \left( 1 +
\frac {m_{WN} \delta} {m_N E_R} \right) ,
\label{eq:vmin}
\end{equation} 
and it is always $\ge v_{thr}$.

Finally, one can write the energy distribution of the recoil rate ($R$) in the form
\begin{eqnarray}  
\frac{dR}{dE_R} & = & N_{T}\frac{\rho_{W}}{m_W}\int^{v_{max}}_{v_{min}(E_{R})}
\frac{d\sigma}{dE_{R}}(v,E_{R}) v f(v) dv = \nonumber \\
& & N_{T} \frac{\rho_{0} \cdot m_N}{2 m_W \cdot m_{Wp}^2}
\cdot A^2 \xi\sigma_{p} F_{SI}^2(E_R) \cdot
I(E_R).
\label{labelt1}
\end{eqnarray}  

Moreover, as derived in the case discussed in the previous subsection,
also in the present case a generalized SI point-like WIMP-nucleon
cross section: 
$\sigma_{p} = \frac{4}{\pi} G_F^2 m_{Wp}^2 g^2$, can be defined.
Finally, the extension of formula (\ref{labelt1}) e.g. to detectors with multiple
nuclei can be easily derived.

In this scenario the modulated and the unmodulated components of the signal
are function of $\xi\sigma_p$, $m_W$ and $\delta$.

\subsubsection{The halo models}
\label{sc:halo}

As discussed above, the expected counting rate for the WIMP elastic 
scattering depends on the local WIMP density, $\rho_W$, and on the 
WIMP velocity distribution, $f(v)$, at Earth's position. 
The experimental observations regarding the dark halo of our Galaxy
do not allow to get information on them without introducing 
a model for the Galaxy matter density. An extensive discussion about 
models has been reported in ref. \cite{Hep}. Here we present a brief 
introduction on this argument both to allow the reader to understand the 
complexity of this aspect and 
in the light of the
results given in 
following subsections on the discussed 
quests.

Important information on the dark halo in the Galaxy can be derived
from measurements of the rotational velocities of objects bounded
in the gravitational galactic field. In fact, 
the following relation between the rotational velocity
of an object placed at distance $r \equiv |\vec{r}|$ from the center of
the Galaxy and the total mass, $M_{tot}(r) = \int_{r'<r}d^3 r' \rho_{tot}(\vec{r'})$, 
inside the radius $r$ can be obtained from the virial theorem:
\begin{equation}
v^2_{rot}(r) = \frac{GM_{tot}(r)}{r}
\label{eq:hh1}
\end{equation}
where $G$ is the Newton's constant.
The total mass density, $\rho_{tot}(\vec{r})$, can be expressed as the sum of 
the mass density of the dark halo,
$\rho_{DM}(\vec{r})$, and the mass density of the visible component,
$\rho_{vis}(\vec{r})$
that constitutes the bulge and the disk of the Galaxy.
The gravitational potential $\Psi(\vec{r})$ is related 
to $\rho_{tot}$ through the Poisson's equation:
\begin{equation}
\nabla^2 \Psi = -4 \pi G \rho_{tot}.
\label{eq:pois}
\end{equation}
The halo density profile $\rho_{DM}(\vec{r})$ can 
also be expressed in term of the  distribution function 
of the WIMP in the six-dimensional phase space $F(\vec{r},\vec{v})$:
\begin{equation}
\rho_{DM}(\vec{r}) \equiv \int F(\vec{r},\vec{v}) \, d^3v
\label{eq:rodm}
\end{equation}
where $\vec{r}$  and $\vec{v}$ represent the position and velocity vectors 
in the rest frame of the Galaxy respectively. 

Inverting eq. (\ref{eq:rodm}) and taking into account observational data
it is, in principle, possible to calculate $F(\vec{r},\vec{v})$ and, then, the WIMP
velocity distribution function at the Earth position in the Galaxy:
\begin{equation}
f(\vec{v}) \equiv F(\vec{R_0},\vec{v}) \label{eq:df}
\end{equation}
where $R_0 \simeq 8.5$ kpc corresponds to the Earth distance 
from the center of the Galaxy along the direction of the galactic 
plane ($\vec{R_0} \equiv (R_0,0,0)$). 
The inversion of eq. (\ref{eq:rodm}) presents degeneracy problem
that can be solved only requiring some degree of symmetry 
for the system. The velocity distribution function 
represents therefore an important source of uncertainties 
in the evaluation of expected counting rate for the WIMP component in the galactic halo.

As well as the velocity distribution function, two other quantities 
are of great importance to estimate the expected counting rate:
the WIMP local velocity, $v_0 = v_{rot}(\vec{R_0})$, and the 
local halo density $\rho_0 \equiv \rho_{DM}(\vec{R_0})$  that appears
as a multiplicative factor in the formula giving the expected counting rate.
Since we are interested in the evaluation of local and rotational
velocity of the WIMP at distance $r = R_0$, it is not necessary for
our purpose a detailed description of the inner part of the Galaxy
($r \ll R_0$) where disk and bulge are dominant. Thus, it is possible to
consider the bulge as a spherical density distribution with relevant
contribution up to about 1 kpc and truncated at $r \simeq 2$ kpc from
the center of Galaxy, and the disk as an exponential distribution
which (according to most of the models) decreases up to 4 kpc where it
can safely be neglected. Therefore, it is generally assumed that for $r \gsim
R_0$ the
dark matter is the dominant component. 

The contribution of the visible matter has been considered in the
calculation of the WIMP local velocity: 
\begin{equation}
v_0^2 = v_{rot}^2(R_0)=\frac{G}{R_0} \left[  M_{vis}(R_0) + M_{halo}(R_0)\right].
\end{equation}
A \textit{maximal halo}, $\rho_0^{max}$, occurs when $M_{vis}(R_0) \ll
M_{halo}(R_0)$; in this case the contribution to the rotational velocity
is due to the halo; on the other hand, when for $M_{vis}$ the maximum
value compatible with observations is considered,  a \textit{minimal
halo}, $\rho_0^{min}$, occurs and only a fraction of $v_0$ is
supported by the dark halo.

The WIMP halo can be represented as a collisionless 
gas of particles whose distribution function satisfy the Boltzmann
equation  \cite{Bi87}. In general case the Boltzmann equation cannot
be solved without reducing the complexity of the system. 
It is possible to consider models based on the Jeans
or on the virial equations that describe a wide range of systems, but one
cannot be sure that these models describe systems with realizable 
equilibrium configuration \cite{Bi87}.

The dark halo model widely used in the calculations carried out in the WIMP 
direct detection approaches is the simple isothermal sphere 
that corresponds to a spherical infinite system with a flat rotational
curve. The halo density profile is:
\begin{equation}
\rho_{DM}(r)= \frac{v_0^2}{4 \pi G} \frac{1}{r^2}
\label{eq:sf_ro}
\end{equation}
corresponding to the following potential:
\begin{equation}
 \Psi_0(r) = -\frac{v_0^2}{2} \log{(r^2)}.
\label{eq:sf_pot}
\end{equation}
In this case, when a maximal halo density is considered, 
the WIMP velocity distribution is the Maxwell function:
\begin{equation}
f(v) = N {\rm exp} \left(-\frac{3v^2}{2v_{rms}^2}\right)
\end{equation}
where $N$ is the normalization constant.
The mean square velocity results: $v^2_{r.m.s.} = (3/2) v_0^2$;
this relation descends from the hypothesis of an halo formed by
particles in hydrostatic equilibrium with an isotropic  
velocity distribution.
Despite the simplicity of this model has favoured its wide use 
in the calculation of expected rate of WIMP-nucleus  
interaction, it doesn't match with astrophysical observations
regarding the sphericity of the halo and the absence of rotation,
the flatness of the rotational curve and the isotropy of the
dispersion tensor, and it presents unphysical behavior: 
the density profile, in fact, has a singularity in the origin and
implies a total infinite mass of the halo unless introducing 
some cut-off at large radii.

\begin{table}[!hbt]
\begin{center}
\vspace{-0.7cm}
\caption{\label{tb:models}
  Summary of the consistent halo models considered in the analysis of ref. \cite{Hep}
  and in the following.
  The labels in the first column identify the models. 
  In the third column the values of the related considered parameters are reported \cite{Hep};
  Other choices are also possible as well as other halo models.
  In the last column references to the corresponding equations in the text are listed.
  The models of the Class C have also been considered including
  possible co--rotation and counter-rotation of the dark halo (see eq. (\ref{eq:corot}).)
}
\begin{tabular}{|c|l|c|c|}
\hline\hline
\multicolumn{3}{|l|}{{\bf Class A:  spherical $\bf \rho_{DM}$,
isotropic velocity dispersion}} & eq. \\
\hline
A0 & {\rm ~Isothermal Sphere}   &     &     (\ref{eq:sf_ro}) \\
A1 & {\rm ~Evans' logarithmic} \cite{Ev93} & $R_c=5$ kpc & (\ref{eq:r_iso}) \\
A2 & {\rm ~Evans' power-law} \cite{Ev94} & $R_c=16$ kpc, $\beta=0.7$
 & (\ref{eq:r_iso2}) \\
A3 & {\rm ~Evans' power-law} \cite{Ev94} & $R_c=2$ kpc, $\beta=-0.1$ &
 (\ref{eq:r_iso2}) \\
A4 & {\rm ~Jaffe} \cite{Ja83}              & $\alpha=1$, $\beta=4$,
$\gamma=2$, $a=160$ kpc & (\ref{eq:univ_rho}) \\ 
A5 & {\rm ~NFW} \cite{nfw}                 & $\alpha=1$, $\beta=3$,
$\gamma=1$, $a=20$ kpc  & (\ref{eq:univ_rho}) \\ 
A6 & {\rm ~Moore et al.} \cite{moore}      & $\alpha=1.5$, $\beta=3$,
$\gamma=1.5$, $a=28$ kpc  & (\ref{eq:univ_rho}) \\ 
A7 & {\rm ~Kravtsov et al.} \cite{kravtsov}& $\alpha=2$, $\beta=3$,
$\gamma=0.4$, $a=10$ kpc   & (\ref{eq:univ_rho}) 
\\
\hline
\multicolumn{3}{|l|}{{\bf Class B: spherical $\bf \rho_{DM}$,
non--isotropic velocity dispersion    }} & \\
\multicolumn{3}{|l|}{{\bf (Osipkov--Merrit, $\bf \beta_0=0.4$)}} &\\
\hline
B1 & {\rm ~Evans' logarithmic} & $R_c=5$ kpc & (\ref{eq:r_iso})(\ref{eq:beta0}) \\
B2 & {\rm ~Evans' power-law} & $R_c=16$ kpc, $\beta=0.7$  & (\ref{eq:r_iso2})(\ref{eq:beta0}) \\
B3 & {\rm ~Evans' power-law} & $R_c=2$ kpc, $\beta=-0.1$ &
(\ref{eq:r_iso2})(\ref{eq:beta0}) \\
B4 & {\rm ~Jaffe}           & $\alpha=1$, $\beta=4$,
$\gamma=2$, $a=160$ kpc  & (\ref{eq:univ_rho})(\ref{eq:beta0}) \\
B5 & {\rm ~NFW}             & $\alpha=1$, $\beta=3$,
$\gamma=1$, $a=20$ kpc   & (\ref{eq:univ_rho})(\ref{eq:beta0})  \\
B6 & {\rm ~Moore et al.}    & $\alpha=1.5$, $\beta=3$,
$\gamma=1.5$, $a=28$ kpc   & (\ref{eq:univ_rho})(\ref{eq:beta0}) \\
B7 & {\rm ~Kravtsov et al.} &  $\alpha=2$, $\beta=3$,
$\gamma=0.4$, $a=10$ kpc    & (\ref{eq:univ_rho})(\ref{eq:beta0}) \\
\hline
\multicolumn{3}{|l|}{{\bf Class C:  Axisymmetric $\bf \rho_{DM}$}} &\\
\hline
C1 & {\rm ~Evans' logarithmic} & $R_c=0$, $q=1/\sqrt{2}$ &
(\ref{eq:pot_ax})(\ref{eq:ro_ax}) \\
C2 & {\rm ~Evans' logarithmic} & $R_c=5$ kpc, $q=1/\sqrt{2}$ &
(\ref{eq:pot_ax})(\ref{eq:ro_ax}) \\
C3 & {\rm ~Evans' power-law} & $R_c=16$ kpc, $q=0.95$, $\beta=0.9$
& (\ref{eq:pot_pl})(\ref{eq:ro_pl}) \\
C4 & {\rm ~Evans' power-law} & $R_c=2$ kpc, $q=1/\sqrt{2}$, $\beta=-0.1$
& (\ref{eq:pot_pl})(\ref{eq:ro_pl}) \\
\hline
\multicolumn{3}{|l|}{{\bf Class D: Triaxial $\bf \rho_{DM}$ \cite{Ev00}
  ($\bf q=0.8$, $\bf p=0.9$)}} & \\
\hline
D1 & {\rm ~Earth on maj. axis, rad. anis.}    & $\delta=-1.78$  &
(\ref{eq:potential_triaxial})(\ref{eq:rho_triaxial}) \\
D2 & {\rm ~Earth on maj. axis, tang. anis. }    &   $\delta=16$ &
(\ref{eq:potential_triaxial})(\ref{eq:rho_triaxial}) \\
D3 & {\rm ~Earth on interm. axis, rad. anis.}  &  $\delta=-1.78$ &
(\ref{eq:potential_triaxial})(\ref{eq:rho_triaxial}) \\
D4 & {\rm ~Earth on interm. axis, tang. anis.} & $\delta=16$ &
(\ref{eq:potential_triaxial})(\ref{eq:rho_triaxial}) \\
\hline\hline
\end{tabular}
\label{tb:halo}
\end{center}
\vspace{-0.5cm}
\end{table}

In the ref. \cite{Hep} the analysis of the first 4
DAMA/NaI annual cycles in a particular case for a SI coupled WIMP candidate 
has been extended by considering a large number
of self-consistent galactic halo models, in which the variation of the
velocity distribution function is originated from the change of the 
halo density profile or of the potential. The different models have been
classified in 4 classes according to the symmetry properties of the density profile
or of the gravitational potential and of the velocity distribution function.
The same strategy has been followed to obtain the new cumulative results given later.
The considered halo model classes correspond to: spherically symmetric matter density with
isotropic velocity dispersion (A); spherically symmetric matter density with
non-isotropic velocity dispersion (B); axisymmetric models (C);
triaxial models (D). The models are summarized in Table \ref{tb:models}
where, according to ref. \cite{Hep}, are identified by a label.

We will present briefly in the following these models since they will be considered 
in the new results on the quest for  possible candidate particle given in 
the following subsections. For a detailed
discussion refers to the ref. \cite{Hep}. 

\vspace{0.3cm}
\noindent{\em I. Spherical halo models with isotropic velocity
dispersion (A)} 
\vspace{0.3cm} 

The first class groups models with spherical density profile;
for these models $\rho(\vec{r})=\rho(r)$ and
$f(\vec{v})=f(v)$. The first type of model is a generalization of the
spherical isothermal sphere in which a core radius $R_c$ is
introduced. The density profile becomes (model $A1$):
\begin{equation}
 \rho_{DM}(r) = \frac{v_0^2}{4 \pi G} \frac{3R_c^2 + r^2}{(R_c^2 + r^2)^2} ,
\label{eq:r_iso}
\end{equation}
with, in case of maximal halo, corresponding potential:
\begin{equation}
 \Psi_0(r) = -\frac{v_0^2}{2} \log{(R_c^2 + r^2)}.
 \label{eq:pot_iso}
\end{equation}
In the limit $R_c \rightarrow 0$ the profile (\ref{eq:sf_ro}) and the potential (\ref{eq:sf_pot}) is obtained.
These models are also named {\em logarithmic} because of the analytic form of the 
potential.

A second class of spherical models are defined by the following matter
density profile (models $A2$ and $A3$):
\begin{equation}
 \rho_{DM}(r) = \frac{\beta \Psi_a R_c^{\beta}}{4 \pi G} \;
 \frac{3R_c^2 + r^2(1-\beta)}{(R_c^2 + r^2)^{(\beta + 4)/2}} ,
\label{eq:r_iso2}
\end{equation}
and potential for a maximal halo:
\begin{equation}
 \Psi_0(r) = \frac{\Psi_a R_c^{\beta}}{(R_c^2 + r^2)^{\beta/2}}
  \qquad (\beta \neq 0).
 \label{eq:pot_iso2}
\end{equation}
We will refer to these models as {\em power-law} halo models. They represent the
spherical limit of the more general axisymmetric model discussed 
later. When the parameter $\beta \rightarrow 0$, the 
logarithmic models are obtained.

The last family of spherical models is described by the matter density
distribution (models $A4$ -- $A7$):
\begin{equation}
 \rho_{DM}(r) = \rho_0 \left( \frac{R_0}{r} \right)^{\gamma}
 \left[  \frac{1+(R_0 / a)^{\alpha}}{1+ (r/a)^{\alpha}} \right]^{(\beta -\gamma)/\alpha}.
\label{eq:univ_rho}
\end{equation}
The different choice of the parameters: $\alpha, \beta, \gamma$ and
$a$, used in the calculations given later are reported in Table
\ref{tb:models}; other choices are possible. 
The density profile of these  models, except for the
Jaffe case, has been obtained from  numerical simulations of Galaxy
evolution.   

\vspace{0.3cm}
\noindent{\em II. Spherical halo models with non-isotropic velocity
dispersion (B)} 
\vspace{0.3cm}       

These models have been studied in the simple case in which the 
velocity distribution function depends on the two integrals of motion
energy and angular momentum vector ($L = |\vec{L}|$) only through the
so called Osipkov-Merrit variable \cite{Bi87,Osi79}:
\begin{equation}
  Q = \epsilon - \frac{L^2}{2r_a^2},
\label{eq:ospm}
\end{equation}
where the parameter $r_a$ appears in the definition of the $\beta_0$,
the degree of anisotropy of the velocity dispersion tensor on the
Earth's position \cite{Osi79}: 
\begin{equation}
  \beta_0 = 1- \frac{\overline{v_{\phi}}^2 } { \overline{v_{r}}^2}
  = \frac{R_0^2}{R_0^2 + r_a^2}.
\label{eq:beta0}
\end{equation}
In this definition the velocity is expressed in spherical
coordinates and $\overline{v_{\phi}} = \overline{v_{\theta}} \neq \overline{v_{r}}$
(con $\overline{v_i}^2 \equiv <v_i^2> - <v_i>^2, i=\phi,\theta,r$).

The considered models are the same as in the isotropic case and
the velocity distribution function has been calculated introducing the
Osipkov-Merrit term in the equations. The degree of anisotropy of the
models depends on the $\beta_0$ values; for $\beta_0 \rightarrow 1$,
or $\overline{v_{\phi}}^2 = \overline{v_{r}}^2$, the distribution
function becomes isotropic.

\vspace{0.3cm}
\noindent{\em III. Axisymmetric models (C)}
\vspace{0.3cm}  

In these models the velocity distribution
depends in general at least on the energy $\epsilon$ and
on the component $L_z$  of the angular momentum  along the axis of
symmetry. The velocity distribution can be written as the sum of an even and an odd
contribution with respect to $L_z$. It can be shown \cite{Ev93,Ev94} that the
$\rho_{DM}$ depends only on the even part and the velocity distribution can be calculated
up to an arbitrary odd part. The 
axisymmetric generalizations of the Evans' logarithmic and power-law
models have been considered in ref. \cite{Hep}. 
For these models the velocity distribution has been calculated
analytically by
Evans \cite{Ev93,Ev94} and corresponds to a maximal halo. 
The axisymmetric logarithmic potential (models $C1$ and $C2$) is:
\begin{equation}
 \Psi_0(R,z) = -\frac{v_0^2}{2}\log \left( R_c^2+R^2+\frac{z^2}{q^2}\right),
\label{eq:pot_ax}
\end{equation}
where $R = (x^2+y^2)$, is the radial coordinate along the galactic
plane and $R_c$ is the core radius; $q$ is the flatness
parameter. The corresponding matter density distribution results:
\begin{equation}
   \rho_{DM}(R,z) = \frac{v_0^2}{4\pi Gq^2} \;
   \frac{(2q^2 + 1)R_c^2+R^2+(2-q^{-2})z^2}{(R_c^2+R^2+z^2q^{-2})^2}.
\label{eq:ro_ax}
\end{equation}

If an asymptotically non-flat rotational curve is considered, the
axisymmetric power-law potential is obtained \cite{Ev94} (models $C3$ and $C4$):
\begin{equation}
\Psi_0(R,z)=\frac{\Psi_a R_c^{\beta}}{(R_c^2+R^2+z^2q^{-2})^{\beta/2}}
\;\;\; (\beta\ne 0).
\label{eq:pot_pl}
\end{equation}
with the distribution function:
\begin{equation}
\rho_{DM}(R,z) = \frac{\beta \Psi_a R_c^{\beta}}{4 \pi G q^2}
~\frac{(2 q^2+1)R_c^2+(1-\beta q^2) R^2+[2-q^{-2}(1+\beta)]z^2}
{(R_c^2+R^2+z^2q^{-2})^{(\beta+4)/2}}.
\label{eq:ro_pl}
\end{equation}

The related velocity distribution functions
for these two cases can be found in ref. \cite{Hep}.

\vspace{0.4cm}
\noindent{\em IV. Co-rotating and counter-rotating halo models} 
\vspace{0.4cm}    

In the case of axisymmetric models it is possible to include an halo
rotation considering that the velocity distribution function
is known up to an arbitrary odd
component. An odd component of velocity distribution function
can easily be defined starting from an
even solution. The velocity distribution function, 
linear combination of even and odd function, is able to describe 
an halo configuration where a particle population moves 
clockwise around the axis of symmetry and a population moves 
in opposite sense. In this case the velocity distribution
can be written as \cite{Hep}:
\begin{equation}
F(\epsilon,L_z)= \eta F_{right}(\epsilon,L_z)+(1-\eta) F_{left}(\epsilon,L_z).
\label{eq:corot}
\end{equation}
The $\eta$ parameter ranges from 1 (maximal co-rotation) to 0 (maximal
counter-rotation) and it is related to the dimensionless spin
parameter $\lambda$ of the Galaxy by: $\lambda=0.36 |\eta-0.5|$ \cite{Kam98}.

Considering the limit $\lambda<0.05$ obtained from numerical work on
Galaxy formation \cite{War92}, $\eta$ can range in the interval $0.36\lsim \eta
\lsim 0.64$. For the Evans' axisymmetric models of class C we have also 
considered possible co--rotation and counter--rotation of the halo
assuming $\eta=0.36$ and $\eta=0.64$.

\vspace{0.5cm}
\noindent{\em V. Triaxial models (D)} 
\vspace{0.3cm} 

The models, belonging to this class, arise from the triaxial potential
discussed in ref. \cite{Ev00}:
\begin{equation}
\Psi_0(x,y,z)=-\frac{1}{2} v_0^2 \log \left (x^2+\frac{y^2}{p^2}+
\frac{z^2}{q^2}\right),
\label{eq:potential_triaxial}
\end{equation}
This potential, in the case of a maximal halo, corresponds to the density profile:
\begin{equation}
\rho_{DM}(x,y,z)=\frac{v_0^2}{4 \pi G}
\frac{A x^2+B y^2/p^2+C z^2/q^2}
{\left (x^2+ y^2/p^2+z^2/q^2\right )^2} ,
\label{eq:rho_triaxial}
\end{equation}
where $A=(p^{-2}+q^{-2}-1)$, $B = (1+q^{-2}-p^{-2})$ and $C = (1+p^{-2}-q^{-2})$.
In ref.\cite{Ev00}, the velocity distribution is approximated by a Gaussian with
$\bar{v}^2_r$, $\bar{v}^2_{\theta}$, $\bar{v}^2_{\phi}$ depending on $v_0$, $\delta$, $p$ and
$q$
parameters an on the Earth position: i) Earth on the major axis of the equipotential ellipsoid
(models $D1$ and $D2$); ii) Earth on the intermediate axis (Models $D3$ and $D4$). 
In these cases the free parameter $\delta$ appears; this parameter,
in the limit of spherical halo ($p=q=1$), measures the degree of
anisotropy of the velocity dispersion tensor:
$\frac{\bar{v}_{\phi}^2}{\bar{v}_r^2}=\frac{2+\delta}{2}$.

\vspace{0.9cm}
\noindent{\em VI. Constraining the models} 
\vspace{0.3cm} 

The parameters of each halo model, given above, have been chosen in ref. \cite{Hep}
taking into account the available observational data. Anyhow, information 
on galactic dark halo can be obtained only in indirect way
\cite{Deh98,Gat96} and considering hypotheses on its form and
characteristic. 

The allowed range of values for the WIMP local velocity has been
estimated there considering the information coming from the rotational curve of
our Galaxy. The considered interval is:
\begin{equation}
v_0=(220 \pm 50)\;\; {\rm km \; s^{-1}} \qquad  {\rm (90\% \;\; C.L.}),
\label{eq:v0}
\end{equation}
that conservatively relies on purely dynamical observations
\cite{koc96}. Similar estimates of the $v_0$ central value with
smaller uncertainty have been obtained studying the proper motion of
nearby stars in the hypothesis of circular orbit of these objects
\cite{Fea97}. In the analyses given in the following, for simplicity, 
we have considered the three
representative values of local velocity: $v_0$ = 170, 220, 270 km/s.

\begin{table}[!h]
\begin{center}
\vspace{-0.6cm}
\caption{Allowed intervals of $\rho_0$ for the
halo models quoted in Table \ref{tb:models} as evaluated in ref. \cite{Hep}.
The $\rho_0^{max}$ and $\rho_0^{min}$ values (in GeV cm$^{-3}$) are here used in the quests for a
candidate particle (see later) for the class $A$ and $B$,
while only the case of $\rho_0^{max}$ is used for the class $C$
and $D$.}
\label{tb:rho}
\vspace{0.3cm}
\begin{tabular}{|c|c|c|c|c|c|c|}
\hline\hline
& \multicolumn{2}{c|} {$v_0 =170$ km s$^{-1}$} & \multicolumn{2}{c|} {$v_0 =220$ km s$^{-1}$}
& \multicolumn{2}{c|} {$v_0 =270$ km s$^{-1}$} \\
\hline
Model &  $\rho_0^{min}$  & $\rho_0^{max}$ &  $\rho_0^{min}$  & $\rho_0^{max}$ &  $\rho_0^{min}$  & $\rho_0^{max}$
\\
\hline
  A0           & 0.18 & 0.28 & 0.30 & 0.47  & 0.45 & 0.71 \\
  A1 ,  B1     & 0.20 & 0.42 & 0.34 & 0.71  & 0.62 & 1.07 \\
  A2 ,  B2     & 0.24 & 0.53 & 0.41 & 0.89  & 0.97 & 1.33 \\
  A3 ,  B3     & 0.17 & 0.35 & 0.29 & 0.59  & 0.52 & 0.88 \\
  A4 ,  B4     & 0.26 & 0.27 & 0.44 & 0.45  & 0.66 & 0.67 \\
  A5 ,  B5     & 0.20 & 0.44 & 0.33 & 0.74  & 0.66 & 1.11 \\
  A6 ,  B6     & 0.22 & 0.39 & 0.37 & 0.65  & 0.57 & 0.98 \\
  A7 ,  B7     & 0.32 & 0.54 & 0.54 & 0.91  & 0.82 & 1.37 \\
  C1           & 0.36 & 0.56 & 0.60 & 0.94  & 0.91 & 1.42 \\
  C2           & 0.34 & 0.67 & 0.56 & 1.11  & 0.98 & 1.68 \\
  C3           & 0.30 & 0.66 & 0.50 & 1.10  & 0.97 & 1.66 \\
  C4           & 0.32 & 0.65 & 0.54 & 1.09  & 0.96 & 1.64 \\
  D1 ,  D2     & 0.32 & 0.50 & 0.54 & 0.84  & 0.81 & 1.27 \\
  D3 ,  D4     & 0.19 & 0.30 & 0.32 & 0.51  & 0.49 & 0.76 \\
\hline\hline
\end{tabular}
\end {center}
\vspace{-0.2cm}
\end{table}

Moreover, in the analyses given in the following, 
for each model --  after fixing the local
velocity -- the
local density $\rho_0$ has been assumed to be 
equal either to the $\rho^{min}_0$ or to the $\rho^{max}_0$ value as obtained in ref. \cite{Hep} 
when imposing the following physical constraints: i) the amount of flatness of the rotational
curve of our Galaxy, considering conservatively $0.8 \cdot v_0  \lsim
v_{rot}^{100}  \lsim  1.2 \cdot v_0$, where $v_{rot}^{100}$ is the
value of rotational curve at distance of 100 kpc from the galactic
center; ii) the maximal non dark halo components in the Galaxy, considering
conservatively $1 \cdot 10^{10} M_{\odot}  \lsim  M_{vis}  \lsim  6
\cdot 10^{10} M_{\odot}$ \cite{Deh98,Gat96}. In particular, 
the allowed intervals for $\rho_0$ are reported in Table
\ref{tb:rho}. For the models of class C and D, as discussed
before, only the case of maximal halo (which correspond to $M_{vis} =
0$) has been considered in the analyses discussed in the following.

\subsubsection{The form factors}
\label{sc:ff}

In order to take into account the finite dimension of the nucleus in the
scattering processes, it is necessary to introduce the nuclear form factor, $F$, that 
generally also depends on the nature of the interaction.

In the case of SI interactions, we can factorize the total cross
section, pointing out the contribution given by the form factor:
\begin{equation}
\sigma (q) = \sigma_{SI} F^2_{SI}(q),
\label{eq:fatt03}
\end{equation}
here $\sigma_{SI}$ is the total cross section when the transfer momentum $q$
is 0. When neglecting possible neutron and proton differences, 
the nuclear form factor can be reasonably described by the Fourier transform of charge density
$\rho(r)$ in the nucleus:

\begin{eqnarray}
F_{SI}(q) = \frac{1}{A} \int \rho (r)
e^{i \vec{q} \cdot \vec{r}} d^3r =  
\frac{1}{A} \frac{4 \pi}{q} \int^{\infty}_0 r   \sin(qr) \rho(r) dr .
\label{eq:effedq}
\end{eqnarray}

In the theoretical calculations and data analysis DAMA/NaI has adopted, for the SI form
factor, the expression suggested by Helm in \cite{Helm56}:
\begin{equation}
F_{SI}(q) = \frac{3j_1(qr_0)}{qr_0} \exp \left[
-\frac{1}{2}s^{2}q^{2} \right],
\label{eq:equ08}
\end{equation}
where $r_0 = \sqrt{r_n^2 - 5s^2}$, $r_n$ is the effective nuclear radius,
$s$ $\simeq$ 1 fm is a parameter that allows to
take into account the thickness of the nuclear surface and
$j_1(qr_0)$ is the spherical Bessel
function of index 1.
We remind that this expression of SI form factor is derived assuming 
a Fermi distribution for the nuclear charge.

\begin{figure}[!ht]
\centering
\vspace{-0.5cm}
\hbox{\psfig{figure=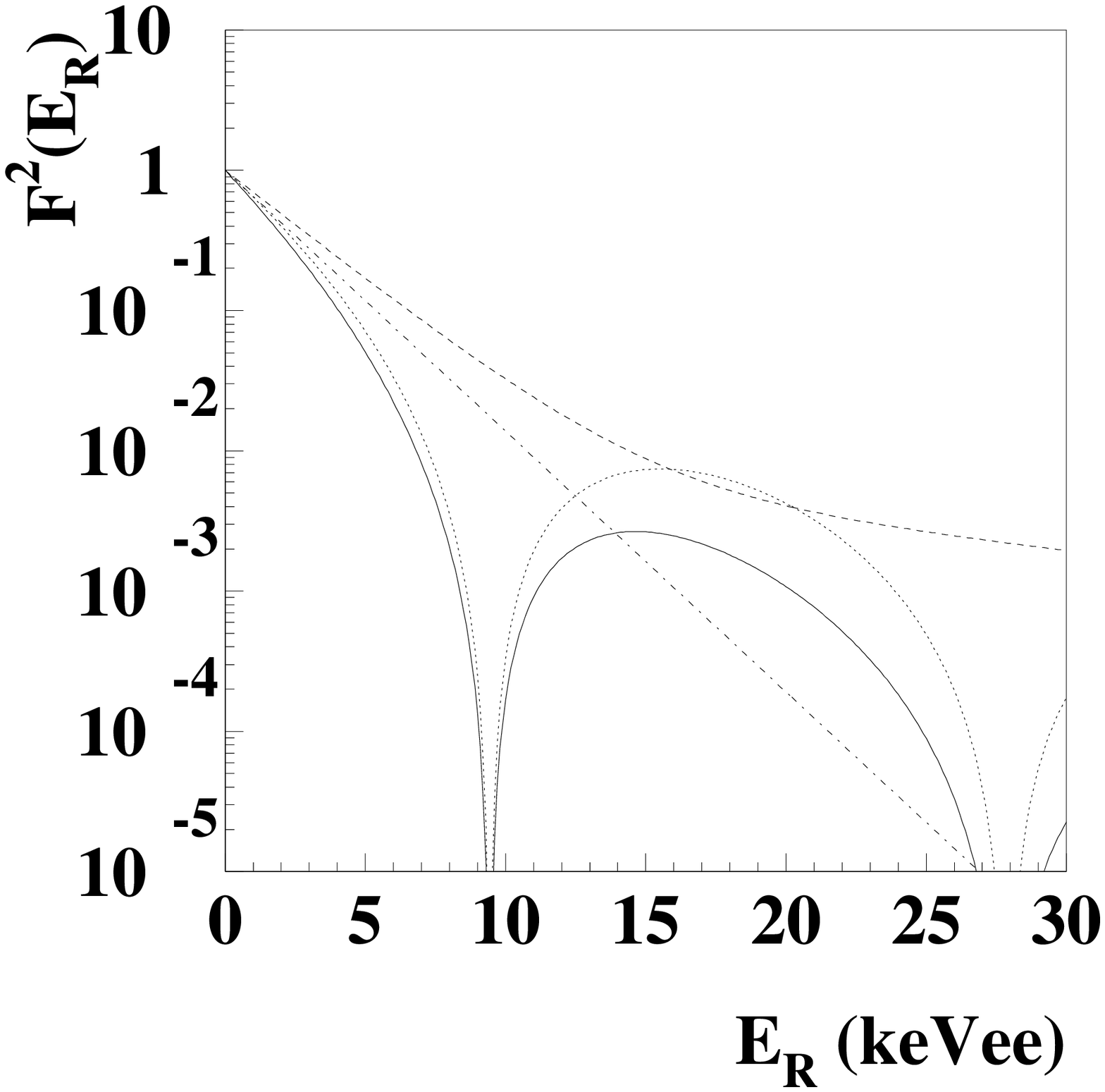,height=6.5cm}
\hspace{-.5cm}
\psfig{figure=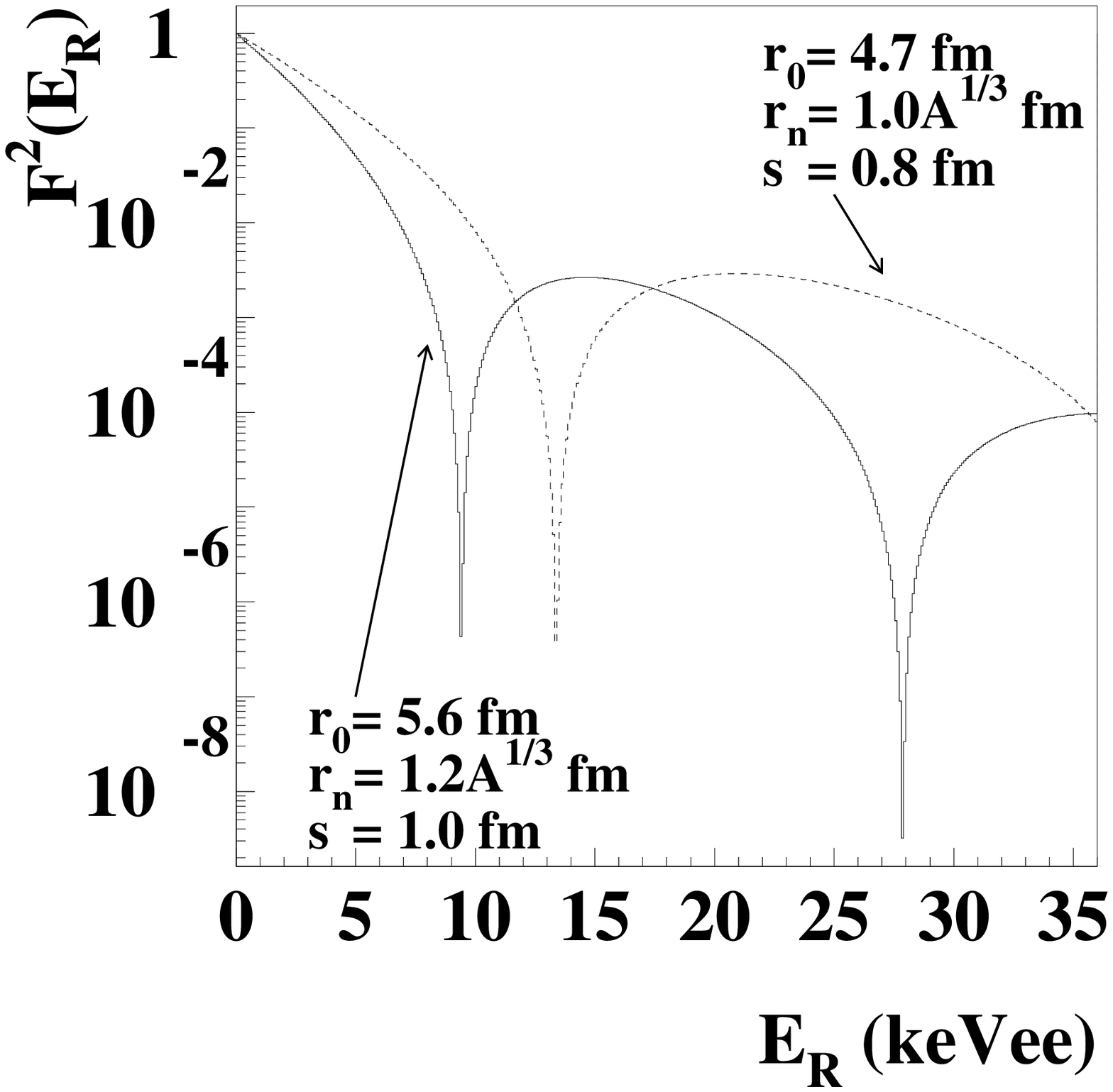,height=6.5cm}}
\vspace{-0.5cm}
\caption{Left panel: some SI nuclear form factors considered so far
in the literature (e.g. in ref. \cite{Smi96}) calculated for the Iodine nucleus.
Right panel: example of the variation of the SI nuclear
form factor, calculated according to ref. \cite{Helm56}, when some uncertainties on
its parameters are considered; as it can be seen,
even a relative small variation 
can produce sizeable change in the behaviour of the nuclear form factor and,
therefore, in the expected SI signal rate. Note that here 
the notation keVee is explicitely mentioned in the panels to indicate
keV electron equivalent in order to remind that the recoil 
energy has been quenched there by using the quenching factor value 
measured by DAMA in ref. \cite{Psd96}.}
\label{fg:ffsi}
\end{figure}

In Fig. \ref{fg:ffsi} on the right, the effect of a relatively small variation
(20\%) of the 
nuclear radius, $r_n$, and the nuclear surface thickness parameter, $s$, in the Helm 
SI form factor is shown; as it can be seen,   
even a relative small variation of these parameters can produce sizeable change 
in the behaviour of the form factor and, therefore, in the expected SI signal 
rate and in the final result.

In addition, also other expressions have been considered in
literature for the SI nuclear form factors of the various nuclei.
Just as an example, in Fig. \ref{fg:ffsi} on the left, SI
form factors discussed e.g. in ref. \cite{Smi96}
are depicted for the Iodine nucleus; the continuous    
line represents the nuclear form factor obtained when using expression
(\ref{eq:equ08}) at the fixed assumed values for the related parameters.        

As mentioned, in the analyses given in the following the most cautious Helm SI form factor
has been adopted taking into account some uncertainties on the nuclear radius and on 
the nuclear surface thickness parameters.
As it can be seen, this form factor is the less favourable one for Iodine
and requires larger SI cross sections for a given signal rate.
For example, in case the other form factor profiles considered in literature
would be used, the allowed regions given in the following sections would extend to
lower cross sections.  

\vspace{0.3cm}

In the case of SD interactions an analytical universal expression for the
form factor does not exist. In fact, 
in this case, the internal degrees of the WIMP particle model (e.g. 
supersymmetry in the case of neutralino) cannot be completely decoupled 
from the nuclear ones. 
Therefore, in order to take into account the property of the nucleus 
interested in the interaction, we have to refer not only to a particular
nuclear model but also to a particular particle physics model.
As an example, if we consider the case of the neutralino in the MSSM model, 
the differential SD cross section can be written from eq. (\ref{eq:prosezdiff}): 
\begin{equation}
\left( \frac{d \sigma}{dE_R} \right)_{SD} = \frac{16G^2_F}{\pi v^2
}\;m_N \; \Lambda^2 \; J(J+1)   \; F^2_{SD}(E_R), \label{eq:sezdf}
\end{equation}
where $F_{SD}(E_R)$ is the SD nuclear form factor defined (see e.g. ref. \cite{res97})
as $F^2_{SD}(E_R) 
= S(q)/S(0)$ with $\pi S(0) = (2J+1)\Lambda^2 J(J+1)$ and:
\begin{equation}
S(q) = a_0^2S_{00}(q) + a_1^2S_{11}(q) + a_0a_1S_{01}(q),
\label{eq:esseq}
\end{equation}
with $a_0 = a_p +a_n$ and $a_1 = a_p - a_n$, where $a_p$ and $a_n$ are
the effective neutralino-nucleon SD coupling strengths defined in \S\ref{sc:rate1}.
The functions $S_{ij}(q)$ generally depend on the considered nuclear model.
For nuclei of interest for Dark Matter detection, (as the $^{127}$I and $^{23}$Na), 
the $S_{ij}(q)$  have been parameterized in ref. \cite{res97} as 
function of the variable $y=(qb/2)^2$ (where $b$ is a parameter of the 
theoretical model, $b \simeq A^{1/6}$ fm):
\begin{equation}
S_{ij} (q) = e^{-2y} \sum_{l=0}^8 C_ly^l,
\end{equation}
where the coefficients $C_l$ generally depend on the adopted nucleon-nucleon potential.
For the $^{23}$Na nucleus only the first three orders of the previous expression have been 
considered in ref. \cite{res97}, while for the $^{127}$I nucleus two different nuclear potentials 
(the Nijmegen II and the Bonn-A ones) have been used in the
evaluation of ref. \cite{res97}.

\begin{figure}[!hb]
\centering
\vspace{-0.5cm}
\hbox{\psfig{figure=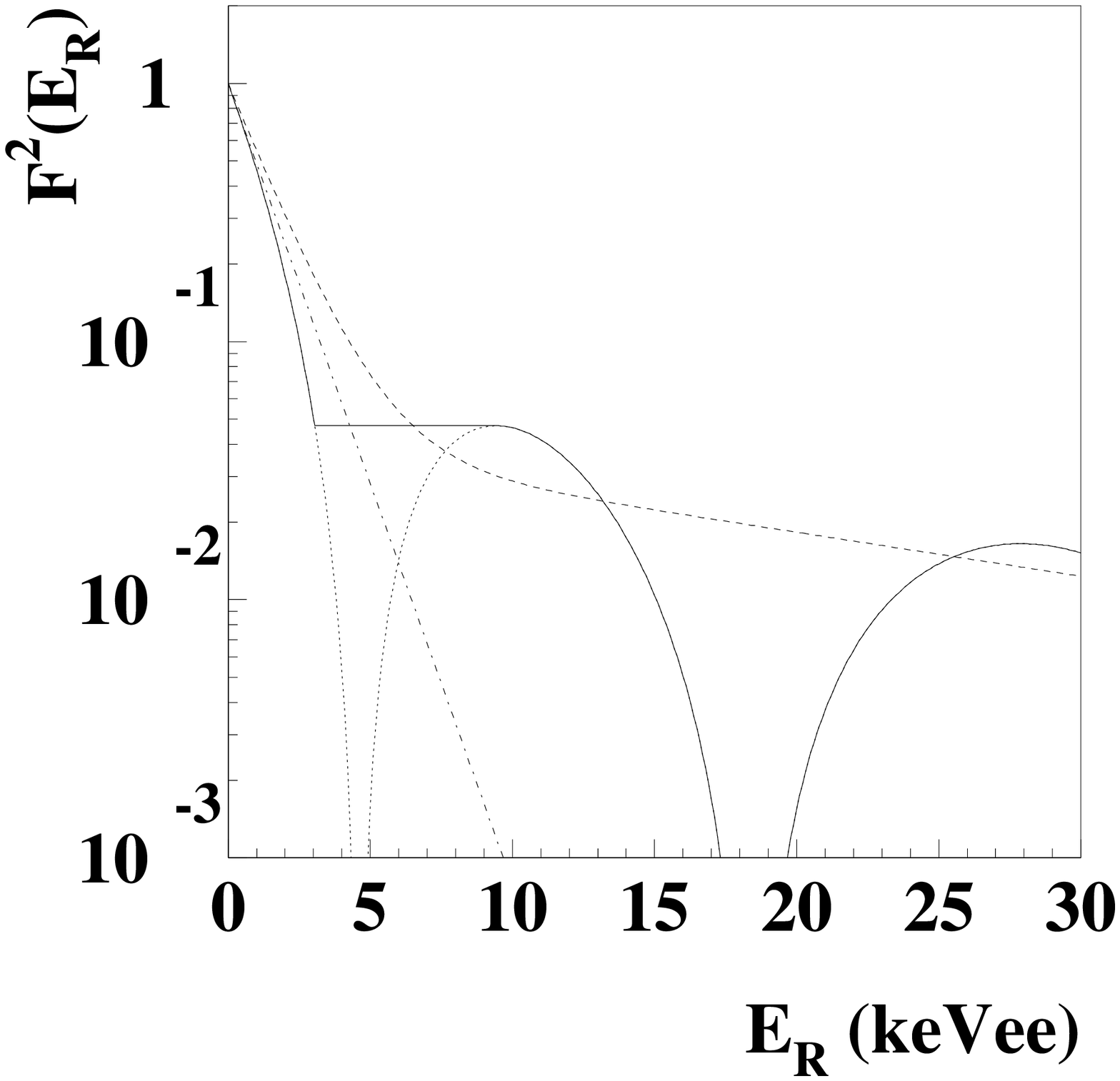,height=6.5cm}
\hspace{-.5cm}
\psfig{figure=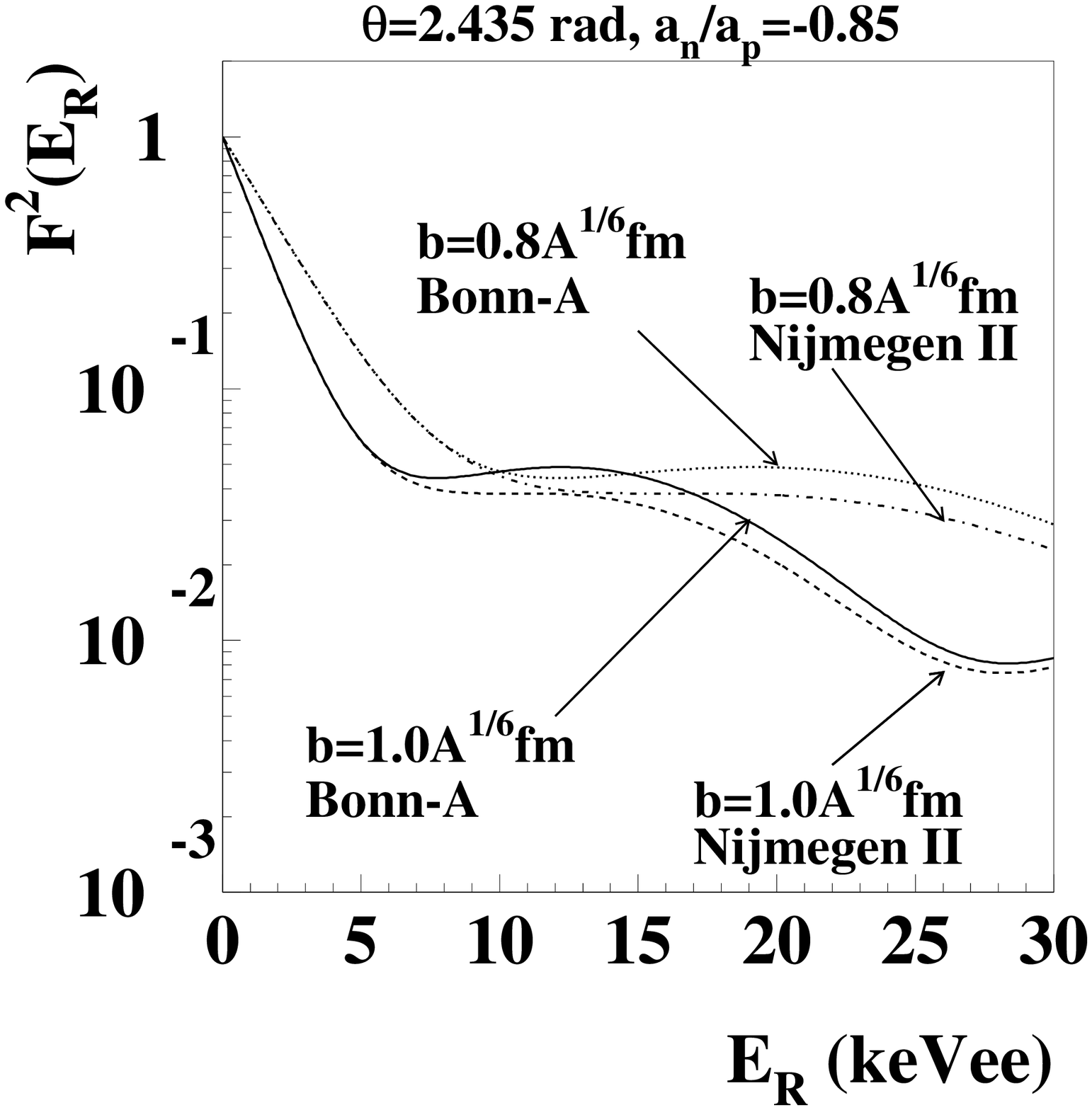,height=6.5cm}}
\vspace{-0.5cm}
\caption{Left panel: 
some SD nuclear form factors considered so far
in the literature (e.g. in ref. \cite{Smi96}) calculated for the Iodine nucleus. 
Right panel: example of variation of the Iodine SD nuclear form factor, calculated 
according to ref. \cite{res97} for two different choices of the nucleon-nucleon 
potential and including also a possible small variation (20\%) of its $b$ parameter. 
As it can be seen, even a relative small variation either of this parameter or 
of the nuclear potential can produce sizeable change in the behaviour of the 
nuclear form factor and, therefore, in the expected SD signal rate.
Note that here
the notation keVee is explicitely mentioned in the panels to indicate
keV electron equivalent in order to remind that the recoil
energy has been quenched there by using the quenching factor value
measured by DAMA in ref. \cite{Psd96}.}
\label{fg:ffsd}
\end{figure}

It is worth to note that the SD form factor depends on the nature 
of the interacting WIMP and on the nuclear potential. Therefore, the SD form factor is 
an important source of uncertainties in the calculation of the expected rate.
As an example, in Fig. \ref{fg:ffsd} on the right the effect of the different choice
of 
the nucleon-nucleon potential and of a relatively small variation (20\%)
of the $b$ parameter in the Iodine SD nuclear form factor, 
calculated according to ref. \cite{res97}, is shown 
(we report here, for simplicity, only the case with 
$\theta = 2.435$ or $a_n/a_p=-0.85$, that is pure $Z_0$ coupling).
Similar uncertainties are present for every nucleus. 

There are other expressions considered in 
literature for the SD nuclear form factor; they can be very 
different. Just as an example Fig. \ref{fg:ffsd} on the left
shows some SD
form factors discussed e.g. in ref. \cite{Smi96}
calculated for the Iodine nucleus.

In the analyses, presented in the following , the SD form factors in the neutralino case
for Sodium and Iodine nuclei calculated by \cite{res97} using -- for the Iodine case --
the nuclear potential by Nijmegen II have been adopted. Analogously, as 
for the SI case, here uncertainties on the $b$ parameter have been 
included in the evaluations of the results.

\vspace{0.3cm}

For the sake of completeness, let us remind that typically 
only purely SI or purely SD $Z_0$ coupling WIMP interactions 
are considered  in the evaluations of the results in this field
among all the wide available  possibilities and that 
the uncertainty on the form factors also largely affects comparisons among  results
obtained by using  
different target nuclei.

\subsubsection{The spin factors}
\label{sc:sf}

Further significant uncertainties in the evaluation of the SD interaction rate arise 
also from the adopted spin factor for the single target-nucleus \cite{res97,Ell91}. 
As an  example, the spin factors of some target-nuclei 
calculated in different models 
are reported in Table \ref{tb:spfac}. Moreover, also for a fixed 
nuclear model, differences arise from the use -- in the calculations --
of different nuclear potentials as it is e.g.  
the case of the 
$^{129}$Xe and of the $^{127}$I nuclei, for which similar calculations
are already available.

As it can be noted in Table \ref{tb:spfac}, since the spin factor is
a multiplicative factor in the expected SD signal rate, its value can
\begin{table}[!ht]
\vspace{-0.3cm}
\caption{Some spin factors estimates assuming simple different models. The values given in this Table
are $\Lambda^2 J(J+1)/a_x^2$, where $a_x$ is either $a_p$ or $a_n$ depending on the unpaired
nucleon.}
\vspace{-0.2cm}
\begin{center}
\begin{tabular}{|c|c|c|c|}
\hline \hline
& & & \\
Target-Nucleus &single particle & odd group & Comment \\
& & & \\
\hline \hline 
$^{29}$Si      & 0.750 & 0.063 & Neutron is   \\
$^{73}$Ge      & 0.306 & 0.065 & the unpaired  \\
$^{129}$Xe     & 0.750 & 0.124 & nucleon      \\
$^{131}$Xe     & 0.150 & 0.055 &              \\
\hline 
$^{1}$H        & 0.750 & 0.750 &              \\
$^{19}$F       & 0.750 & 0.647 &              \\
$^{23}$Na      & 0.350 & 0.041 & Proton is    \\
$^{27}$Al      & 0.350 & 0.087 & the unpaired  \\
$^{69}$Ga      & 0.417 & 0.021 & nucleon      \\
$^{71}$Ga      & 0.417 & 0.089 &              \\
$^{75}$As      & 0.417 & 0.000 &              \\
$^{127}$I      & 0.250 & 0.023 &              \\
\hline \hline
\end{tabular}
\end{center}
\label{tb:spfac}
\vspace{-0.3cm}
\end{table}
also drastically affect the expectations in direct search experiments and,
therefore, also the inferred exclusion plots or allowed regions
can largely vary as well as the results of any comparison.

Moreover, for a complete analysis of a SD component it is worth to remind that
$\theta$ (whose tangent is the ratio between the SD WIMP-neutron and
SD WIMP-proton effective strengths; see \S \ref{sc:rate1}), can continuously 
assume values in the range 0 to $\pi$. 
For example, 
in Table \ref{tb:spfac2} spin factors 
calculated on the basis of ref. \cite{res97} are given for some $\theta$ values 
considering few target nuclei and two different nuclear potentials.
 
\begin{table}[!ht]
\caption{Spin factors
calculated on the basis of ref. \cite{res97} for some of the possible $\theta$ values
considering some target nuclei and two different nuclear potentials.
The values given in this
Table
are $\Lambda^2 J(J+1)/{\bar a}^2$, where ${\bar a}^2$ has been defined in \S \ref{sc:rate1}.}
\vspace{-0.2cm}
\begin{center}
\begin{tabular}{|c|c|c|c|c|}
\hline \hline
& & & &\\
Target-Nucleus /& $\theta$=0 & $\theta$=$\pi$/4 & $\theta$=$\pi$/2  & $\theta$=2.435 \\
nuclear potential & & & &(pure $Z_0$ \\
& & & & coupling) \\
& & & &\\
\hline \hline 
$^{23}$Na              & 0.102 & 0.060 & 0.001          & 0.051       \\
$^{127}$I/Bonn A       & 0.134 & 0.103 & 0.008          & 0.049        \\
$^{127}$I/Nijmegen II  & 0.175 & 0.122 & 0.006          & 0.073        \\
$^{129}$Xe/Bonn A      & 0.002 & 0.225 & 0.387          & 0.135      \\
$^{129}$Xe/Nijmegen II & 0.001 & 0.145 & 0.270          & 0.103       \\
$^{131}$Xe/Bonn A      & 0.000 & 0.046 & 0.086          & 0.033        \\
$^{131}$Xe/Nijmegen II & 0.000 & 0.044 & 0.078          & 0.029        \\
$^{125}$Te/Bonn A      & 0.000 & 0.124 & 0.247          & 0.103         \\
$^{125}$Te/Nijmegen II & 0.000 & 0.156 & 0.313          & 0.132         \\
\hline \hline
\end{tabular}
\end{center}
\label{tb:spfac2}
\vspace{-0.2cm}
\end{table}

In conclusion, not only large differences in the rate can be expected when using
target nuclei sensitive to the SD component of the interaction (such as e.g.
$^{23}Na$ and $^{127}I$) with respect to those largely insensitive to such a
coupling 
(such as e.g. $^{nat}Ge$ and $^{nat}Si$), but also when using different 
target nuclei although all -- in principle --
sensitive to such a coupling (compare e.g. the Xenon and Tellurium case 
with the Sodium and Iodine case 
in Table \ref{tb:spfac2}). 

Moreover, other nuclear models and calculations beyond those reported here
are possible, introducing large uncertainties in the right estimate of the 
used spin factor for each given target-nucleus.

\subsubsection{The quenching factors}
\label{sc:qf}

\begin{figure}[ht]
\vspace{0.6cm}
\centerline{\hbox{ \psfig{figure=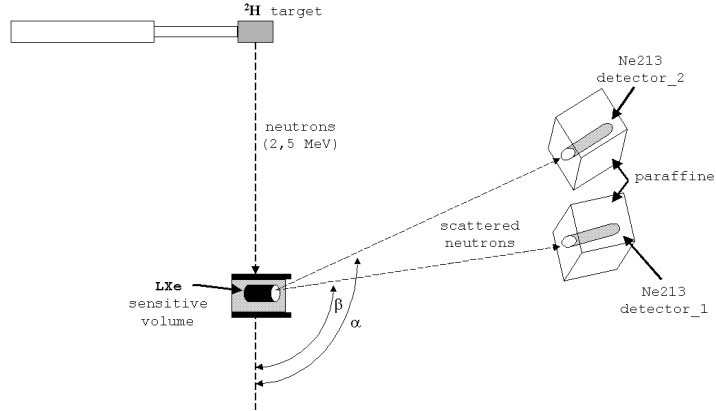,height=5.5cm} }}
\vspace{0.3cm}
\caption{Example of a set-up for the measurement of the 
recoil/electron response ratio of a detector. This example is taken from ref.
\cite{Qf01}.}
\label{qf_setup}
\end{figure}

The proper knowledge of other quantities is also necessary for
a WIMP direct search such as e.g. the recoil/electron response ratio
for the given nucleus in the given detector (named {\em quenching factor, $q$}).
The  recoil/electron response ratio
can be measured with a neutron source or at a neutron generator (see as
an example ref. \cite{Qf01}).         
In the latter case a set-up similar to that reported in Fig. \ref{qf_setup}
can be used on a 
quasi-monochromatic neutron beam tagging the scattered neutrons.
Let us comment that while this configuration works out well for the
measurements of recoil/electron response ratio, it is unsuitable 
to determine the absolute detection efficiency for recoils
because of the uncertain knowledge of the elastic cross sections, 
of the background due to inelastic 
scatterings, of the electromagnetic and hadronic background from neutron interaction
in the environment, of the cuts for noise rejection in a very high rate environment, 
of the duty cycle (which is very small), etc. 

Of course, significant differences are often present
in literature for the measured value of this recoil /electron response
ratio for the same nucleus 
in similar detectors as it can be clearly deduced from Table \ref{tb:cli1},
where the quenching
factors measured for the detectors most commonly used in this field are reported.
\begin{table}[!ht]
\caption{Quenching factor, $q$, values
measured by using neutron sources or neutron beams for some detectors and nuclei.
When a significant $q$ dependence on the energy has been explicitely given,
only the maximum and the 
minimum central values are reported for the given 
energy interval without quoting here -- for simplicity -- the associated errors.}
\vspace{-0.2cm}
\begin{center}
\begin{tabular}{|c|c|c|c|}
\hline \hline
& & &\\
 Nucleus/Detector & Recoil Energy (keV) & $q$ & Reference \\
& & &\\
\hline 
NaI(Tl)    &  (6.5-97) & (0.30 $\pm$ 0.01) for Na   & \cite{Psd96}  \\
           &  (22-330) & (0.09 $\pm$ 0.01) for I    & \cite{Psd96}  \\
           &  (20-80)  & (0.25 $\pm$ 0.03) for Na   & \cite{Ger99}  \\
           &  (40-100) & (0.08 $\pm$ 0.02) for I    & \cite{Ger99}  \\
           &  (4-252)  & (0.275 $\pm$ 0.018) for Na & \cite{Spo98} \\
           &  (10-71)  & (0.086 $\pm$ 0.007) for I  & \cite{Spo98} \\
           &  (5-100)  & (0.4   $\pm$ 0.2)   for Na & \cite{Fu93} \\
           &  (40-300) & (0.05  $\pm$ 0.02)  for I  & \cite{Fu93} \\
\hline 
CaF$_2$(Eu)&  (30-100) & (0.06-0.11) for Ca         & \cite{Spo98} \\
           &  (10-100) & (0.08-0.17) for F          & \cite{Spo98} \\
           &  (90-130) & (0.049 $\pm$ 0.005) for Ca & \cite{Caf94} \\
           &  (75-270) & (0.069 $\pm$ 0.005) for F  & \cite{Caf94} \\
           &  (53-192) & (0.11-0.20) for F  & \cite{Haz01} \\
           &  (25-91)  & (0.09-0.23) for Ca & \cite{Haz01} \\
\hline 
CsI(Tl)    &  (25-150) & (0.15-0.07)                & \cite{Pe99} \\
           &  (10-65)  & (0.17-0.12)                & \cite{Ku00} \\
           &  (10-65)  & (0.22-0.12)                & \cite{Kim00} \\
\hline 
CsI(Na)    &  (10-40)  & (0.10-0.07)                & \cite{Kim00} \\
\hline 
Ge         &  (3-18)   & (0.29-0.23)& \cite{Mes95} \\
           &  (21-50)  & (0.14-0.24) & \cite{Sa66} \\
           &  (10-80)  & (0.18-0.34) & \cite{Cashman} \\
           &  (20-70)  & (0.24-0.33) & \cite{Shu} \\
\hline 
Si         &  (5-22)   & (0.23-0.42) & \cite{Ge90} \\
           &    22     & (0.32 $\pm$ 0.10) & \cite{Sa65} \\
\hline 
Liquid Xe  &  (30-70)  & ( 0.46 $\pm$ 0.10) & \cite{Qf01} \\
           &  (40-70)  & (0.18 $\pm$ 0.03) & \cite{Arn00} \\
           &  (40-70)  & (0.22 $\pm$ 0.01) & \cite{Aki01} \\
\hline 
Bolometers &   -        & none available & \\
 &     & at time of writing this paper & \\
&& (assumed 1) &             \\
\hline \hline
\end{tabular}
\label{tb:cli1}
\end{center}
\vspace{-0.6cm}
\end{table}
As it can be seen, significant differences in the measured values are
present also 
for the same nucleus in the same kind of detectors. 
This is generally
due to different peculiarities of the detectors themselves, 
besides possible
additional experimental uncertainties. 
For example, in doped scintillator it can depend on the dopant 
concentration, in liquid Xenon on the residual trace 
contaminants due to specific experimental features (such as
the initial purity of the used Xenon gas, the inner surface treatment,
the level of vacuum reached before filling, the used purification
line components and the degassing/release features of all the materials 
of the inner vessel) and in Ge or Si on
impurities, etc.
Moreover, some dependence of the recoil /electron response
ratio on the energy has been quoted in several cases (see Table \ref{tb:cli1}).

\vspace{0.3cm}

As far as regards the bolometers, no direct measurement of the recoil
energy of the target-nucleus has been reported up to now by any of the groups
involved in this activity, although several bolometers have been irradiated with
neutrons along the past decade.
For the sake of completeness, we remind that
a measurement of the response of a TeO$_2$ bolometer to surface $^{224}$Ra 
recoiling nuclei has been reported in ref. \cite{Cuore}; 
this measurement, although its importance, does not represent a determination of the
quenching factor of the target- (either Te or O) nuclei of the TeO$_2$ bolometer.
In fact, the recoiling nuclei are not the Te and O ones and  
the ``external'' recoils are generated on the detector surface 
where the sensors are located and, thus, do not
involve the response of the whole bulk of the target-detector.
Anyhow, these values cannot of course be extended to whatever kind of bolometer.

\subsubsection{Some miscellaneous}
\label{sc:misc}

Besides the uncertainties already
discussed, there exist a large number
of experimental details that have to be considered in 
the calculations of the signal expectations.
For example, it must be suitably studied, checked, monitored and discussed
the role played by the stability of the energy scale,
by its right evaluation (see also \S \ref{sc:qf}), 
by a reliable identification of the energy threshold
and of the residual noise above it (as it can be
effectively done in
NaI(Tl) detectors with adequate number of photoelectrons/keV,
see e.g. \cite{Nim98}), by external veto (especially when 
``high'' rate anticoincidences are used) and by the rejection procedures 
used in some experiments to filter the data, etc..
In the case of DAMA/NaI details have been given along the last about ten years
and published (see e.g. \cite{Nim98,Mod2,Mod3,Sist}).

Finally, let us remind that --
when results obtained by using different target nuclei are considered --
also the effect of the uncertainty on the scaling laws of the nuclear cross sections
to the {\it reference} ones (e.g., generally, on nucleon) must be taken into account.
In fact, it is common practice to use the WIMP-nucleon cross sections $\sigma_{SI}$ 
and $\sigma_{SD}$ and the scaling laws reported in \S \ref{sc:rate1}; but --
in principle -- other scaling laws cannot be excluded at the present 
knowledge of the real nature of a WIMP candidate. This can be an additional
uncertainty in quests for a candidate and in comparisons among experiments using 
different target nuclei.

\subsubsection{Priors}
\label{sc:pri}

It is common practice in extracting physical information from the data to account for 
related priors. 

In particular, in the quest for the candidate particle two main priors have been considered
in the DAMA/NaI first quests for a candidate \cite{Mod1,Mod2,Ext,Mod3,Sisd,Inel,Hep}.

The first prior, which has been properly included in ref. \cite{Mod3,Sisd,Inel,Hep} and also considered in the
following, accounts for the upper limits
measured on the recoil fractions in the data of the DAMA/NaI-0
running period
\cite{Psd96,Sist}, which was carefully and especially devoted to such an investigation. 

The second prior regards the  mass limit for supersymmetric candidates, achieved
-- within some assumed model
frameworks -- by experiments at accelerators.
In particular, because of this prior, WIMP masses above 30 GeV (25 GeV in ref.
\cite{Mod1}) have been investigated
in refs. \cite{Mod2,Mod3,Sisd,Inel,Hep} for few (of the many possible) model frameworks.
Specifically, it accounted for the lower bound on the
neutralino mass as derived from the LEP data in the adopted supersymmetric
schemes based on GUT assumptions \cite{Dpp0}. However, other model assumptions are possible and
would imply significant variations of some accelerators bounds. As an example, we mention
the recent ref. \cite{Bo03} where the assumption on the gaugino-mass
unification at GUT scale has been released \footnote{
In this case also neutralino masses down to $\simeq$ 6 GeV are possible, this lower bound being
determined by current upper limit on relic abundance for cold dark matter (somewhat  higher values
of 15-18 GeV are obtained for the neutralino mass lower bound, if Higgs masses are assumed large
(~ 1 TeV) \cite{lowm,bbpr}.}.
The development of these schemes  is very interesting since -- as well known -- DAMA/NaI is
intrinsically sensitive both to low and high WIMP mass having both a light
(the $^{23}$Na) and a heavy (the $^{127}$I) target-nucleus.

However, still following the present model dependent results quoted by LEP in the
supersymmetric schemes based on GUT assumptions 
the considered lower bound is at present
37 GeV \cite{Dpp} \footnote{Higher limits are available for other
more constrained models. These latter ones 
as well as possible future increase of the present 37 GeV lower bound 
would further select - for these scenarios - the possible models
in the quest for the candidate from the DAMA/NaI data
favouring, in particular, halo models with small local velocity and/or co-rotation.}.
It worth to note that this mass limit selects the 
WIMP-Iodine elastic scattering as dominant because of the adopted
scaling laws and of kinematical arguments. 

\vspace{-0.5cm}
\subsection{Results on the quest for a candidate in some of the possible 
model frameworks}
\label{sc:que}
\vspace{-0.2cm}

Just as a corollary of the model independent result given in \S \ref{sc:evi},
in the following some of the many possible model dependent quests for a WIMP candidate
is carried out using the data
collected during all the seven annual cycles and 
considering all the halo models summarized in \S \ref{sc:halo} for three of the
possible values 
of the local velocity $v_0$: 170 km/s, 220 km/s and 270 km/s.
The used halo density follows the prescriptions of \S \ref{sc:halo}.VI.
The escape velocity has been maintained at the fixed value: 650 km/s;
of course, it is worth to note that the
present existing uncertainties affecting the knowledge of the escape velocity will significantly
extend allowed regions e.g. in the cases of {\em
preferred inelastic} WIMPs and of light mass 
WIMP candidates; its effect would be instead marginal at large WIMP masses (see
e.g. the case
for exclusion plots given in Fig. \ref{fg:caf94}).

In particular, possible scenarios have been
exploited for the halo models described in \S \ref{sc:halo} in 
some discrete cases either considering
the mean values of the parameters of the used nuclear form factors and 
of the measured quenching factors (case $A$) or adopting the same procedure as in
refs. 
\cite{Sisd,Inel} \footnote{that is, by varying either: i) the mean values of the 
measured $^{23}$Na and
$^{127}$I quenching factors \cite{Psd96} 
up to +2 times the errors; ii) the nuclear radius, $r_n$,
and the nuclear surface thickness parameter, $s$, in the SI Form 
Factor \cite{Helm56} from their central values down to -20\%;
iii)  the $b$ parameter in the considered SD form factor
from the given value \cite{res97} down to -20\%.} (case $B$) or
in one of the possible more extreme cases where the Iodine nucleus
parameters are fixed at the values of case $B$, while 
for the Sodium nucleus one considers:
i) $^{23}$Na quenching factor at the lowest value measured in literature (see Table \ref{tb:cli1}); 
ii) the nuclear radius, $r_n$,
and the nuclear surface thickness parameter, $s$, in the SI Form
Factor \cite{Helm56} from their central values up to +20\%;
iii)  the $b$ parameter in the considered SD form factor
from the given value \cite{res97} up to +20\%
(case $C$). 

In the following sections, for simplicity, the results of these corollary quests for a
candidate particle
is presented in terms of allowed regions
obtained as superposition of the configurations corresponding
to likelihood function values {\it distant} more than $4\sigma$ from
the null hypothesis (absence of modulation) in each of the several 
(but still a limited number) of the possible 
model frameworks considered here. Priors have been discussed in \S \ref{sc:pri}.

Obviously, larger  
sensitivities than those reported in the following would be reached when including
the effect of other
existing 
uncertainties on assumptions and related parameters, as it can be also inferred from the previous
sections.

\subsubsection{WIMPs with mixed SI\&SD interaction in some of the possible model frameworks}
\label{sc:sisd}

The most general scenario of WIMP nucleus elastic interaction, to which the DAMA/NaI target 
nuclei are fully sensitive,
is the one  where both the SI and the SD components of the cross section (see \S \ref{sc:rate1}) 
\begin{figure}[!ht]
\begin{center}
\vspace{-0.8cm}
\epsfig{figure=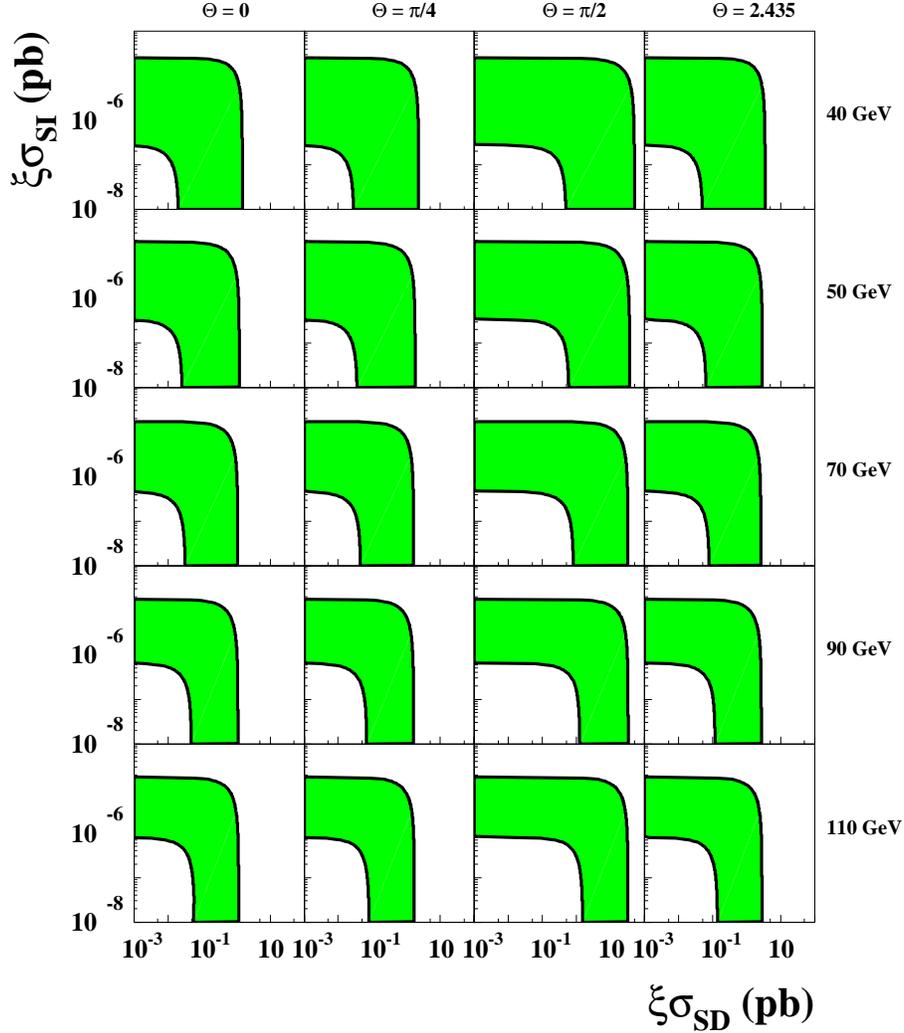, height=14cm}
\end{center}  
\vspace{-0.6cm}
\caption{{\it A case of a WIMP with mixed SI\&SD interaction in the model
frameworks given in the text}. Colored areas: example of slices (of
the allowed volume) in the plane 
$\xi \sigma_{SI}$ vs
$\xi \sigma_{SD}$ for some of the possible $m_W$ and $\theta$ values.
See \S \ref{sc:que}.
Inclusion of other existing uncertainties on parameters and models (as previously 
discussed to some extent in this paper) would further extend the regions; for example,
the use of more favourable form factors than those we considered here (see \S \ref{sc:ff}) 
alone would move them towards lower cross sections.}
\label{fg:pan_sisd}
\vspace{-0.3cm}
\end{figure}
are present. Thus, as first we introduce here the case for a
candidate with both SI and SD couplings to ordinary matter similarly as in ref. \cite{Sisd}.

As already described in \S \ref{sc:rate1}, 
in this most general scenario the space of the free parameters is 
a 4-dimensional volume defined by $m_W$, $\xi \sigma_{SI}$,
$\xi \sigma_{SD}$ and $\theta$ (which varies from 0 to $\pi$).
Thus, the general solution would be a four dimensional allowed volume
for each considered model framework.
Since the graphic representation of this allowed volume is quite difficult, 
we show in Fig. \ref{fg:pan_sisd} the obtained regions in the plane 
$\xi \sigma_{SI}$ vs $\xi \sigma_{SD}$ for some of the possible 
$\theta$ and $m_W$ values in the model frameworks considered here.
In particular, we report just four couplings, which correspond to the
following values of the mixing angle $\theta$: i)  $\theta$ = 0 ($a_n$
=0 and $a_p \ne$ 0 or  $|a_p| >> |a_n|$) corresponding to a particle
with null SD coupling to neutron; ii) $\theta = \pi/4$ ($a_p = a_n$)
corresponding to a particle with the same SD coupling to neutron and
proton; iii)  $\theta$ = $\pi/2$ ($a_n \ne$ 0 and $a_p$ = 0 
or  $|a_n| >> |a_p|$) corresponding to a particle with null SD
couplings to proton; iv) $\theta$ = 2.435 rad ($ \frac {a_n} {a_p}$
= -0.85) corresponding to a particle with SD coupling through $Z_0$
exchange. The case $a_p = - a_n$ is nearly similar to the case iv).  

\begin{figure}[!ht]
\begin{center}
\vspace{-1.3cm}
\epsfig{figure=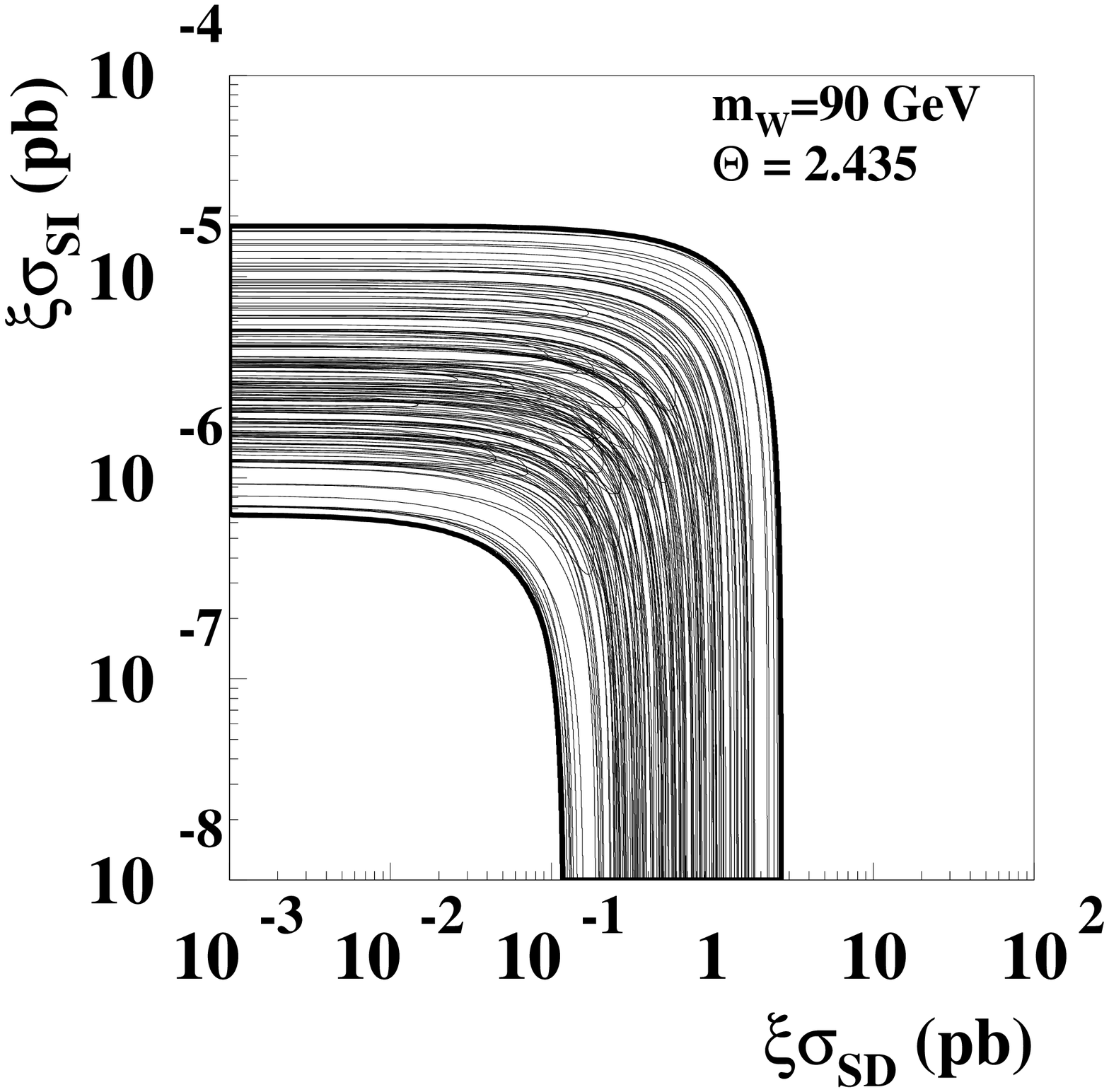,height=7.cm}
\end{center} 
\vspace{-0.8cm}
\caption{This figure explicitely shows all the slices (of the allowed volume) in the plane
$\xi \sigma_{SI}$ vs $\xi \sigma_{SD}$ obtained for
$m_W = 90$ GeV and $\theta = 2.435$ (pure $Z_0$ coupling)
for each one of the considered model frameworks (see \S \ref{sc:que}).
The region included in between the two extreme lines 
is that shown for the same $m_W$ and
$\theta$ values in Fig. \ref{fg:pan_sisd} (see also the related caption). }
\label{fg:fig_exsd}
\vspace{-0.2cm}
\end{figure}

To offer an example of how the allowed regions have been built, 
Fig. \ref{fg:fig_exsd} shows explicitely the superposition of the slices 
obtained for each one of the model frameworks considered here 
in the particular case of  $m_W = 90$ GeV and 
$\theta = 2.435$ (pure $Z_0$ coupling). 

From the given figures it is clear that at present either a purely SI or a purely 
SD or a mixed SI\&SD configurations
are supported by the experimental data of the seven annual cycles. 

Some other comments related to the effect of a SD component different from zero will be 
also addressed in the following.

\subsubsection{WIMPs with dominant SI interaction in some of the possible model frameworks}

Generally, mainly the case of purely
spin-independent coupled WIMP is considered in literature. In fact, often the 
spin-independent interaction
with ordinary matter is assumed to be dominant since e.g.
most of the used target-nuclei are practically
not sensitive to SD interactions (as on the contrary
$^{23}$Na and $^{127}$I are) and the theoretical calculations 
are even much more complex and uncertain.

Thus, following an analogous procedure as for the previous case, we have exploited for the same model
frameworks the purely SI scenario alone. 
In this case the free parameters are two: $m_W$ and $\xi \sigma_{SI}$. 

\begin{figure}[!ht]
\vspace{-0.2cm}
\begin{center}
\epsfig{figure=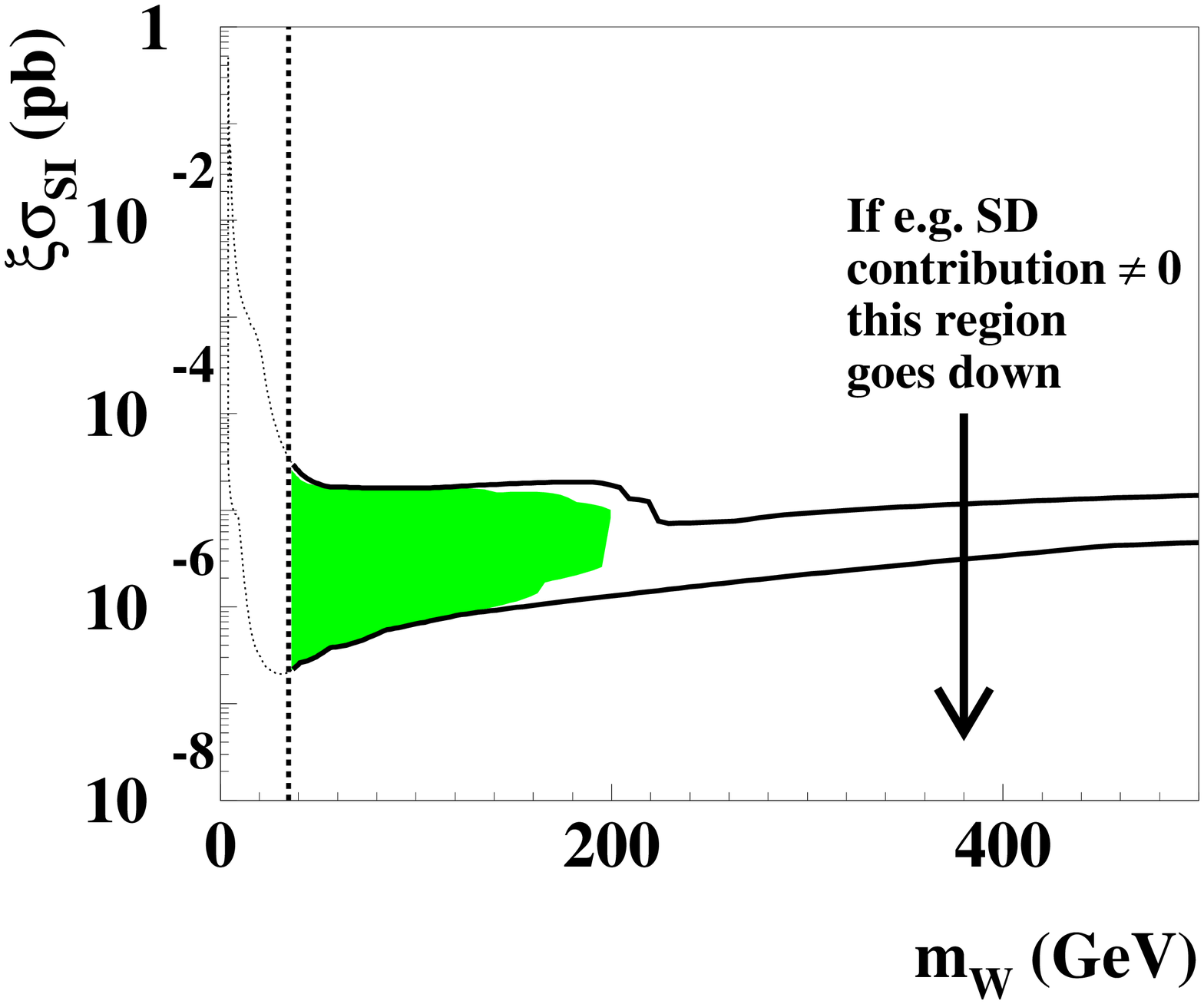, height=6.cm}
\end{center}  
\vspace{-0.8cm}
\caption{{\it Case of a WIMP with dominant SI interaction for the model
frameworks given in the text.} Region allowed in the plane ($m_W$, $\xi
\sigma_{SI}$).
See \S \ref{sc:que}; the vertical dotted line represents the model dependent prior discussed in 
\S \ref{sc:pri}. 
The area at WIMP masses above 
200 GeV is allowed for low local velocity -- $v_0$=170km/s -- and all considered
sets of parameters
by the Evans' logarithmic $C1$
co-rotating halo model, by the Evans' logarithmic $C2$
co-rotating halo model, by the triaxial $D2$ and $D4$ non-rotating halo models and
also by the Evans power-law $B3$ model with parameters of the set A).
The inclusion of other existing uncertainties on parameters and models (as previously 
discussed to some extent in this paper) would further extend the region; for example, 
the use of more favourable SI form factor for Iodine (see \S \ref{sc:ff})
alone would move it towards lower cross sections.}
\label{fg:fig_puresi}
\vspace{-0.1cm}
\end{figure}

In Fig. \ref{fg:fig_puresi}  the region allowed in the plane $m_W$ and $\xi
\sigma_{SI}$ for the considered model frameworks is reported. 
The vertical dotted line represents the model dependent prior discussed in 
\S \ref{sc:pri}, that is
the present lower bound on supersymmetric candidate
as derived from the LEP data in
supersymmetric scheme with gaugino-mass unification
at GUT (see \S \ref{sc:pri}). The configurations below the vertical line can be of interest 
for neutralino when other schemes are considered (see \S \ref{sc:pri}) and for generic WIMP 
candidate.
As shown in Fig. \ref{fg:fig_puresi}, also WIMP masses above 
200 GeV are allowed, in particular, for every set of parameters' values when
considering low local velocity
and: i) the Evans' logarithmic $C1$ and $C2$
co-rotating halo models; ii) the triaxial $D2$ and $D4$ non-rotating halo models;
iii) the Evans power-law $B3$ model, but only with parameters as in set A). 

Of course, best fit values of cross section and WIMP mass 
span over a large range depending on the model framework. Just as an example,
in the triaxial D2 halo model with maximal $\rho_0$,
$v_0 = 170$ km/s and parameters as in the case $C$,
the best fit values are $m_W = (74^{+17}_{-12})$ GeV
and $\xi \sigma_{SI} = (2.6 \pm 0.4) \cdot 10^{-6}$ pb.

\vspace{0.4cm}
\noindent{\em \bf Effect of a SD component different from zero on allowed SI regions}
\vspace{0.4cm}

\begin{figure}[htbp]
\begin{center}
\vspace{-0.5cm}
\epsfig{figure=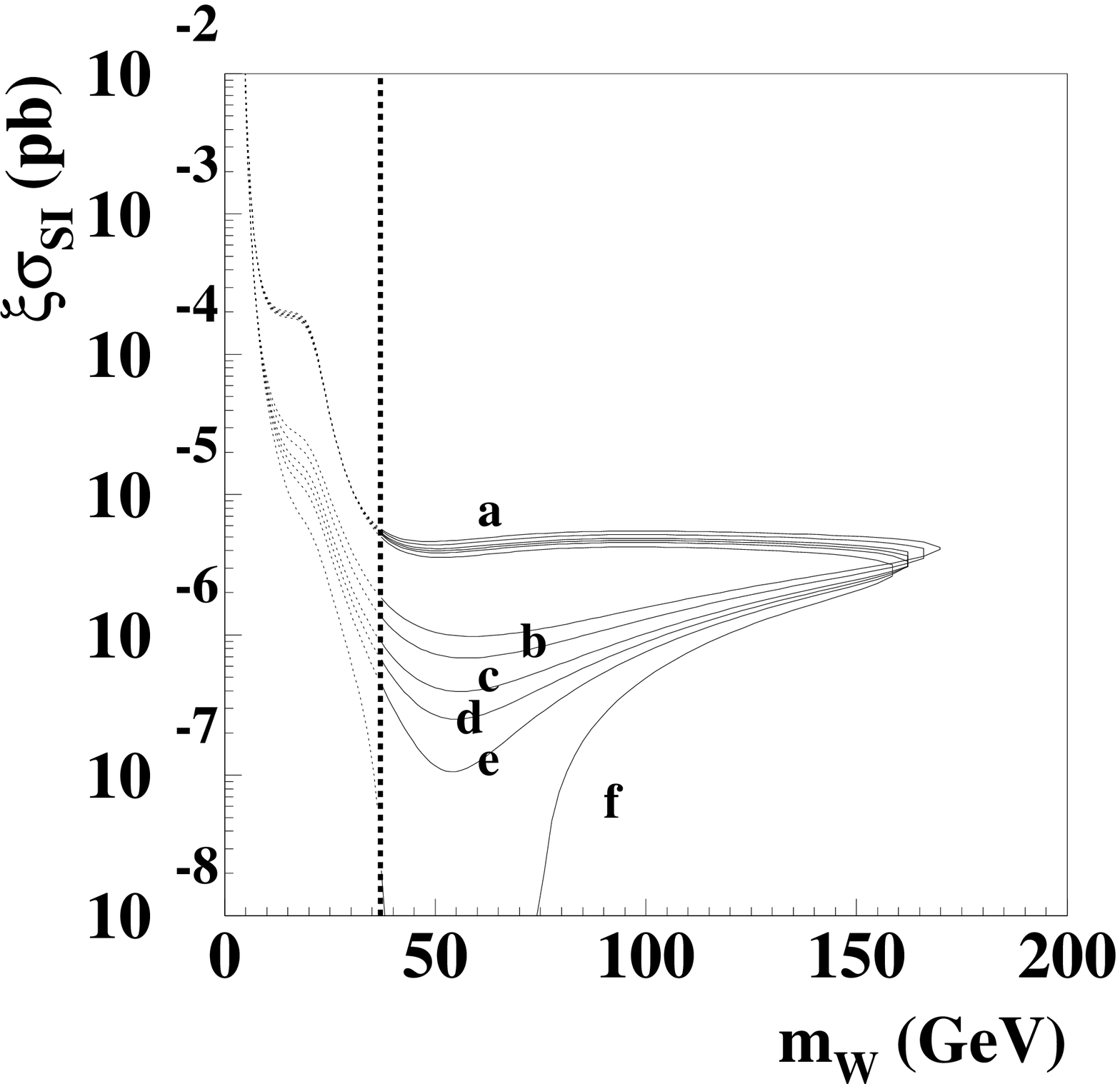, height=6.5cm}
\end{center}  
\vspace{-0.8cm}
\caption{Example of the effect induced by the inclusion of a SD component
different from zero on the 
allowed regions given in the plane $\xi\sigma_{SI}$ vs $m_W$.
In this example the
Evans' logarithmic axisymmetric $C2$ halo model with
$v_0 = 170$ km/s, $\rho_0$ equal to the maximum value for this model 
(see Table \ref{tb:rho}) and the set of parameters $A$ have been considered. 
The different regions refer to different SD contributions for the particular case of 
$\theta = 0$:
$\sigma_{SD}=$ 0 pb (a), 0.02 pb (b), 0.04 pb (c), 0.05 pb (d),
0.06 pb (e), 0.08 pb (f). See \S \ref{sc:que}; the vertical dotted line represents the model dependent 
prior discussed in \S \ref{sc:pri}.}
\label{fg:fig_sifsd}
\vspace{-0.2cm}
\end{figure}

Let us now point out, in addition, that 
configurations with $\xi \sigma_{SI}$ even much lower than those shown in
Fig. \ref{fg:fig_puresi} would be accessible also if an even small SD contribution would
be present in the interaction as described in \S \ref{sc:sisd}. 
This possibility is clearly pointed out
in Fig. \ref{fg:fig_sifsd} where an example of  regions in the plane
($m_W$, $\xi \sigma_{SI}$) corresponding to different SD contributions
are reported for the case $\theta =0$.
In this example the
Evans' logarithmic axisymmetric $C2$ halo model with
$v_0 = 170$ km/s, $\rho_0$ equal to the maximum value for this model 
(see Table \ref{tb:rho}) and the set of parameters $A$ have been considered.
The values of $\xi\sigma_{SD}$ 
range there from 0 to 0.08 pb. As it can be seen,
increasing the SD contribution the regions allowed in the ($m_W$, $\xi \sigma_{SI}$) plane involve
SI cross sections much lower than 
$1 \times 10^{-6}$ pb. It can be noted that for $ \sigma_{SD} \geq 0.08$
pb the annual modulation effect observed is also compatible -- for $m_W \simeq 40-75$ GeV -- with a
WIMP
candidate with no SI interaction at all in this particular model framework. 

These arguments clearly show that also a relatively small SD contribution 
can drastically change the allowed region in the  ($m_W$, $\xi \sigma_{SI}$) 
plane; therefore, e.g. there is not meaning in the bare comparison between 
regions allowed in experiments that are also sensitive to SD coupling 
and exclusion plots achieved by experiments that are not.
The same is when comparing regions allowed by experiments whose target-nuclei have unpaired proton
with exclusion plots quoted by experiments using target-nuclei with unpaired neutron 
when the 
SD component of the WIMP interaction would correspond 
either to $\theta \simeq 0$ or $\theta \simeq \pi.$

\subsubsection{WIMPs with dominant SD interaction in some of the possible model frameworks}

Let us now focus on the case of a candidate with purely spin-dependent 
\begin{figure}[!ht]
\begin{center}
\vspace{-0.8cm}
\epsfig{figure=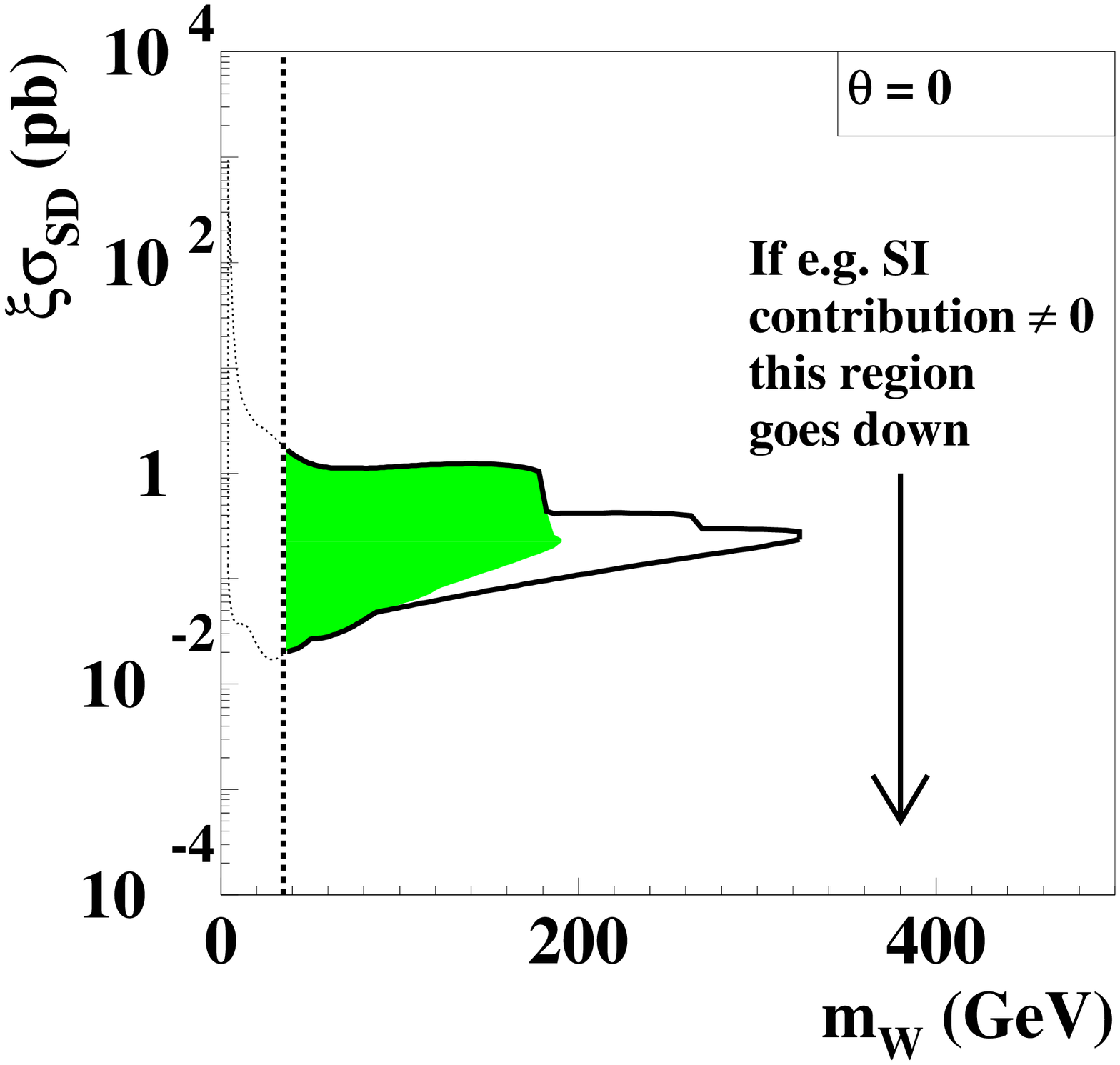, height=6cm}
\epsfig{figure=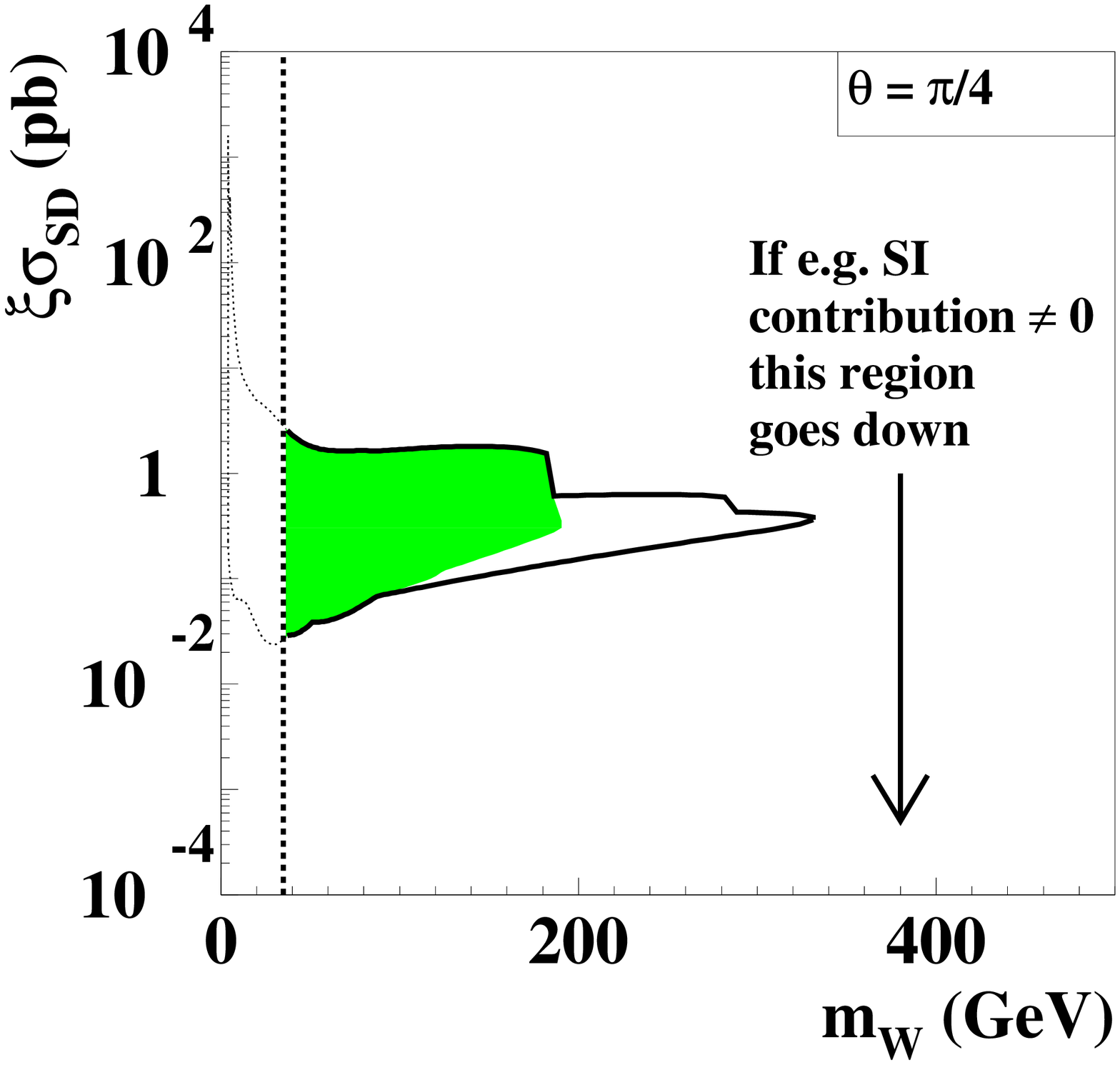, height=6cm}
\epsfig{figure=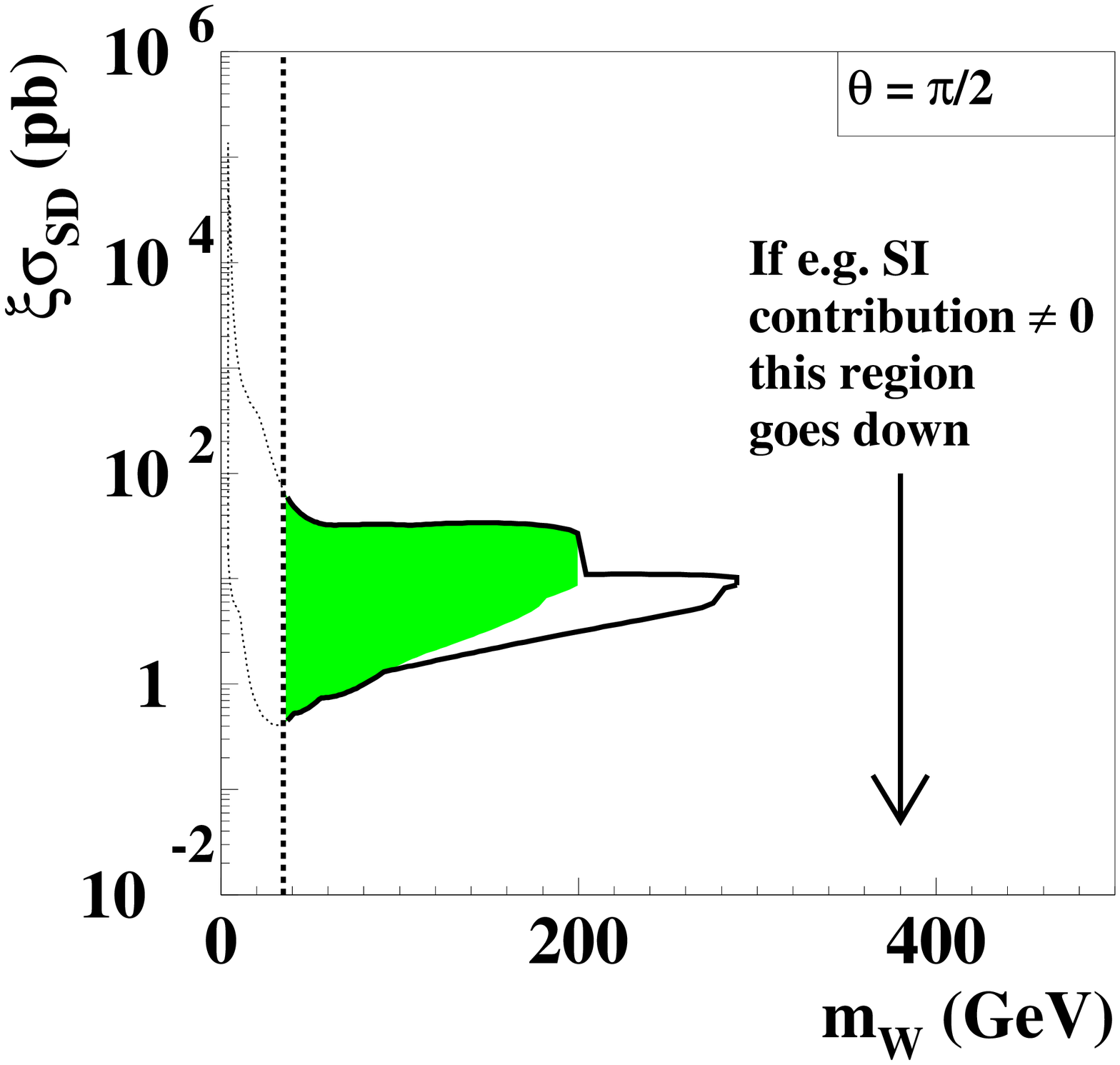, height=6cm}
\epsfig{figure=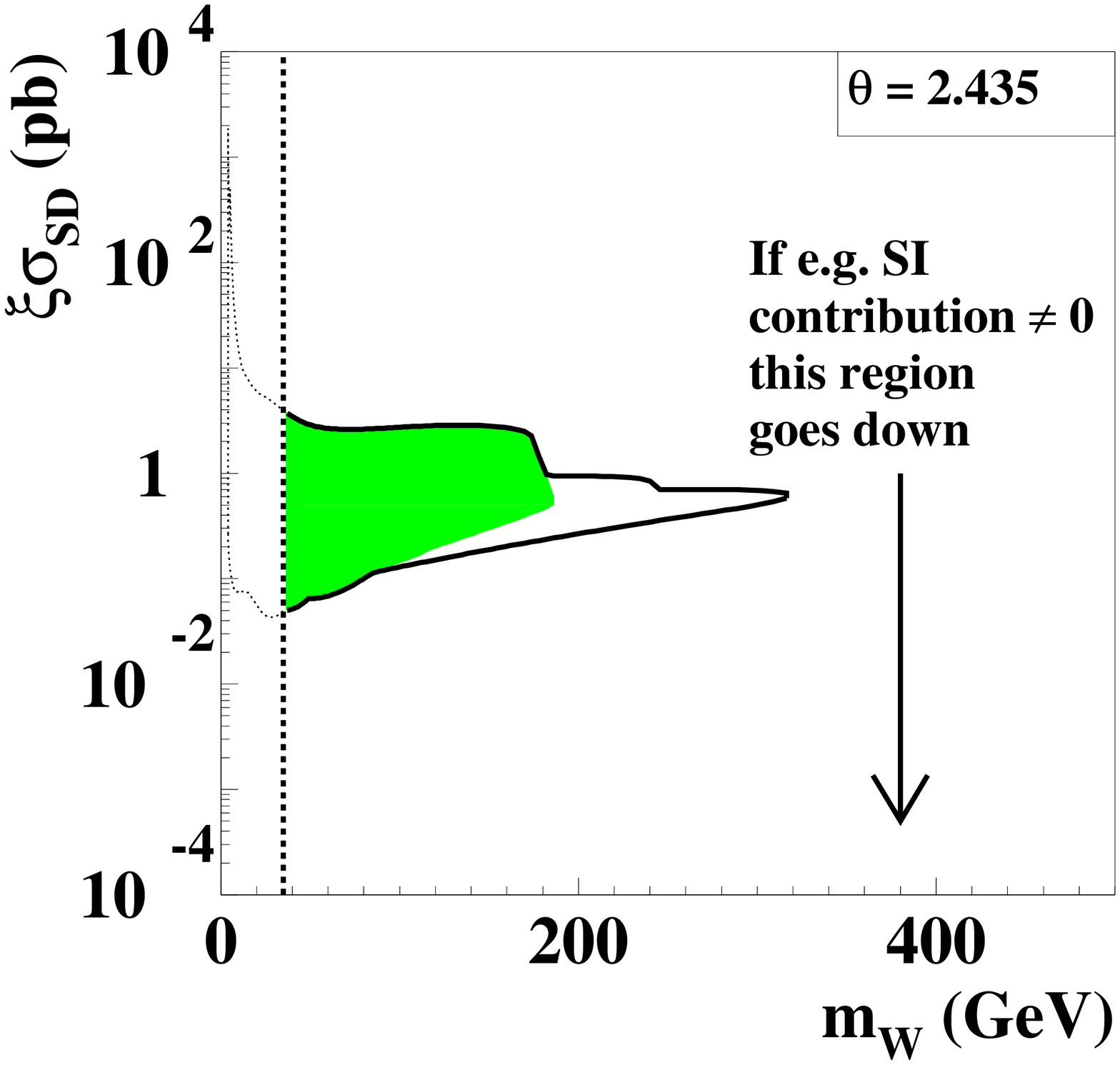, height=6cm}
\end{center}  
\vspace{-0.8cm}
\caption{{\it Case of a WIMP with dominant SD interaction in the model
frameworks given in the text.} 
Regions allowed in the plane ($m_W$, $\xi \sigma_{SD}$).
See \S \ref{sc:que}; the vertical dotted line represents the model dependent prior discussed in
\S \ref{sc:pri}.
The panels refer to only few particular cases for 
$\theta$ (which can instead vary between 0 and $\pi$). 
The area at WIMP masses above
200 GeV is allowed for low local velocity -- $v_0$=170km/s -- and all considered
sets of parameters
by the Evans' logarithmic $C2$
co-rotating halo model.
Inclusion of other existing uncertainties on parameters and models (as previously 
discussed to some extent in this paper) would further extend the regions; for example, 
the use of more favourable SD form factors (see \S \ref{sc:ff})
alone would move them towards lower cross sections.}
\label{fg:fig_puresd}
\end{figure}
coupling to which DAMA/NaI is 
-- as mentioned --  fully sensitive.

When the SD component is different from zero, 
a very large number of possible configurations is available
(see \S \ref{sc:rate1}).
In fact, in this scenario the space of free parameters 
is a 3-dimensional volume defined by $m_W$, $\xi \sigma_{SD}$ and 
$\theta$ (which can vary from 0 to $\pi$).
Here, for simplicity as already done in \S \ref{sc:sisd},
we show the results obtained 
only for 4 particular couplings, which correspond to the
following values of the mixing angle $\theta$: i)  $\theta$ = 0 ($a_n$
=0 and $a_p \ne$ 0 or  $|a_p| >> |a_n|$; ii) $\theta = \pi/4$ ($a_p = a_n$); 
iii)  $\theta$ = $\pi/2$ ($a_n \ne$ 0 and $a_p$ = 0 
or  $|a_n| >> |a_p|$; iv) $\theta$ = 2.435 rad ($ \frac {a_n} {a_p}$
= -0.85). 

Fig. \ref{fg:fig_puresd} shows the regions allowed in the plane ($m_W$,
$\xi\sigma_{SD}$) for the same model frameworks quoted above;
other configurations are possible varying the $\theta$ value.
The area at WIMP masses above
200 GeV is allowed for low local velocity -- $v_0$=170km/s -- and all considered
sets of parameters
by the Evans' logarithmic $C2$
co-rotating halo model.  

Moreover, the accounting
for
the uncertainties e.g. on the spin factors as well as
different possible formulations of the SD form factors would extend the allowed regions, e.g. towards
lower
$\xi\sigma_{SD}$ values.

\begin{figure}[ht]
\begin{center}
\vspace{-0.5cm}
\epsfig{figure=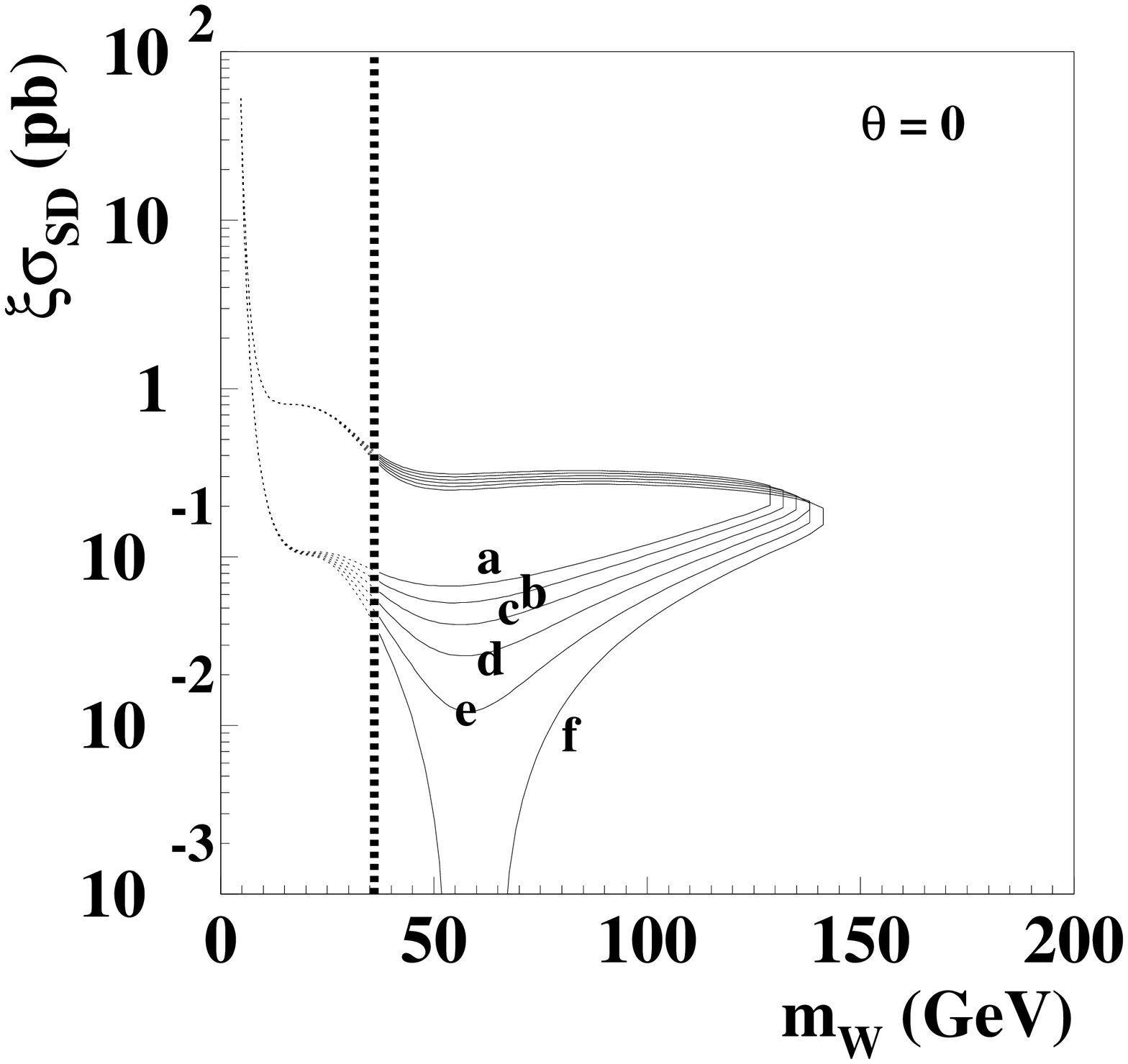, height=6.5cm}
\end{center} 
\vspace{-0.5cm}
\caption{Example of the effect induced by the inclusion of a SI component
different from zero on the
allowed regions in the plane $\xi\sigma_{SD}$ vs $m_W$.
In this example the
Evans' logarithmic axisymmetric $C2$ halo model with
$v_0 = 170$ km/s, $\rho_0$ equal to the maximum value for this model 
(see Table \ref{tb:rho}) and the set of parameters $A$ for $\theta = 0$
have been considered.
The different regions refer to different SI contributions with:
$\sigma_{SI}=$ 0 pb (a), $2 \times 10^{-7}$ pb (b), $4 \times 10^{-7}$ pb (c), 
$6 \times 10^{-7}$ pb (d), $8 \times 10^{-7}$ pb (e), $10^{-6}$ pb (f). 
See \S \ref{sc:que}; the vertical dotted line represents the model dependent prior discussed in
\S \ref{sc:pri}.}
\label{fg:fig_puresd2}
\end{figure}

Finally, $\xi \sigma_{SD}$ lower than those corresponding to the
regions 
shown in Fig. \ref{fg:fig_puresd} are possible also e.g. in case of an even small
SI contribution, as shown in Fig. \ref{fg:fig_puresd2}.

\subsubsection{WIMPs with {\em preferred} inelastic interaction in some of the possible model
frameworks}

An analysis considering the same model frameworks  has been carried out 
for the case of WIMPs with {\em preferred} inelastic interaction (see \S \ref{sc:inel}).

\begin{figure}[!htb]
\begin{center}
\vspace{-0.6cm}
\epsfig{figure=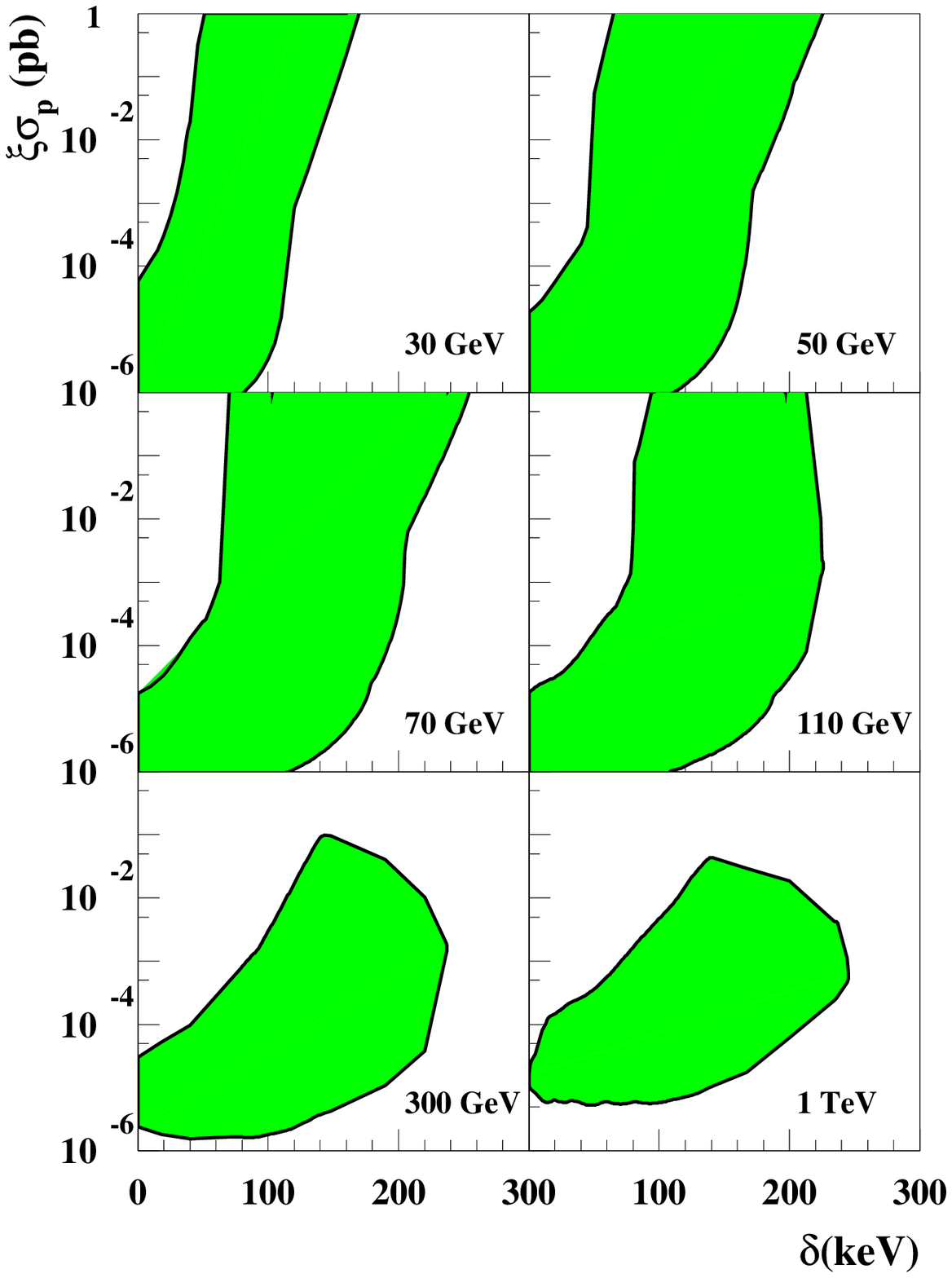, height=9cm}
\end{center}  
\vspace{-0.5cm}
\caption{{\it Case of a WIMP with preferred inelastic interaction
in the model frameworks given in the text.} Examples of
slices (colored areas) of the allowed volumes
($\xi \sigma_p$, $\delta$, $m_W$) for some $m_W$ values for the considered model frameworks
for WIMP with {\em preferred} inelastic interaction. 
See \S \ref{sc:que}. In these calculations $v_{esc}$
has been assumed at fixed value, while its present uncertainties would play a significant
role in the obtained results. Inclusion of other existing uncertainties on parameters and models (as previously 
discussed to some extent in this paper) would further extend the regions; for example, already
the use of a more favourable SI form factor for Iodine (see \S \ref{sc:ff})
alone would move them towards lower cross sections.}
\label{fg:fig_inel}
\end{figure}

In this inelastic Dark Matter scenario an allowed volume 
in the space ($\xi \sigma_p$,$m_W$,$\delta$) 
is obtained. For simplicity, Fig. \ref{fg:fig_inel} shows slices of such an allowed 
volume at some given WIMP masses.

There the superpositions
of the allowed regions obtained, when varying the model framework
within the considered set, are shown for each $m_W$. 
As a consequence, the cross section value 
at given $\delta$ can span over several orders of magnitude.
The upper border of each region is reached when 
$v_{thr}$ approximates the maximum WIMP velocity in the Earth frame
for each considered model framework. It can also be noted that when $m_W \gg m_N$, 
the expected differential energy spectrum is trivially dependent on $m_W$ 
and, in particular, it is proportional to the ratio between $\xi \sigma_p$ 
and $m_W$; therefore for very high mass the allowed region can be obtained straightforward.
We remind that in these calculations
$v_{esc}$
has been assumed at fixed value, while its present uncertainties can play a significant 
role in the scenario of WIMP with {\em preferred inelastic} scattering as mentioned in
\S \ref{sc:inel}. 

Note that each set of values (within those allowed by 
the associated uncertainties) for the previously mentioned 
parameters gives rise to a different expectation, thus to a different 
best fit values. 
As an example we mention the best fit values for 
$m_W$ = 70 GeV in the NFW B5 halo model with $v_0$ = 170 km/s,
maximal $\rho_0$ in this model and parameters as in case B): $(\delta = 86
^{+6}_{-8})$ keV and
$\xi \sigma_p = (1.2 \pm 0.2) \times 10^{-5}$ pb.

\subsubsection{Conclusion on the quest for a candidate in some of the possible model frameworks}

In this section the possible nature of a candidate, which could
account for the observed model independent evidence, has been investigated
by exploring -- 
as already done on the partial statistics
\cite{Mod1,Mod2,Ext,Mod3,Sist,Sisd,Inel,Hep} -- various kinds of possible
couplings 
and some (of the many) possible model frameworks.

We stress that, although several scenarios have been investigated, the 
analyses are not exhaustive at all of the existing possibilities because of 
the poor present knowledge on many astrophysical, nuclear and particle 
physics assumptions and related parameters as well as of the existing uncertainties in the
determination of some
experimental parameters which are necessary in the calculations.
For example, other parameters values can be considered for the
investigated halo models as well as other different halo models too, other form factors and related parameters,
other spin factors etc..
We remind that analogous uncertainties are present in every model dependent result
(such as e.g. exclusion plots and WIMP parameters from indirect searches);
thus, intrinsically, bare comparisons 
have always only a very relative meaning.

The discussion, carried out in this section, has also
allowed to introduce the main general arguments related to the model dependent 
calculations in WIMP direct searches.

\section{Conclusion}

In this paper general aspects of the Dark Matter direct search have been reviewed in the light
of the activity and results achieved by the DAMA/NaI experiment at the Gran Sasso National 
Laboratory of I.N.F.N..
DAMA/NaI has been a pioneer experiment running at LNGS for several years and 
investigating as first the WIMP annual modulation signature 
with suitable sensitivity and control of the running parameters. During seven independent 
experiments of one year each one, it has pointed out the presence of a modulation
satisfying the many peculiarities of a WIMP induced effect, reaching a significant evidence.
As a corollary result, it has also pointed out the complexity of the quest for a WIMP
candidate because of the present poor knowledge on the many astrophysical,
nuclear and particle physics aspects.

As regards other experiments  --
to have a realistic comparison -- experiments investigating with the same sensitivity
and control of the running condition the annual modulation signature are necessary.  
Of course, the target nuclei also play a crucial role,
since they can offer significantly different
sensitivities depending e.g. on the nature
of the WIMP particle and on their nuclear properties. 

The growing in the field of serious and independent efforts searching for WIMP
model independent signatures 
will certainly contribute to increase the knowledge in the field as well as efforts to more
deeply investigate models and parameters.

Some of the most competitive activities for the near future, exposing a significantly
large target-mass,
are starting at the Gran Sasso National Laboratory:  CUORICINO, 
GENIUS--TF (which will also be devoted to the investigation of double beta decay processes) 
and our new experiment DAMA/LIBRA. In fact, 
on our behalf, after the completion of the data taking of the $\simeq 100$ kg NaI(Tl)
set-up (on July 2002), as a result of our 
continuous efforts toward the creation of ultimate radiopure 
set-ups, the new DAMA/LIBRA 
has been installed (see ref. \cite{web}).
The LIBRA set-up is made by 25 NaI(Tl) detectors, 9.70 kg each one.
The new detectors have been realised thanks to a second generation R\&D
with Crismatec/SaintGobain company, by exploiting in particular new radiopurification
techniques of the NaI and TlI selected powders.
In the framework of this R\&D new materials have been selected, prototypes have been built and
devoted protocols have been fixed and used. The whole installation has largely been modified.
This new DAMA/LIBRA set-up, having a larger exposed mass 
and an higher overall radiopurity, will offer a significantly increased 
sensitivity to contribute to further efforts in improving the understanding 
of this field.

\section{Acknowledgements}

The authors take this opportunity to thank those who significantly contributed to the realization of the DAMA/NaI
experiment. In particular, they thank
the INFN Scientific Committee II for the effective support and control and 
the INFN -- Sezione Roma2, the INFN -- Sezione Roma, the Gran Sasso National Laboratory
and the IHEP/Beijing for the continuous assistance. They are also indebted to the referees of the experiment
in that Committee, that allowed them several times to improve the
quality of their efforts, and to the Directors and to the coordinators of the INFN involved
units for their support. 
They thank Dr. C. Arpesella, Ing.
M. Balata, Dr. M. Laubenstein, Mr. M. De Deo, Prof. A. Scacco and Prof. L. Trincherini,  for
their
contribution to sample measurements and related
discussions and Prof. I. R. Barabanov and Prof. G. Heusser for
many useful suggestions on the features of low radioactive detectors. They also 
wish to thank Dr. M. Amato, Prof. C. Bacci, Mr. V.
Bidoli, Mr. F. Bronzini, Dr. D.B. Chen,
Prof. L.K. Ding,  Dr. W. Di Nicolantonio, Dr. H.L. He, Dr. G. Ignesti, Dr. V.
Landoni, Mr. G. Ranelli, Dr. X.D. Sheng, Dr. G.X. Sun and Dr. Z.G. Yao for
their contribution
to the collaboration efforts in various periods and Dr. M. Angelone, Dr.
P. Batistoni and Dr. M. Pillon for their effective collaboration in the
neutron measurements at ENEA-Frascati. They thank Dr. R. McAlpine and Dr.
T. Wright, from EMI-THORN/Elec\-tron-Tu\-bes, for their competent assistance and the
Cris\-ma\-tec
com\-pa\-ny for the devoted efforts in the realization of the low background NaI(TI)
crystals. They also thank Mr.
A. Bussolotti and A. Mattei for their qualified technical
help and the LNGS, INFN -- Sezione di Roma and INFN -- Sezione di Roma2
mechanical and electronical staffs for support as well as 
the ACF, GTS, SEGEA staffs for the effective support in hardware works and
assistance. They thank Prof. A.
Bottino, Dr. F. Donato, Dr. N. Fornengo and Dr. S. Scopel for 
useful discussions on theoretical aspects.
Finally they are grateful to the dark matter community for the
continuous discussions about their work and to their families
for the patience and forbearance demonstrated in helping to manage them.

\end{document}